\documentclass[aps,prb,reprint,groupedaddress]{revtex4-2}

\usepackage{bm}
\usepackage{physics}
\usepackage{graphicx}
\usepackage{amssymb,comment}
\usepackage{mathrsfs}
\usepackage{xcolor}
\definecolor{forestgreen}{rgb}{0.13, 0.55, 0.13}
\definecolor{deepyellow}{rgb}{0.85, 0.7, 0.0}
\usepackage{longtable}
\usepackage[colorlinks=true,
            citecolor=blue,
            linkcolor=blue,
            urlcolor=blue]{hyperref}
\usepackage[normalem]{ulem}
\newcommand{\red}[1]{\textcolor{red}{#1}}

\definecolor{vividorange}{rgb}{1.0,0.5,0.0}

\definecolor{RED}{rgb}{1,0,0}

\begin{document}

%\preprint{}

\title{Order parameter scaling of chirality in structural phase transitions}
%\title{Chirality in Structural Phase Transitions}

\author{Keita Matsubara}
\email[]{matsubara-keita@ed.tmu.ac.jp}
\author{Kazumasa Hattori}
%\thanks{}
\affiliation{Department of Physics, Tokyo Metropolitan University,
1-1, Minami-osawa, Hachioji, Tokyo 192-0397, Japan}


\date{\today}

\begin{abstract}
Chirality, defined by the absence of improper spatial symmetries, has attracted considerable attention as a symmetry principle that enables characteristic electronic, transport, and lattice responses, including current-induced magnetization (CIM) and chiral-phonon splitting (CPS). Recent symmetry-based approaches have shown that structural chirality can be characterized by electric toroidal (ET) multipoles, with the ET monopole $G_0$ and quadrupole $G_u$ playing central roles in cubic and noncubic systems, respectively. In this paper, we investigate achiral-to-chiral (AtC) structural phase transitions driven by atomic displacements and construct $G_{0,u}$ as explicit functions of the displacement order parameter $\bm{\eta}$ using a group-theoretical approach. We show that the leading-order form of $G_{0,u}(\bm{\eta})$ is determined by the symmetry of the parent structure and the character of the displacive mode. This provides a symmetry-based classification of AtC transitions in which chirality is treated not merely as a binary property but as an induced structural order parameter with a characteristic dependence on $\bm{\eta}$. Through a systematic analysis of various types of $G_{0,u}(\bm{\eta})$, we find $|G_{0,u}(\bm{\eta})| \propto |\bm{\eta}|^n$ with $n=1,2,3,\cdots$, including high-order multipolar chirality such as $n=3,4$, and 6. We also discuss a distinct class of chirality, termed $k$-arm chirality, in which $G_{0,u}$ is formed from unequal invariant weights $|\bm{\eta}_{\bm{q}}|^2$ among different arms of the ordering wave vector $\bm{k}$. We demonstrate that, when the response is governed by the ET-multipole coupling, the leading displacement dependences of CIM and CPS follow the same power law as that of $G_{0,u}(\boldsymbol{\eta})$. We further clarify the microscopic mechanism by which AtC transitions generate chiral phonons and the associated CPS, and show that folded optical modes can also exhibit nonchiral symmetry-lowering splittings whose displacement dependence differs from that of $G_{0,u}(\boldsymbol{\eta})$. Since $\boldsymbol{\eta}$ evolves with temperature across the transition, the functional form of $G_{0,u}(\boldsymbol{\eta})$ governs the temperature dependence of ET-multipole-controlled CIM and CPS, providing a route to identify the symmetry of the underlying AtC transition.
\end{abstract}
%\keywords{}

\maketitle

\section{\label{sec:intro}Introduction}
Chirality refers to a property of an object whose mirror image cannot be superimposed on the original by proper rotations, as exemplified by the right and left hands of a human~\cite{Kelvin}. In crystalline solids, chirality is characterized by the absence of improper spatial symmetries, such as mirror and inversion operations. In particular, structural chirality is ubiquitous in nature and materials, ranging from molecular systems such as tartaric acid~\cite{Pasteur1848,Pasteur1848_full} and the double helix of DNA~\cite{WATSON1953} to inorganic crystals such as $\alpha$-quartz~\cite{Huggins_quartz,Bragg_quartz} and elemental Te~\cite{Bradley}. In condensed matter systems, chirality has long been discussed in relation to optical activity~\cite{BarronBook,Barron1986}, and has recently attracted renewed interest as a symmetry principle that enables cross-correlated responses among electronic, magnetic, transport, and lattice degrees of freedom~\cite{Fiebig2016,Kusunose_2022,Bousquet_2025_review}. Representative examples include magnetochiral anisotropy~\cite{Rikken1997MChD,Rikken2001EMChA,Pop2014EMChA}, chirality-induced spin selectivity (CISS)~\cite{Naaman2012CISS,Dieny2020}, current-induced magnetization (CIM) in chiral crystals~\cite{CIM_nature,CIM,Inui2020ChiralSpinPolarization,Kubo_murakami,roy2022long}, and electric-field-induced mechanical rotation and enantioselection in chiral crystals~\cite{oiwa_rot}. Interfaces between opposite chiral domains can also affect spin transport through spin-flip scattering~\cite{Matsubara2025}. These studies demonstrate that structural chirality is not merely a crystallographic classification but an active symmetry source that governs physical responses.

The lattice sector provides a particularly direct arena in which structural chirality becomes dynamical. In chiral crystals, phonon modes can carry angular momentum and exhibit handedness-dependent dispersions~\cite{Zhang_EdH,Kishine2020ChiralPhonon}. The resulting chiral-phonon splitting (CPS), i.e., the lifting of degeneracy between phonon modes with opposite circular polarization, has been theoretically studied in structurally chiral systems~\cite{tsunetsugu_kusunose,kusunose_cubic,Tateishi2025JPSJ}. Experimentally, chiral phonons have been identified in genuinely chiral bulk crystals by circularly polarized Raman scattering in $\alpha$-HgS \cite{Ishito} and Te~\cite{Ishito_Te}, and by circularly polarized resonant inelastic x-ray scattering in quartz~\cite{Ueda2023}. These experiments established chiral phonons as dynamical carriers of structural chirality in bulk crystals. Moreover, phonon angular momentum can couple to electronic spin and orbital degrees of freedom, providing a route for angular-momentum transfer between lattice vibrations and electrons~\cite{Hamada,Tateishi2025JPSJ}. Chiral phonons are also relevant from a topological perspective, through nontrivial band topology in phonon spectra~\cite{murakami_chiral_weyl}. Thus, chiral phonons provide a platform for connecting structural chirality, angular momentum, and topology in lattice dynamics.

Despite these advances, a fundamental issue remains unresolved: how to quantify chirality in a manner directly connected to structural degrees of freedom and experimentally accessible order parameters. The idea of describing chirality by continuous chiral parameters has a long history~\cite{Harris1999ChiralParameters}, and geometric chirality measures~\cite{CCM,Hausdorff,Buda1992QuantifyingChirality,Zabrodsky1995CCM,Helicity} provide useful ways to characterize the handedness of atomic arrangements. In parallel, symmetry-based multipole theory has provided a systematic description of chirality and related cross correlations~\cite{Toroidal,Multipole_review}. In particular, electric toroidal (ET) multipoles provide symmetry-adapted quantities for structural chirality, with the ET monopole $G_0$ and quadrupole $G_u$ playing central roles in cubic and noncubic systems, respectively~\cite{Inda,Oiwa_Kusunose_ET,Kishine2022DefinitionChirality}. A distinct but related development concerns relativistic electronic chirality, where chirality encoded in electronic wave functions is described using Dirac-field-based quantities~\cite{Hoshino_chiral} and evaluated quantitatively by first-principles calculations~\cite{Miki2025ElectronicAsymmetry}.

These electronic approaches clarify how chirality can be encoded in electronic wave functions, whereas the present work focuses on structural chirality generated by lattice distortions. Thus, how structural order parameters generate symmetry-adapted measures of chirality, such as ET multipoles, and how these quantities govern chirality-induced responses remain to be clarified. Recent theoretical work has further shown that chirality-related responses can even arise from purely electronic order without structural chirality~\cite{IshitobiHattori2026PEC}, highlighting the need to distinguish structural, electronic, and dynamical origins of chirality.

Structural achiral-to-chiral (AtC) phase transitions provide a natural setting for addressing this issue, because chirality emerges from symmetry-lowering lattice distortions, including polar, antipolar, and rotational modes. Such transitions have been discussed in systems such as K$_3$NiO$_2$~\cite{KNO_experi}, CsCuCl$_3$~\cite{CsCuCl}, MgTi$_2$O$_4$~\cite{MgTiO}, and Ba(TiO)Cu$_4$(PO$_4$)$_4$~\cite{Hayashida2021Ferrochiral}. In parallel, systematic searches for displacive pathways from achiral parent structures to chiral phases have begun to clarify how structural chirality can emerge in periodic solids~\cite{PathwaysChirality2025}. Recent studies of the $\beta$-pyrochlore oxide CsW$_2$O$_6$ and lacunar spinel GaNb$_4$Se$_8$ further illustrate that structural chirality can emerge in a subtle manner while retaining cubic symmetry~\cite{Okamoto2020,Kitou_chiral}. These examples show that chirality is often generated by collective structural degrees of freedom. Nevertheless, the functional relation between chirality and the structural order parameter remains largely unexplored, especially in connection with measurable responses such as CPS and CIM.

In this work, we address this issue by formulating the ET monopole $G_0$ and quadrupole $G_{u}$ directly in terms of the atomic displacements associated with structural AtC transitions. This formulation extends previous ET-multipole-based descriptions of structural chirality~\cite{Inda,Oiwa_Kusunose_ET,Kishine2022DefinitionChirality} by explicitly incorporating the displacement order parameter and also connects to related pseudoscalar descriptions of structural chirality, often discussed in terms of helicity~\cite{Moffat1992,Moffat2014,Helicity}, which have been employed in model systems undergoing AtC transitions~\cite{KNO_experi,CsCuCl,MgTiO}. In addition, it is complementary to recent structural approaches that quantify chirality or identify displacive pathways toward chiral phases in solids~\cite{Helicity,Bousquet_2025_review,PathwaysChirality2025}.
%It also connects to related pseudoscalar descriptions of structural chirality, often discussed in terms of helicity~\cite{Moffat1992,Moffat2014,Helicity}, which have been employed in model systems undergoing AtC transitions such as K$_3$NiO$_2$~\cite{KNO_experi}, CsCuCl$_3$~\cite{CsCuCl}, and MgTi$_2$O$_4$~\cite{MgTiO}. It is also complementary to recent structural approaches that quantify chirality or identify displacive pathways toward chiral phases in solids~\cite{GomezOrtiz2024SolidsChirality,Bousquet_2025_review,PathwaysChirality2025}. 
The present framework thereby establishes a direct link between crystallographic degrees of freedom and chirality-induced responses, enabling a unified analysis of lattice and electronic phenomena such as CPS~\cite{tsunetsugu_kusunose,kusunose_cubic} and CIM~\cite{CIM_nature,CIM}. Because the relevant structural order parameters can be determined by diffraction-based probes, as demonstrated in ferrochiral and cubic chiral systems~\cite{Hayashida2021Ferrochiral,Kitou_chiral}, the present framework provides an experimentally accessible way to quantify structural chirality.

%Since atomic displacements can be determined by x-ray and neutron diffraction~\cite{Cullity,Squires}, it provides an experimentally accessible way to quantify structural chirality.

A central finding of this study is that chirality is not merely a binary property, i.e., present or absent, but exhibits a characteristic functional dependence on the structural order parameter dictated by symmetry. More generally, the relation between a primary order parameter and a secondary or induced order parameter is itself a nontrivial problem in statistical mechanics. While some induced quantities follow scaling laws governed by the operator content and symmetry-allowed couplings of the primary critical mode~\cite{PrimarySecondaryOP2024}, others can exhibit behavior not simply determined by the primary order parameter, as discussed for induced multipolar moments~\cite{HattoriTsunetsugu2016}. In the present context, $G_{0,u}$ plays the role of such an induced structural order parameter generated by the displacement order parameter of an AtC transition. While previous studies have established that structural chirality can induce CPS and CIM~\cite{tsunetsugu_kusunose,kusunose_cubic,Tateishi2025JPSJ,CIM_nature,CIM}, the way in which the symmetry of an AtC transition determines the leading order of $G_{0,u}$ in the displacement has not been systematically addressed. We show that, depending on the symmetry of the parent structure and the nature of the structural transition, the leading contribution to $G_{0,u}$ can appear at different orders in the displacement order parameter. As a consequence, chirality-induced responses inherit nontrivial scaling behaviors from the displacement dependence of $G_{0,u}$, which is reflected in their temperature dependences.
%across a structural phase transition, these scaling laws are reflected in their temperature dependences. 
Moreover, by comparing the displacement dependence of $G_{0,u}$ with that of the phonon-band splitting across AtC transitions, we find that not every splitting is chirality-induced, i.e., constitutes CPS. This demonstrates that the displacement dependence of $G_{0,u}$ is not merely a classification scheme but a key perspective for understanding chirality-induced phenomena.

We demonstrate these principles by analyzing representative AtC transitions in both noncubic and cubic systems. %The resulting symmetry-controlled order-parameter dependence reveals how chirality-induced phenomena encode detailed information about the symmetry of the underlying structural transition.
Since the symmetry of the parent structure and the ordering mode fix how $G_{0,u}$ grows with the displacement, measuring the temperature dependence of CPS and CIM diagnoses the symmetry class of an AtC transition, beyond the binary question of whether the low-temperature phase is chiral.

This paper is organized as follows. In Sec.~\ref{sec:toroidals}, we present a general framework for describing chirality in terms of ET multipoles and derive their displacement dependence for representative AtC transitions. In Sec.~\ref{sec:numerical}, we numerically evaluate CPS and CIM and clarify their relationship to the functional form of $G_{0,u}$. In Sec.~\ref{sec:discussion}, we provide a microscopic analysis of the CPS mechanism based on the dynamical matrix, discuss the nonchiral origin of sound-velocity splitting, and examine the temperature dependence of CPS and CIM. Finally, Sec.~\ref{sec:conclusion} summarizes our results and presents an outlook. Throughout this paper, we use the Boltzmann constant $k_B=1$ and the Dirac constant $\hbar=1$.

\section{\label{sec:toroidals}Construction of electric toroidal monopole and quadrupole}
We begin by considering structural phase transitions driven by symmetry-lowering lattice displacements. Structural phase transitions can arise from various mechanisms, including charge ordering~\cite{Tokura_review}, lattice distortions~\cite{Cowley}, and electronic ordering involving orbital and multipolar degrees of freedom~\cite{tokura_nagaosa_orbital,Santini}. Among these, those driven by lattice displacements represent one of the most fundamental phenomena in crystalline solids. Although we primarily focus on continuous phase transitions, most of our conclusions extend to discontinuous cases. 

We first formulate the general structure of the free energy and examine how chirality---namely, ET monopoles and quadrupoles---evolves as a function of the displacements, i.e., the order parameter of the structural phase transition, in Sec.~\ref{subsec:General}.  We then provide general conditions for classifying the functional form for these quantities in Sec.~\ref{subsec:conditions} and present representative examples in Sec.~\ref{subsec:D2d}.
As the simplest example, Sec.~\ref{subsubsec:D2d}  discusses an AtC structural phase transition in which the displacement order parameter $\boldsymbol{\eta}$ itself transforms as $G_{0,u}$. This occurs when the structural order parameter belongs to a one-dimensional pseudoscalar irrep. In Secs.~\ref{subsec:D6h}, \ref{subsec:rhombo}, and \ref{subsec:fcc} we discuss AtC transitions induced by nonuniform displacements with the ordering vector $\bm{q}\ne\bm{0}$ for simple systems including trigonal Te and a cubic AtC transition in the fcc lattice. For more complicated systems, we summarize the analyses of GaNb$_4$Se$_8$~\cite{Kitou_chiral}, 
 K$_3$NiO$_2$~\cite{KNO_experi,KNO}, and some cubic AtC transitions in Appendix~\ref{subsec:others}.

\subsection{\label{subsec:General}Free energy and electric toroidal monopole/quadrupole}
The order parameter of a structural phase transition at $T=T_c$ is the lattice displacement $\bm{d}_s(\bm{r}_i) = (d_{x,s}(\bm{r}_i), d_{y,s}(\bm{r}_i), d_{z,s}(\bm{r}_i))$, defined with respect to the high-temperature equilibrium position. The atomic position in the high-temperature phase is written as $\tilde{\bm{r}}_{s,i} \equiv \bm{r}_i + \bm{t}_s$, and the displacement shifts it to $\tilde{\bm{r}}_{s,i} + \bm{d}_s(\bm{r}_i)$. Here, $\bm{r}_i$ denotes the coordinate of the $i$th unit cell, and $\bm{t}_s$ specifies the position of the $s$th sublattice within the unit cell.

These displacements originate from the condensation of phonon eigenmodes $\epsilon_{\gamma,\bm{q}_n}$ belonging to an irreducible representation (irrep) $\gamma$ of the high-temperature phase at ordering wave vectors $\bm{q}_n$ ($n=1,2,\cdots,\mathcal{N}$) in the first Brillouin zone (BZ). We assume that the $\bm{q}_n$ belong to the same star, because eigenmodes associated with different stars generally have different energies and therefore do not condense simultaneously.

We consider a structural phase transition characterized by the coefficient of the eigenmodes $\eta_{\gamma,\bm{q}_n}$ within a given star at $\bm{q}_n=\bm{q}_1,\bm{q}_2,\cdots,\bm{q}_{\mathcal N}$, where $-\bm{q}_n$ is not included among the $\bm{q}_n$. The lattice displacement $d_{\mu,s}(\bm{r}_i)$ is then expressed as
\begin{align}
d_{\mu,s}(\bm{r}_i)=\sum_{n=1}^{\mathcal N}\sum_{\gamma}(\boldsymbol{\epsilon}^s_{\gamma,\bm{q}_n})_{\mu}\eta_{\gamma,\bm{q}_n}e^{i\bm{q}_n\cdot \bm{r}_i}+{\rm H.c.}, \label{eq:disp_general}
\end{align}
where $\gamma$ labels the irreps and 
$(\boldsymbol{\epsilon}^s_{\gamma,\bm{q}_n} )_{\mu}$ denotes the $\mu$ ($\mu=x,y,z$) component of the displacement 
$\boldsymbol{\epsilon}^s_{\gamma,\bm{q}_n}$ on the $s$th sublattice for the eigenmode $\epsilon_{\gamma,\bm{q}_n}$ and it is normalized as $\sum_{s}(\boldsymbol{\epsilon}^s_{\gamma,\bm{q}_n})^* \cdot \boldsymbol{\epsilon}^s_{\gamma',\bm{q}_n}=\delta_{\gamma\gamma'}$. The summation over $\gamma$ is required when the irrep is multidimensional. Conversely, $\eta_{\gamma,\bm{q}_m}$ is obtained as
\begin{align}
\eta_{\gamma,\bm{q}_n}= \sum_{\mu,s}(\boldsymbol{\epsilon}^s_{\gamma,\bm{q}_n})^*_{\mu} d_{\mu,s,\bm{q}_n},
\label{eq:Psi_d_FT}
\end{align}
where $d_{\mu,s,\bm{q}_n}$ is the Fourier component of $d_{\mu,s}(\bm{r}_i)$ defined as $d_{\mu,s,\bm{q}_n} = \frac{1}{N}\sum_{\bm{r}_i}d_{\mu,s}(\bm{r}_i)e^{-i\bm{q}_n\cdot \bm{r}_i}$, where $N$ is the number of unit cells, 
and $d_{\mu,s,-\bm{q}_n}=d_{\mu,s,\bm{q}_n}^*$. This convention implies that the overall sign of the eigenmode for each $\gamma$ and $\bm{q}_n$ is encoded in $\eta_{\gamma,\bm{q}_n}$. Consequently, when $\eta_{\gamma,\bm{q}_n}\to -\eta_{\gamma,\bm{q}_n}$ for all $\gamma$ and $\bm{q}_n$, the lattice displacement transforms as $d_{\mu,s}(\bm{r}_i)\to -d_{\mu,s}(\bm{r}_i)$ for all $\mu,s,\bm{r}_i$.

When a structural phase transition is described by a single-$\bm{q}$ order, only one component of $\eta_{\gamma,\bm{q}_n}$ is nonzero. In contrast, for multiple-$\bm{q}$ orders, several $\bm{q}_n$ components are finite. In either case, multiple symmetry-equivalent domains are generally formed in actual phase transitions. In this study, however, we assume that the coefficients $\eta_{\gamma,\bm{q}_n}$ are fixed to those of a reference domain for simplicity.

Within the framework of Landau theory, structural phase transitions are described by the Landau free energy expressed in terms of $\eta_{\gamma,\bm{q}_n}$,
\begin{align}
\mathcal{F} =&\frac{a(T)}{2}\sum_{\gamma}\sum_{n=1}^{\mathcal N}
|\eta_{\gamma,\bm{q}_n}|^2+\cdots, \label{eq:Freeenergy}
\end{align}
where $a(T)=a_0(T-T_c)$ with $a_0>0$. Because $\bm{q}_1,\cdots,\bm{q}_{\mathcal N}$ belong to the same star and $\gamma$ labels the components of the relevant irrep, the coefficient of the quadratic term $a(T)$ is independent of $n$ and $\gamma$. In contrast, higher-order terms allow couplings between $\eta_{\gamma,\bm{q}_n}$ with different $\gamma$'s, which determines the anisotropy in the multidimensional irrep and whether the transition occurs at single-$\bm{q}$ or multiple-$\bm{q}$.

When the structural phase transition is of the AtC type, the ET monopole $G_{0}$ or quadrupole $G_{u}$ also acquires a finite value in the low-temperature symmetry-broken phase, together with the displacements. Here, $G_0$ and $G_u$ are pseudoscalar quantities, i.e., even under all proper rotations while odd under all improper operations of the high-temperature phase. We note that $G_{0,u}$ should be uniform since it describes the macroscopic chirality, corresponding to the modulation wave vector $\bm{q}=\bm{0}$ defined with respect to the high-temperature phase. To discuss the emergence of chirality through AtC transitions, it is essential to identify the form of secondary order parameters at $\bm{q}=\bm{0}$ that transform according to the same irreps as $G_{0,u}$. This can be carried out in a standard manner once the irreps of $\eta_{\gamma,\bm{q}_n}$ and the symmetry of the high-temperature phase are specified.

Our central goal is to express $G_{0,u}$ in terms of $\eta_{\gamma,\bm{q}_n}$ for $T\lesssim T_c$. Accordingly, we expand $G_{0,u}$ in powers of the small order parameters $\eta_{\gamma,\bm{q}_n}$ as
\begin{align}
G_{0,u}(\bm{\eta}) = &\sum_{n=1}^{\infty} G_{0,u}^{(n)}(\bm{\eta}),\\
G^{(n)}_{0,u}(\bm{\eta}) 
\equiv&
\sum_{\gamma_1,\cdots,\gamma_n}
\sum_{\bm{p}_1=\pm \bm{q}_1}^{\pm\bm{q}_{\mathcal N}}
\sum_{\bm{p}_2=\pm \bm{q}_1}^{\pm\bm{q}_{\mathcal N}}\cdots
\sum_{\bm{p}_n=\pm \bm{q}_1}^{\pm\bm{q}_{\mathcal N}}\notag\\
&\times C^{\gamma_1,\cdots,\gamma_n}_{\bm{p}_1,\cdots,\bm{p}_n}
\eta_{\gamma_1,\bm{p}_1}
\cdots
\eta_{\gamma_n,\bm{p}_n}
\delta_{\sum_i \bm{p}_i,\bm{G}},
\label{eq:G0_n}
\end{align}
where $\bm{\eta} = (\eta_{\gamma_1,\bm{q}_1},\eta_{\gamma_2,\bm{q}_2},\cdots)$ symbolically denotes the set of all $\eta_{\gamma,\bm{q}_n}$, $C^{\gamma_1,\cdots,\gamma_n}_{\bm{p}_1,\cdots,\bm{p}_n}$ are expansion coefficients, and the Kronecker delta 
$\delta_{\sum_i \bm{p}_i,\bm{G}}$ 
ensures that the total wave vector equals a reciprocal lattice vector $\bm{G}$ of the high-temperature phase, so that $G^{(n)}_{0,u}$ is spatially uniform.

We note that when $G_{0,u}(\bm{\eta})$ contains only either even- or odd-order contributions $G_{0,u}^{(n)}(\bm{\eta})$, reversing the sign of all displacements $\bm{\eta}\rightarrow -\bm{\eta}$ leads to $G_{0,u}(-\bm{\eta})= \pm G_{0,u}(\bm{\eta})$, where the ``$+$'' (``$-$'') sign corresponds to the even (odd) case. The relation $G_{0,u}(-\bm{\eta})=G_{0,u}(\bm{\eta})$ implies that the sign (i.e., direction) of the order parameters $\bm{\eta}$ does not affect the chirality, indicating that the chirality is governed by internal degrees of freedom of $\bm{\eta}$. In contrast, when $G_{0,u}(-\bm{\eta})=-G_{0,u}(\bm{\eta})$, the chirality directly depends on the direction of the displacements. These two cases represent qualitatively distinct classes and constitute important characteristics of AtC transitions. In both cases, the two structures $\bm{\eta}$ and $-\bm{\eta}$ must be energetically equivalent, since $|G_{0,u}(\pm \bm{\eta})|$ is invariant and the two configurations are either mirror images of each other or identical up to appropriate translations and rotations, i.e., related by a symmetry operation of the high-temperature phase. This implies that odd-order terms are absent in the Landau free energy~(\ref{eq:Freeenergy}).

When both even- and odd-order terms are present in $G_{0,u}(\bm{\eta})$, the two states corresponding to $\bm{\eta}$ and $-\bm{\eta}$ must be inequivalent, since $|G_{0,u}(\bm{\eta})|\neq |G_{0,u}(-\bm{\eta})|$. This implies that the structures associated with $\bm{\eta}$ and $-\bm{\eta}$ are not symmetry equivalent and they cannot be interchanged by symmetry operations of the high-temperature phase. 
Within the Landau free energy~\eqref{eq:Freeenergy}, this situation corresponds to the presence of odd-order terms in $\mathcal{F}$. One should note that this does not necessarily imply a first-order transition beyond the Landau framework.

\subsection{\label{subsec:conditions}Conditions for the leading-order terms of $G_{0,u}(\boldsymbol{\eta})$}
Let us now discuss the conditions for some of the leading-order terms of $G_{0,u}(\bm{\eta})$ to be finite, using a group-theoretical approach. We focus on the leading contributions, which dominate near the transition temperature $T\lesssim T_c$. We first provide the conditions for the $\bm{q}=\bm{0}$ case, and then we extend our discussion to the $\bm{q}\ne\bm{0}$ case. 

The results of this subsection can be summarized as follows. For $\bm{q}=\bm{0}$, a linear contribution appears only when the structural mode itself transforms as the pseudoscalar irrep, i.e., when $\gamma$ is a one-dimensional irrep. For multidimensional irreps, the lowest order $G_{0,u}$'s are identified for all achiral point groups. For $\bm{q}\neq\bm{0}$, the second-order contribution is allowed only when the ordering mode and the little-group symmetry satisfy Eq.~\eqref{eq:character_condition}, or when chirality is generated from inequivalent but symmetry-related star arms. Otherwise, the leading term is fixed by the lowest Umklapp order. These rules determine the exponent $n$ in $G_{0,u}\sim \eta^n$. To examine which chiral space group is derived by a displacement irrep at wave vector $\bm{p}$, it is useful to use ISODISTORT~\cite{ISODISTORT,ISODISPLACE} together with REPRES~\cite{REPRES}.

\subsubsection{\label{subsubsec:condition_uniform}Uniform transitions at $\boldsymbol{q}=\boldsymbol{0}$}
For the transitions at $\bm{q}=\bm{0}$, the Kronecker delta in Eq.~\eqref{eq:G0_n} is always satisfied, and thus, $G_{0,u}$ can appear at any order in $\bm{\eta}$. In particular, the first-order term $G_{0,u}^{(1)}(\bm{\eta})$ is unique to the transitions at $\bm{q}=\bm{0}$ since this term is prohibited for $\bm{q}\ne\bm{0}$ due to the Kronecker delta. The first-order term arises when the eigenmode $\bm{\epsilon}^s_{\gamma,\bm{q}=\bm{0}}$ belongs to the same irrep as $G_{0,u}$, which requires the eigenmode to belong to a one-dimensional irrep corresponding to the pseudoscalar irrep $\Gamma_{\text{ps}}$ of the high-temperature phase. Moreover, one can prove that only the odd-order terms are allowed since the square of a one-dimensional irrep is invariant and does not break any symmetries. Therefore, the AtC transition driven by a one-dimensional irrep at $\bm{q}=\bm{0}$ is always described as
\begin{align}
G_{0,u}(\bm{\eta}) = C^{(1)}\,\eta_{\Gamma_{\text{ps}},\bm{q}=\bm{0}} +C^{(3)}\, \eta^3_{\Gamma_{\text{ps}},\bm{q}=\bm{0}} +\cdots,
\end{align}
where $C^{(1)},\,C^{(3)},\cdots$ are constants.
The irreps $\Gamma_\text{ps}$ leading to a finite $G_{0,u}^{(1)}(\bm{\eta})$ for the achiral point groups $\mathcal G$ are listed in Table~\ref{tab:pseudoscalar}. 

\begin{table}[bt]
\centering
\caption{
List of achiral point groups $\mathcal{G}$ and the labels of the one-dimensional pseudoscalar irreps $\Gamma_{\rm ps}$ with representative basis functions.
}
\vspace{2mm}
\begin{tabular}{llc}
\hline\hline
$\mathcal{G}$ & $\Gamma_{\rm ps}$ & basis function \\
\hline
C$_{\rm i}$ & $A_{\rm u}$ & $z$ \\
C$_{\rm s}$ & $A''$ & $z$ \\
C$_{2{\rm h}}$ & $A_{\rm u}$ & $z$ \\
C$_{3{\rm h}}$ & $A''$ & $z$ \\
C$_{4{\rm h}}$ & $A_{\rm u}$ & $z$ \\
C$_{6{\rm h}}$ & $A_{\rm u}$ & $z$ \\
S$_4$ & $B$ & $z$ \\
S$_6$ & $A_u$ & $z$ \\
C$_{2{\rm v}}$ & $A_2$ & $xy$ \\
C$_{3{\rm v}}$ & $A_2$ & $y(3x^2-y^2)$ \\
C$_{4{\rm v}}$ & $A_2$ & $xy(x^2-y^2)$ \\
C$_{6{\rm v}}$ & $A_2$ & $xy(x^2-3y^2)(3x^2-y^2)$ \\
D$_{2{\rm h}}$ & $A_{\rm u}$ & $xyz$ \\
D$_{3{\rm h}}$ & $A_1''$ & $yz(3x^2-y^2)$ \\
D$_{4{\rm h}}$ & $A_{1{\rm u}}$ & $xyz(x^2-y^2)$ \\
D$_{6{\rm h}}$ & $A_{1{\rm u}}$ & $xyz(x^2-3y^2)(3x^2-y^2)$ \\
D$_{2{\rm d}}$ & $B_1$ & $x^2-y^2$ \\
D$_{3{\rm d}}$ & $A_{1{\rm u}}$ & $x(x^2-3y^2)$ \\
T$_{\rm h}$ & $A_{\rm u}$ & $xyz$ \\
T$_{\rm d}$ & $A_2$ & $(x^2-y^2)(y^2-z^2)(z^2-x^2)$ \\
O$_{\rm h}$ & $A_{1{\rm u}}$ & $xyz(x^2-y^2)(y^2-z^2)(z^2-x^2)$ \\
\hline\hline
\end{tabular}
\label{tab:pseudoscalar}
\end{table}

\begin{table*}[bt]
\centering
\scriptsize
\caption{
List of parity- or mirror-odd multidimensional real irreps $\Gamma$ of crystallographic point groups $\mathcal G$, where $\mathcal G$ belongs to an achiral point group. The lowest-order ET monopole (quadrupole) $G_{0} (G_{u})$ polynomial constructed from a
single copy of the primary order parameter $\boldsymbol{\eta}=(\eta_1,\eta_2,\cdots)$.
The subgroup $\mathcal H(\boldsymbol{\eta})$ denotes the isotropy subgroup
$\mathcal H(\boldsymbol{\eta})=\mathrm{Stab}_G(\boldsymbol{\eta})$ of the primary order parameter $\boldsymbol{\eta}$, not the
isotropy subgroup of $G_{0,u}$ itself. Entries marked ``none'' mean that no $G_{0,u}$ is generated from symmetric powers of a single copy of that irrep, usually because inversion or horizontal-mirror parity forbids it. In the third column, examples of $\boldsymbol{\eta}$ are shown, where $x,y$, and $z$ are real-space coordinates, and $R_{\mu}(\mu=x,y,z)$ is the $\mu$ component of an axial vector. Note that apart from the point groups listed in the first column, C$_\text{i}$, C$_{\rm s}$, C$_{2\text{v}}$, C$_{2\text{h}}$, and D$_{2\text{h}}$ belong to achiral point groups and do not have multidimensional irreps, some of which appear in the fourth and the seventh columns as $\mathcal H(\boldsymbol{\eta})$. In the list, we use notations $u\equiv 2z^2-x^2-y^2$, $v=\sqrt{3}(x^2-y^2)$, $v_y=\sqrt{3}(z^2-x^2)$, $v_x=\sqrt{3}(y^2-z^2)$, $\eta_{123}=\eta_1\eta_2\eta_3$, and $\eta_{12;23;31}=(\eta_1^2-\eta_2^2)(\eta_2^2-\eta_3^2)(\eta_3^2-\eta_1^2)$.
}
\vspace{2mm}

\label{tab:G0uq_zero}
\resizebox{\textwidth}{!}{%
\begin{tabular}{lllclll}
%\toprule
\hline
\hline
$\mathcal G$ &
$\Gamma$ & $\boldsymbol{\eta}=(\eta_1,\eta_2,\cdots)$
 &
$\mathcal{H}(\boldsymbol{\eta})$ &
$G_{0,u}$ &
lowest-order polynomial & $\mathcal H(\boldsymbol{\eta})$ along special directions\\[0.5mm]
\hline

C$_{3\rm{i}}$ &
$E_{\rm{u}}$ &
$(x,y)$ &
C$_1$ &
$A_{\rm{u}}$ &
$3\eta_1^2\eta_2-\eta_2^3$, $\eta_1^3-3\eta_2^2\eta_1$ 
& \\

C$_{3\rm{h}}$ &
$E'$ &
$(x,y)$ &
C$_{\rm s}$ &
$A''$ &
none & \\

C$_{3\rm{h}}$ &
$E''$ &
$(R_x,R_y)$, $(zx,yz)$ &
C$_1$ &
$A''$ &
$3\eta_1^2\eta_2-\eta_2^3$, $\eta_1^3-3\eta_2^2\eta_1$ 
& \\[0.5mm]

C$_{3\rm{v}}$ &
$E$ &
$(x,y)$, $(R_y,-R_x)$ &
C$_1$ &
$A_2$ &
$3\eta_1^2\eta_2-\eta_2^3$ 
& C$_{\rm{s}}[\boldsymbol{\eta}\parallel(1,0)]$\\[0.5mm]

D$_{3\rm{d}}$ &
$E_{\rm{u}}$ &
$(x,y)$, $z(2xy,x^2-y^2)$&
C$_1$ &
$A_{1\rm{u}}$ &
$\eta_1^3-3\eta_1\eta_2^2$ 
& C$_2[\boldsymbol{\eta}\parallel(1,0)]$, C$_s[\boldsymbol{\eta}\parallel(0,1)]$\\[0.5mm]

D$_{3\rm{h}}$ &
$E''$ &
$(R_x,R_y)$, $(yz,-zx)$ &
C$_1$ &
$A_1''$ &
$3\eta_1^2\eta_2-\eta_2^3$ 
&  C$_2[\boldsymbol{\eta}\parallel(0,1)]$,  C$_s[\boldsymbol{\eta}\parallel(1,0)]$\\[0.5mm]

\hline

S$_4$ &
$E$ &
$(x,y)$, $(R_y,R_x)$&
C$_1$ &
$B$ &
$\eta_1^2-\eta_2^2$ & \\

C$_{4\rm{h}}$ &
$E_{\rm{u}}$ &
$(x,y)$&
C$_{\rm{s}}$ &
$A_{\rm{u}}$ &
none & \\

C$_{4\rm{v}}$ &
$E$ &
$(x,y)$, $(R_y,-R_x)$ &
C$_1$ &
$A_2$ &
$\eta_1\eta_2(\eta_1^2-\eta_2^2)$ &  C$_s[\boldsymbol{\eta}\parallel(1,0),(1,1)]$\\[0.5mm]

D$_{2\rm{d}}$ &
$E$ &
$(x,y)$, $(R_x,-R_y)$ &
C$_1$ &
$B_1$ &
$\eta_1^2-\eta_2^2$ & 
 C$_2[\boldsymbol{\eta}\parallel(1,0)]$,
 C$_{\rm{s}}[\boldsymbol{\eta}\parallel(1,1)]$
\\

D$_{4\rm{h}}$ &
$E_{\rm{u}}$ &
$(x,y)$ &
C$_{\rm{s}}$ &
$A_{1\rm{u}}$ &
none & \\

\hline

C$_{6\rm{h}}$ &
$E_{1\rm{u}}$ &
$(x,y)$ &
C$_{\rm{s}}$ &
$A_{\rm{u}}$ &
none & \\

C$_{6\rm{h}}$ &
$E_{2\rm{u}}$ &
$z(x^2-y^2,2xy)$ &
C$_2$ &
$A_{\rm{u}}$ &
$3\eta_1^2\eta_2-\eta_2^3$, $\eta_1^3-3\eta_1\eta_2^2$  
& \\[0.5mm]

C$_{6\rm{v}}$ &
$E_1$ &
$(x,y)$, $(R_y,-R_x)$ &
C$_1$ &
$A_2$ &
$(3\eta_1^2\eta_2-\eta_2^3)(\eta_1^3-3\eta_1\eta_2^2)$ 
&  C$_{\rm{s}}[\boldsymbol{\eta}\parallel(1,0)]$\\[0.5mm]

C$_{6\rm{v}}$ &
$E_2$ &
$(x^2-y^2,2xy)$ &
C$_2$ &
$A_2$ &
$3\eta_1^2\eta_2-\eta_2^3$ 
& 
C$_{2\rm{v}}[\boldsymbol{\eta}\parallel(1,0)]$\\

D$_{6\rm{h}}$ &
$E_{1\rm{u}}$ &
$(x,y)$ &
C$_s$ &
$A_{1\rm{u}}$ &
none & \\

D$_{6\rm{h}}$ &
$E_{2\rm{u}}$ &
$z(x^2-y^2,2xy)$ &
C$_2$ &
$A_{1\rm{u}}$ &
$3\eta_1^2\eta_2-\eta_2^3$ 
& 
D$_2[\boldsymbol{\eta}\parallel(0,1)]$, 
C$_{2\rm{v}}[\boldsymbol{\eta}\parallel(1,0)]$
\\[0.5mm]

\hline

T$_{\rm{h}}$ &
$E_{\rm{u}}$ &
$xyz(u,v)$ &
D$_2$ &
$A_{\rm{u}}$ &
$\eta_1^3-3\eta_1\eta_2^2$, $3\eta_1^2\eta_2-\eta_2^3$ & \\

T$_{\rm{h}}$ &
$T_{\rm{u}}$ &
$(x,y,z)$ &
C$_1$ &
$A_{\rm{u}}$ &
$\eta_{123}$ & 
C$_3[\boldsymbol{\eta}\!\parallel\!(1,\!1,\!1)]$,
C$_{\rm{s}}[\boldsymbol{\eta}\!\parallel\!(1,\!1,\!0)]$,
C$_{2\rm{v}}[\boldsymbol{\eta}\!\parallel\!(0,\!0,\!1)]$
\\

T$_{\rm{d}}$ &
$E$ &
$(u,v)$ &
D$_2$ &
$A_2$ &
$3\eta_1^2\eta_2-\eta_2^3$ & 
D$_{2\rm{d}}[\boldsymbol{\eta}\parallel (1,0)]$,
D$_{2}[\boldsymbol{\eta}\parallel (0,1)]$
\\

T$_{\rm{d}}$ &
$T_1$ &
$(R_x,R_y,R_z)$ &
C$_1$ &
$A_2$ &
$\eta_{123}$ & 
C$_3[\boldsymbol{\eta}\!\parallel\!(1,\!1,\!1)]$,
C$_{\rm{s}}[\boldsymbol{\eta}\!\parallel\!(1,\!1,\!0)]$,
S$_4[\boldsymbol{\eta}\!\parallel\!(0,\!0,\!1)]$
\\

T$_{\rm{d}}$ &
$T_2$ &
$(x,y,z)$, $(yz,zx,xy)$ &
C$_1$ &
$A_2$ &
$
\eta_{12;23;31}
%(\eta_1^2-\eta_2^2)(\eta_2^2-\eta_3^2)(\eta_3^2-\eta_1^2)
$
& 
C$_{3\rm{v}}[\boldsymbol{\eta}\!\parallel\!(1,\!1,\!1)]$,
C$_{\rm{s}}[\boldsymbol{\eta}\!\parallel\!(1,\!1,\!0)]$,
C$_{2\rm{v}}[\boldsymbol{\eta}\!\parallel\!(0,\!0,\!1)]$
\\

O$_{\rm h}$ &
$E_{\rm u}$ &
$xyz(v,-u)$ &
D$_2$ &
$A_{1\rm{u}}$ &
$3\eta_1^2\eta_2-\eta_2^3$ 
& 
D$_{4}[\boldsymbol{\eta}\parallel (0,1)]$,
D$_{2\rm{d}}[\boldsymbol{\eta}\parallel (1,0)]$
\\

O$_{\rm{h}}$ &
$T_{1\rm{u}}$ &
$(x,y,z)$ &
C$_1$ &
$A_{1\rm{u}}$ &
$%\eta_1\eta_2\eta_3
\eta_{123}
\eta_{12;23;31}
%(\eta_1^2-\eta_2^2)
%(\eta_2^2-\eta_3^2)
%(\eta_3^2-\eta_1^2)
$ 
& 
C$_{3\rm{v}}[\boldsymbol{\eta}\!\parallel\!(1,\!1,\!1)]$,
C$_{2\rm{v}}[\boldsymbol{\eta}\!\parallel\!(1,\!1,\!0)]$,
C$_{4\rm{v}}[\boldsymbol{\eta}\!\parallel\!(0,\!0,\!1)]$
\\

O$_{\rm{h}}$ &
$T_{2\rm{u}}$ &
$
(xv_x,yv_y,zv)$ &
C$_1$ &
$A_{1\rm{u}}$ &
$\eta_{123}$ 
& 
D$_{3}[\boldsymbol{\eta}\!\parallel\!(1,\!1,\!1)]$,
C$_{2\rm{v}}[\boldsymbol{\eta}\!\parallel\!(1,\!1,\!0)]$,
D$_{2\rm{d}}[\boldsymbol{\eta}\!\parallel\!(0,\!0,\!1)]$
\\[0.5mm]
\hline\hline
\end{tabular}%
}
\end{table*}

\color{black}
When the eigenmode $\epsilon_{\gamma,\bm{q}=\bm{0}}$ belongs to a multidimensional irrep, the ET monopole or quadrupole $G_{0,u}$ is generally not a primary order parameter but a secondary quantity induced by the primary structural order parameter. The possible forms of $G_{0,u}$ in terms of $\bm{\eta}$ depend on the point-group symmetry. Nevertheless, the lowest-order symmetry-allowed expression can be constructed systematically. For a finite $\boldsymbol{\eta}$ to drive the system into a chiral phase, the isotropy subgroup of $\boldsymbol{\eta}$ must break all mirror and inversion symmetries. Under this condition, $G_{0,u}$ can be constructed for multidimensional order parameters in each point group, except for the $E'$ irrep in C$_{3\text{h}}$, the $E_{\rm u}$ irrep in C$_{4\text{h}}$, the $E_{\rm u}$ irrep in D$_{4\text{h}}$, the $E_{1{\rm u}}$ irrep in C$_{6{\rm h}}$, and the $E_{1{\rm u}}$ irrep in D$_{6{\rm h}}$. These exceptions originate from the presence of horizontal mirror symmetry in these point groups.

Table~\ref{tab:G0uq_zero} summarizes the lowest-order forms of $G_{0,u}$ in terms of the multidimensional structural order parameter $\boldsymbol{\eta}=(\eta_1,\eta_2,\cdots)$ for achiral point groups $\mathcal{G}$ and the resulting subgroup $\mathcal H(\bm{\eta})$. The entries show that $G_{0,u}$ is of the form $\sim \eta^{2p}$ for the $E$ irrep in S$_4$, C$_{4\text{v}}$, and D$_{2{\rm d}}$, and for the $T_2$ irrep in T$_{\rm d}$, whereas it is of the form $\sim \eta^{2p+1}$ in the other cases, where $p$ is an integer. These two cases constitute qualitatively distinct classes. The cases with the isotropy subgroup $\mathcal H(\bm{\eta}) = \text{C}_1$ correspond to general order-parameter directions for which no symmetry remains; therefore, they are not the main focus of this paper. In the following, we discuss only the transitions resulting in $\mathcal H(\bm{\eta}) \ne \rm C_1$ with $\bm{\eta}$ along the high-symmetry directions, which are presented in the seventh column of Table~\ref{tab:G0uq_zero}.

For the cases with $G_{0,u}\propto \eta^{2p+1}$, the chirality changes sign under $\boldsymbol{\eta}\to -\boldsymbol{\eta}$. In contrast, for $G_{0,u}\propto \eta^{2p}$, as in the $E$ irrep for S$_4$ and D$_{2{\rm d}}$, the sign of $G_{0,u}$ is unchanged under $\boldsymbol{\eta}\to -\boldsymbol{\eta}$. Instead, the sign changes when the structural order parameter is switched between distinct components. For S$_4$ and D$_{2{\rm d}}$, this is achieved by a mirror operation or a fourfold roto-reflection, which transforms $\eta_1\leftrightarrow \pm \eta_2$~\cite{piezochiral}. 
%We discuss analogous properties for finite-$\bm{q}$ structural transitions in Sec.~\ref{sec:q_nonzero}. For the T$_{\rm d}$ case with the $T_2$ irrep, when the order parameter is identified with an electric polarization $\bm{P}=(P_x,P_y,P_z)$, $G_0$ appears as a sixth-order harmonic, $G_0 \sim (P_x^2-P_y^2)(P_y^2-P_z^2)(P_z^2-P_x^2)$. This expression vanishes along the high-symmetry directions $\bm{P}\parallel [111]$, $[110]$, and $[001]$, as listed in the seventh column of Table~\ref{tab:G0uq_zero}. Thus, for a general direction of $\bm{P}$, the isotropy subgroup is $H(\boldsymbol{\eta})=$~C$_1$.

When the system has several copies of the same irrep, denoted by $\boldsymbol{\eta}$, $\boldsymbol{\xi}$, $\boldsymbol{\zeta},\cdots$, additional symmetry-allowed expressions can appear that are not listed in Table~\ref{tab:G0uq_zero}. Here, we regard one of them as the primary order parameter and the others as secondary order parameters. Since it is beyond the scope of the present paper to complete all types of $G_{0,u}$ consisting of such multiple degrees of freedom, we list $G_{0,u}$ consisting of two or three independent copies in Appendix~\ref{app:BiTriG0u}. Table~\ref{tab:multi_copy_G0} in Appendix~\ref{app:BiTriG0u} can be used for analyzing chiral systems with multiple degrees of freedom.

\subsubsection{\label{subsubsec:nonuniform_condition}Second-order $G_{0,u}^{(2)}$ for nonuniform transitions at $\boldsymbol{q}\ne\boldsymbol{0}$}
When a transition occurs at $\bm{q}\ne\bm{0}$, $G_{0,u}$ is always described by the second- or higher-order terms in $\bm{\eta}$ due to the Kronecker delta in Eq.~\eqref{eq:G0_n}. In this section, we discuss conditions under which the second-order term is allowed. The conditions for finite higher-order terms are provided in Sec.~\ref{subsubsec:higher_order_G0}.
%\redsout{Most of the discussion and examples below are AtC transitions with the ordering vector at special $\bm{k}$ points, i.e. high-symmetry points at the BZ boundary, since these cases are expected to occur in realistic situations. In addition to this, we note that the discussion for $2\bm{k}\ne \bm{G}$ with $\bm{G}$ being a reciprocal lattice vector is applicable to cases for high-symmetric line/surface and general $\bm{k}$ points in the BZ.}

%\redsout{We first consider a condition for a finite second-order term $G_{0,u}^{(2)}(\bm{\eta})$. }
According to Eq.~\eqref{eq:G0_n}, the second-order term $G_{0,u}^{(2)}(\bm{\eta})$ needs to be constructed from the two wave vectors $\bm{p}_1$ and $\bm{p}_2$ satisfying $\bm{p}_1+\bm{p}_2=\bm{G}$ due to the Kronecker delta, where $\bm{p}_1$ and $\bm{p}_2$ belong to the same star as they are summed over the ordering vectors $\bm{q}_n$. This condition is only satisfied by $\bm{p}_2=-\bm{p}_1$ or $\bm{p}_1=\bm{p}_2=\frac{\bm{G}}{2}$, 
%\redsout{which indicates that $G_{0,u}^{(2)}(\bm{\eta})$ should be constructed at each $\bm{q}_n$ even for multiple-$\bm{q}$ orders }
and thus, Eq.~\eqref{eq:G0_n} for $n=2$ is reduced to
\begin{align}
G_{0,u}^{(2)}(\bm{\eta}) = \sum_{\bm{p}=\bm{q}_1}^{\bm{q}_\mathcal{N}}\sum_{\gamma_1,\gamma_2}C_{\bm{p}}^{\gamma_1,\gamma^*_2}\eta_{\gamma_1,\bm{p}}\eta_{\gamma^*_2,-\bm{p}}.\label{eq:G0_second}
\end{align}
Here, the sums over $\gamma_1$ and $\gamma_2$ run over all the components of the irrep $\gamma$ of the ordering mode, and hence they are taken only when the irrep is multidimensional. Also, $\gamma_2^*=\gamma_2$ if $\gamma$ is a real irrep. For wave vectors satisfying $\bm{p}_1=\bm{p}_2=\frac{\bm{G}}{2}$, $-\bm{p}\equiv\bm{p}$ holds and only real irreps are allowed.
%\redsout{represent different components of the same irrep $\gamma$, which needs to be a multidimensional irrep. In the following, we will discuss how ET monopole moments are constructed in terms of order parameter $\eta_{\gamma,\bm{p}}$ of structural phase transition for finite $\bm{p}$.} 
When we consider whether $G_{0,u}^{(2)}(\bm{\eta})$ exists in an AtC transition, we distinguish the cases according to whether the little group $\mathscr{G}_{\bm{p}}$ of the ordering vector $\bm{p}$ contains improper operations or not. When $\mathscr{G}_{\bm{p}}$ contains improper operations, i.e., at points with relatively high symmetry, $G_{0,u}^{(2)}$ is constructed as a sum of pseudoscalars of $\mathscr{G}_{\bm{p}}$; that is, $\sum_{\gamma_1,\gamma_2}C_{\bm{p}}^{\gamma_1,\gamma^*_2}\eta_{\gamma_1,\bm{p}}\eta_{\gamma^*_2,-\bm{p}}$ in Eq.~\eqref{eq:G0_second} itself serves as a pseudoscalar of $\mathscr{G}_{\bm{p}}$ at each arm. We discuss this case in (i) below.
In contrast, when $\mathscr{G}_{\bm{p}}$ contains no improper operation, a pseudoscalar cannot be distinguished from a scalar within $\mathscr{G}_{\bm{p}}$. In this case, $G_{0,u}^{(2)}$ is instead constructed as a linear combination of the scalars of $\mathscr{G}_{\bm{p}}$ at the different arms, which is discussed in (ii) below.

{\flushleft (i) {$\mathscr{G}_{\bm{p}}$ with improper operations}}

When $\mathscr{G}_{\bm{p}}$ contains improper operations, we show below that a finite $G_{0,u}^{(2)}$ is allowed only when the character $\chi_\gamma(g)$ of the irrep $\gamma$ of the little group at $\bm{p}$ is real for every proper operation $g$ and purely imaginary (including zero) for every improper operation $g$. This condition excludes one-dimensional real irreps, while it is met by one-dimensional complex irreps as well as by multidimensional irreps whose improper-operation characters vanish or are purely imaginary.

The necessary condition for a finite $G^{(2)}_{0,u}$ is derived as follows. Since the ET monopole/quadrupole transforms as a pseudoscalar, we introduce the one-dimensional representation whose character is $s(g)=+1$ for all proper operations $g$ and $s(g)=-1$ for all improper operations $g$ of the little group $\mathscr{G}_{\bm{p}}$ at $\bm{p}$. For the bilinear form $\sum_{\gamma_1,\gamma_2}C_{\bm{p}}^{\gamma_1,\gamma^*_2}\eta_{\gamma_1,\bm{p}}\eta_{\gamma^*_2,-\bm{p}}$ to be a pseudoscalar, the coefficient matrix $(\hat{C}_{\bm{p}})_{\gamma_1,\gamma^*_2}\equiv C_{\bm{p}}^{\gamma_1,\gamma^*_2}$ must satisfy
\begin{align}
\hat{D}_{\gamma}(g)\,\hat{C}_{\bm{p}}\,\hat{D}^{-1}_{\gamma^*}(g) = s(g)\,\hat{C}_{\bm{p}} \qquad (\forall g),\label{eq:schur_1}
\end{align}
 where $\hat{D}_{\gamma}(g)$ is the representation matrix of the operation $g\in\mathscr{G}_{\bm{p}}$ for $\gamma$. By multiplying Eq.~\eqref{eq:schur_1} from the right by $\hat{D}_{\gamma^*}(g)$, this is rewritten as $\hat{D}_{\gamma}(g)\,\hat{C}_{\bm{p}} = \hat{C}_{\bm{p}}\,[s(g)\hat{D}_{\gamma^*}(g)]$. Since $s(g)$ is the character of the pseudoscalar representation, $\hat{D}^{(s)}_{\gamma'}(g)\equiv s(g)\hat{D}_{\gamma^*}(g)$ is also an irreducible representation. According to Schur's lemma~\cite{gunron}, $\hat{C}_{\bm{p}}$ is an intertwiner between $\hat{D}_\gamma(g)$ and $\hat{D}^{(s)}_{\gamma'}(g)$ that is either zero or invertible ($\text{det}\hat{C}_{\bm{p}}\ne0$). Thus, a nontrivial $\hat{C}_{\bm{p}}$ exists if and only if $\hat{D}_{\gamma}\cong \hat{D}^{(s)}_{\gamma'}$, i.e., the two irreps are isomorphic. This holds if and only if the characters coincide,
 \begin{align}
 \chi_{\gamma}(g)=s(g)\chi_{\gamma^*}(g)\qquad(\forall g).\label{eq:character_condition}
 \end{align}  
 
For real irreps, Eq.~\eqref{eq:character_condition} reduces to $\chi_{\gamma}(g)=s(g)\chi_{\gamma}(g)$, which is automatically satisfied for proper operations owing to $s(g)=1$, whereas for improper operations $s(g)=-1$ enforces $\chi_{\gamma}(g)=0$. This can never be satisfied by a one-dimensional real irrep, and thus, for a real irrep, a finite $G_{0,u}^{(2)}$ requires the irrep to be multidimensional with vanishing characters for all the improper operations.

For complex irreps, Eq.~\eqref{eq:character_condition} can be satisfied even when the ordering mode belongs to a one-dimensional complex irrep. Using $\chi_{\gamma^*}(g) = [\chi_{\gamma}(g)]^*$, Eq.~\eqref{eq:character_condition} reduces to $\chi_{\gamma}(g)=s(g)[\chi_{\gamma}(g)]^*$, i.e., $\chi^2_{\gamma}(g)=s(g)|\chi_{\gamma}(g)|^2$. This condition holds regardless of the dimensionality of the irrep, as long as the characters of the proper and improper operations are real and purely imaginary, respectively. For a one-dimensional complex irrep, $|\chi_{\gamma}(g)|=1$ further reduces it to $\chi^2_{\gamma}(g)=s(g)$, so that the characters of the proper and improper operations are $\pm1$ and $\pm i$, respectively.
 
When these conditions are satisfied, the explicit form of $G_{0,u}^{(2)}$ depends on whether $\gamma$ is real or complex. We first consider a real irrep, for which $\gamma^*=\gamma$. Squaring both sides of Eq.~\eqref{eq:schur_1} then gives
\begin{align}
\hat{D}_{\gamma}(g)\,\hat{C}_{\bm{p}}^2\,\hat{D}^\dagger_{\gamma}(g) = \hat{C}_{\bm{p}}^2\qquad (\forall g).
\end{align}
Schur's lemma implies that $\hat{C}_{\bm{p}}^2 = \bm{1}$ when $\hat{C}_{\bm{p}}$ is normalized, where $\bm{1}$ is the identity matrix. Mathematically, this indicates that $\hat{C}_{\bm{p}}$ can be diagonalized and its eigenvalues are either $+1$ or $-1$. Recalling that $\hat{C}_{\bm{p}}$ also commutes with all the rotational operations because of $s(g)=1$ for such operations in Eq.~\eqref{eq:schur_1}, $\hat{C}_{\bm{p}}$ is diagonalized in the basis of the eigenstates of rotational operations and given by $\hat{C}_{\bm{p}} = \text{diag}(1,1,1,\cdots,-1,-1,-1,\cdots)$. Therefore, by denoting the basis vector for the subspace with the eigenvalue $\pm1$ as $\bm{\eta}^{\pm}$, $G_{0,u}^{(2)}$ is expressed as
\begin{align}
G_{0,u}^{(2)}(\bm{\eta})\propto |\bm{\eta}^{+}|^2 - |\bm{\eta}^{-}|^2. \label{eq:G0_second_gen}
\end{align}
Physically, $\bm{\eta}^{\pm}$ are eigenstates of the rotation (or screw) operations about the principal axes, and $\bm{\eta}^{+}$ and $\bm{\eta}^{-}$ possess opposite crystal angular momenta~\cite{PAM_Zhang}. Thus, $G_{0,u}^{(2)}$ is invariant under the rotation operations, whereas any improper operation interchanges $\bm{\eta}^{+}$ and $\bm{\eta}^{-}$ and reverses its sign. Moreover, as is evident from Eq.~\eqref{eq:G0_second_gen}, the sign of $G_{u}^{(2)}$ is solely determined by which of the two eigenmodes $\bm{\eta}^{\pm}$ orders, and the overall sign (direction) of $\bm{\eta}^{\pm}$ does not affect the chirality, as explained in Sec.~\ref{subsec:General}.

For a complex irrep, $\hat{C}_{\bm{p}}$ intertwines the inequivalent irreps $\gamma$ and $\gamma^*$, so the eigenmode form of Eq.~\eqref{eq:G0_second_gen} does not directly apply. In the simplest case of a one-dimensional complex irrep, $\hat{C}_{\bm{p}}$ reduces to a single phase factor and $G_{0,u}^{(2)}$ takes the form $G_{0,u}^{(2)}(\bm{\eta})\propto \mathrm{Re}(c\,\eta_{\gamma,\bm{p}}\eta_{\gamma^*,-\bm{p}})$ with a complex constant $c$, and multidimensional complex irreps are treated analogously by pairing each component with its conjugate.

From these results, one can predict whether the second-order term appears in the expansion of $G_{0,u}$ simply from the transformation properties of the eigenmode and the symmetry of the little group in the high-temperature phase. When the system is symmorphic, the second-order term $G_{u}^{(2)}$ can be constructed if the eigenmode belongs to a two-dimensional irrep and the little group is isomorphic to one of the point groups C$_{3\text{v}}$, C$_{4\text{v}}$, C$_{6\text{v}}$, S$_4$, D$_{2\text{d}}$, and T$_\text{d}$. However, this does not apply when the system is nonsymmorphic, since the representation matrices and their characters in the little group $\mathscr{G}_{\bm{p}}$ in general differ from those of the point groups; one should therefore calculate them for each nonsymmorphic space group. We show that certain transverse modes of the nonsymmorphic rutile and diamond structures give rise to the second-order $G_u^{(2)}$ and $G_0^{(2)}$ in Appendix~\ref{subsec:others} as representative examples for this type of construction, and we present their representation matrices in Appendix~\ref{apdx:irrep}.
We also note that one needs to take into account the position of the wave vector $\bm{p}$ for deriving the actual form of $G^{(2)}_{0,u}$.

{\flushleft (ii) ${\mathscr G}_{\bm{p}}$ without improper operations: $k$-arm chirality} 

For some space groups ${\mathscr G}$, there are no improper operations in the little group ${\mathscr G}_{\bm{p}}$, while improper operations exist in ${\mathscr G}$. In such cases, one needs to analyze the full induced representation of ${\mathscr G}$, since the previous discussion in (i) does not apply. Since ${\mathscr G}_{\bm{p}}$ consists only of proper operations, a pseudoscalar cannot be formed within a single arm. The only invariant available at each arm is then the modulus $|\bm{\eta}_{\bm{q}_n}|^2$, and the ET monopole/quadrupole must be constructed as a linear combination of such arm-resolved invariants. 
More generally, let $\bm{k}=\bm{q}_1$ be a representative arm of the star, and let ${\mathscr G}_{\bm{k}}$ be its little group. The other arms $\bm{q}_i$ ($i=2,3,\cdots,N_{\text{arm}}$) are generated as $g_i\bm{k}$ by applying representative operations $g$ chosen from the cosets ${\mathscr G}/{\mathscr G}_{\bm{k}}$ to $\bm{k}$. Here, $N_{\text{arm}}$ is the number of distinct arms. When ${\mathscr G}_{\bm{k}}$ contains only proper operations, the arm-resolved invariants $|\bm{\eta}_{g_i\bm{k}}|^2$ can be combined into
\begin{equation}
G^{(2)}_{0,u}(\eta) \propto
\sum_{i=1}^{N_\text{arm}} \det R_{g_i}\,|\eta_{g_i\bm{k}}|^2, \label{eq:k-arm_chi}
\end{equation}
where $R_{g_i}$ is the point part of $g_i$ and $g_1$ is the identity. This expression is independent of the choice of representatives $\lbrace g_i \rbrace$, because replacing $g_i$ by $g_i h$ with $h\in{\mathscr G}_{\bm{k}}$ leaves both the arm and the coefficient unchanged: $\det R_{g_i h}=\det R_{g_i}\det R_h=\det R_{g_i}$. Under a symmetry operation $a\in{\mathscr G}$, the arms are permuted as $g_i\bm{k}\mapsto ag_i\bm{k}$, while the coefficient changes as $\det R_{ag_i}=\det R_a\det R_{g_i}$. Hence $G^{(2)}_{0,u}$ acquires the factor $\det R_a$ under $a$: it is even under proper operations and odd under improper operations. Thus, Eq.~(\ref{eq:k-arm_chi}) transforms as a pseudoscalar of $\mathscr{G}$.

 For P$\bar{4}2_1$m (No.~113) and P$\bar{4}2_1$c (No.~114), an AtC transition to P2$_1$ (No.~4) is driven by a two-dimensional irrep $\boldsymbol{\eta}_{\rm{X,Y}}$ at the X (R) points. Here, X and Y are the two distinct arms in the X (R) star. In this case, the little group ${\mathscr G}_{\bm p}$ at the X or R points has no improper operation and is isomorphic to the chiral point group D$_2$ in contrast to D$_{2\rm d}$ of the space group. Thus, the discussion in (i) is not applicable to this system. Instead, we can construct the trivial invariants $|\boldsymbol{\eta}_{\rm{X}}|^2$ and $|\boldsymbol{\eta}_{\rm{Y}}|^2$ at the two arms under $\mathscr{G}_{\bm{p}}$, where every $g\in \mathscr{G}_{\bm{p}}$ is a proper operation. Noting that improper operations in $\mathscr{G}$ exchange the X and Y, $G^{(2)}_u$ is given by $\propto |\boldsymbol{\eta}_{\rm{X}}|^2-|\boldsymbol{\eta}_{\rm{Y}}|^2$. This combination is odd under the improper operations in D$_{2\text{d}}$ and therefore transforms as the pseudoscalar representation. Note that $G^{(2)}_u$ is finite, i.e., the ordered phase is chiral, whenever the two arms carry unequal weights, $|\boldsymbol{\eta}_{\rm{X}}|^2\ne|\boldsymbol{\eta}_{\rm{Y}}|^2$, and the sign of $G^{(2)}_u$, chirality, is set by which arm dominates. The single-$\bm{q}$ states $(|\boldsymbol{\eta}_{\rm{X}}|^2,|\boldsymbol{\eta}_{\rm{Y}}|^2)\propto (1,0)$ or $(0,1)$ are the two enantiomers with maximal chirality, leading to a $\bm{k}$-arm version of the piezochiral effects recently proposed in Ref.~\cite{piezochiral}. 

This construction of $G^{(2)}_{0,u}$ is possible only for noncentrosymmetric space groups. This is because $|\bm{\eta}_{\bm{p}}|^2$ is invariant under inversion; hence, no inversion-odd (pseudoscalar) combination of the per-arm quantities $|\bm{\eta}_{\bm{p}}|^2$ can be formed. For special $\bm{k}$ points (high-symmetry points at BZ boundaries), in addition to the P$\bar{4}2_1$m and P$\bar{4}2_1$c, there are several $\bm{k}$-arm chirality candidates in monoclinic systems with the crystal point group C$_{\rm s}$, orthorhombic systems with C$_{2\rm{v}}$, and tetragonal systems with S$_4$ and D$_{2{\rm d}}$. These candidates share a common property that the (electric) quadrupole moment breaks all the mirror operations in these point groups~\cite{piezochiral}. We summarize the candidate space groups that can host $\bm{k}$-arm chirality at high-symmetry points at BZ boundaries in Appendix~\ref{apdx:karm}.

%Similar AtC transitions are found also for those driven by the 2-dimensional irrep at the W point in Imma (No.~74) and Ibca(No.~73) to ????. This type of construction of ET monopole/quadrupole can be understood as the

\subsubsection{\label{subsubsec:higher_order_G0}$G_{0,u}^{(n)}$ ($n\ge 3$) for nonuniform transitions at $\boldsymbol{q}\ne\boldsymbol{0}$}

The higher-order terms of $G_{0,u}^{(n)}$ with $n\ge3$ are allowed only when the Umklapp condition expressed by the Kronecker delta in Eq.~\eqref{eq:G0_n}, $\bm{p}_1 + \bm{p}_2 + \cdots +\bm{p}_n=\bm{G}$, is satisfied with $\bm{p}_1,\bm{p}_2,\cdots,\bm{p}_n$ belonging to the same star. 
When the conditions discussed in (i) and (ii), or further conditions due to the space-group symmetry, are not satisfied, the second-order $G^{(2)}_{0,u}$ does not appear and the third- or higher-order terms become the leading contribution. These forms of pseudoscalar are regarded as ``multipolar chirality''. When the third-order term becomes the leading contribution, orderings at high-symmetry points in the first BZ of trigonal and hexagonal systems provide typical examples. The K point $(1/3,1/3,0)$ in the hexagonal (trigonal) system possesses D$_{3\text{h}}$ or C$_{3\text{h}}$ symmetry when the system belongs to the point group D$_{6\text{h}}$ (D$_{3\text{h}}$) or C$_{6\text{h}}$ (C$_{3\text{h}}$). At this point, the ordering vector satisfies $3\bm{q}=\bm{G}$, and assuming that the system is symmorphic, the second-order term is prohibited since the characters of the improper operations are not all zero or purely imaginary, i.e., the condition in (i) is not satisfied. Therefore, the third-order term can become the leading contribution. 
%This is also regarded as a generalization of the $\bm{k}$-arm chirality discussed in (ii) above.

Similarly, the star of the X point in the fcc lattice consists of three arms related by threefold rotation about the $[111]$ axis, which allows $\bm{q}_1+\bm{q}_2+\bm{q}_3=\bm{G}$ for a triple-$\bm{q}$ order. These points possess D$_{4\text{h}}$ symmetry when the crystal structure belongs to the point group O$_\text{h}$, and thus, as long as the system is symmorphic, corresponding to the space groups Fm$\bar{3}$m (No.~225), the second-order term for the two-dimensional transverse modes is prohibited, making the third-order term the leading contribution. We demonstrate that the third-order term appears as the leading contribution in an AtC transition from a simple fcc lattice at the three X points in Sec.~\ref{subsec:fcc}.
In contrast, we show in Appendix~\ref{subsubsec:Breathing} that the same triple-$\bm{q}$ order at the X points can instead give rise to both the second- and third-order terms when the parent structure has a different symmetry: for the breathing pyrochlore lattice, which belongs to the space group F$\bar{4}3$m (No.~216) and transforms into the cubic chiral P$2_1$3 structure as experimentally observed in GaNb$_4$Se$_8$~\cite{Kitou_chiral}, the little group $\mathscr{G}_{\bm{p}}$ at the X point is isomorphic to D$_{2\text{d}}$ rather than D$_{4\text{h}}$, so that condition (i) is satisfied.

In addition, the third-order term serves as the leading contribution in the bcc lattice, where a six-$\bm{q}$ order of a one-dimensional irrep at the N points $(1/2,1/2,0)$ drives an AtC transition. This case is discussed in Appendix~\ref{apdx:G0_6q}. Furthermore, we show in Appendix~\ref{apdx:G0_sixth} that an even higher, sixth-order term becomes the leading contribution when an AtC transition is driven by a one-dimensional irrep at the W points $(1/2,1,0)$ in the fcc lattice.

As an example of AtC transitions with multiple-$\bm{q}$ orders where the sum of the independent $\bm{q}$'s does not coincide with $\bm{G}$, we briefly discuss an AtC transition from the diamond structure into the chiral tetragonal space group P$4_122$ (No.~91) or P$4_322$ (No.~95). This is driven by a double-$\bm{q}$ ordering of a longitudinal mode at two of the three X points of the fcc BZ; $\bm{q}_1=(1,0,0)$ and $\bm{q}_2=(0,1,0)$. Since this mode has real and nonzero characters for improper operations, the second-order term $G_u^{(2)}$ is prohibited. Moreover, the third-order term $G_u^{(3)}$ is not allowed from the Umklapp condition. As a result, the fourth-order term $G_u^{(4)}$, which satisfies $2\bm{q}_1+2\bm{q}_2=\bm{G}$, provides the leading contribution to $G_u$.

We close this section with a remark on the scope of the conditions discussed above. They specify when $G_{0,u}^{(n)}(\bm{\eta})$ is symmetry-allowed for a given eigenmode and ordering vector, but they do not by themselves guarantee that the resulting expression is nonzero for every direction of $\bm{\eta}$ within the irrep. For high-symmetry directions of $\bm{\eta}$, the symmetry-allowed $G_{0,u}^{(n)}$ may vanish, in which case the leading contribution to chirality shifts to a higher-order term; when $G_{0,u}^{(n)}$ vanishes at every order, the ordered phase remains achiral despite the symmetry analysis permitting a pseudoscalar invariant as shown in the seventh column in Table~\ref{tab:G0uq_zero} for the point group classification. The actual leading order of $G_{0,u}$ therefore depends on both the symmetry of the eigenmode and the direction of $\bm{\eta}$ realized in the ordered phase. Finally, we also note that when several degrees of freedom are present, there appear other possibilities beyond the discussions in (i) and (ii), although these are complicated and rare cases.

\subsection{\label{subsec:D2d}Examples of AtC transitions and the expressions of $G_{0,u}$}
In this section, we present several examples of AtC transitions. In Sec.~\ref{subsubsec:D2d}, we consider an AtC transition from the achiral D$_{2\text{d}}$ structure to the chiral D$_2$ structure driven by a one-dimensional irrep at the ordering vector $\bm{q} = \bm{0}$. As explained in Sec.~\ref{subsubsec:condition_uniform}, in such a transition, the eigenmode $\epsilon_{\gamma,\bm{q}=\bm{0}}$ is identical to $G_{0,u}$ itself, and as a result, the leading term is linear in $\bm{\eta}$. Then in Secs.~\ref{subsec:D6h}, \ref{subsec:rhombo}, and \ref{subsec:fcc}, we discuss structural AtC transitions with the ordering vector $\bm{q}\ne \bm{0}$. We begin by discussing a helical structure and then analyze how high-temperature symmetry (space group) affects the form of $G_{0,u}$ even when the low-temperature space group is the same. In the final part, a cubic to cubic AtC structural transition is analyzed.

\subsubsection{\label{subsubsec:D2d} $\boldsymbol{q}=\boldsymbol{0}$ transition with a one-dimensional irrep: D$_{2\rm{d}}$ to D$_2$}
\begin{figure}
\includegraphics[width=0.8\linewidth]{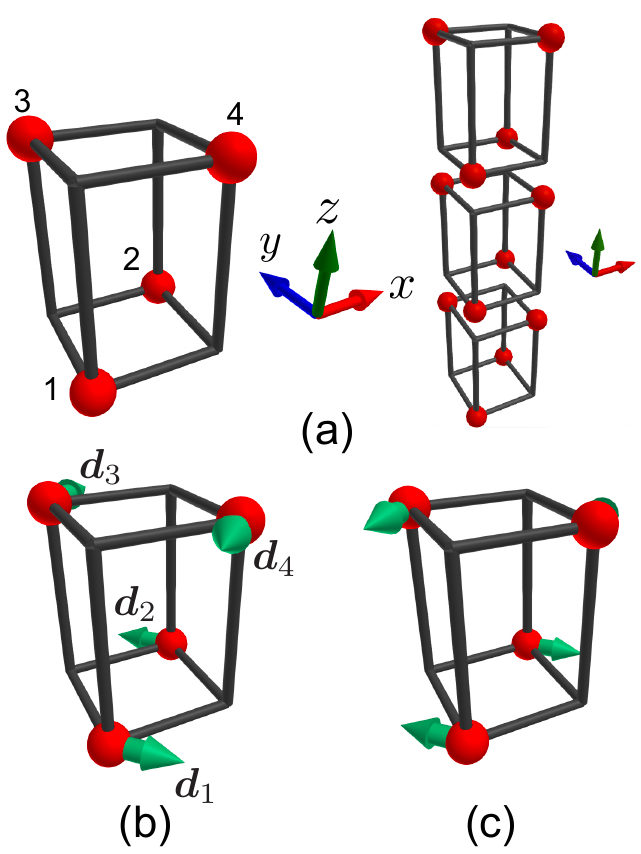}
\caption{\label{fig:D2d_image}(a) Crystal structure of the achiral tetragonal D$_{2\text{d}}$ phase. Displacement patterns that transform the achiral structure into the chiral D$_2$ phase for (b) the RH and (c) the LH structures.}
\end{figure}
Here, we consider an AtC transition that occurs at $\bm{q}=\bm{0}$. AtC transitions at $\bm{q} = \bm{0}$ have been extensively studied in recent years owing to the possibility of controlling chirality with external fields such as laser pulses~\cite{Romao2024,photo_induced_chirality,Juraschek2025}.
Moreover, such transitions can lead to right-handed (RH) and left-handed (LH) structures that belong to the same non-enantiomorphic space group~\cite{gomezortiz2026}.
In the following, we demonstrate this explicitly for the D$_{2\text{d}}$ to D$_2$ transition, where the leading contribution to $G_{0,u}(\bm{\eta})$ is the first-order term $G_u^{(1)}(\bm{\eta})$.
One of the simplest setups is a transition from a tetragonal achiral D$_{2\text{d}}$ structure with four sublattices in the unit cell aligned along the $z$ direction [Fig.~\ref{fig:D2d_image}(a)] to an orthorhombic chiral D$_2$ structure, as shown in Figs.~\ref{fig:D2d_image}(b) and \ref{fig:D2d_image}(c), where the displacements are indicated by green arrows.

In tetragonal systems, $G_0$ and $G_u$ belong to the same irrep, since the quadrupole form $u=2z^2-x^2-y^2$ is fully symmetric under the tetragonal point group, and $G_u$ therefore transforms identically to the pseudoscalar $G_0$; therefore, we focus on $G_u$ in the following. The two displacement patterns in Figs.~\ref{fig:D2d_image}(b) and \ref{fig:D2d_image}(c) correspond to RH and LH structures, respectively. The displacement patterns belong to the one-dimensional $\gamma=B_1$ irrep of the point group D$_{2\text{d}}$ and the modulation is uniform. Using Eq.~\eqref{eq:disp_general}, they are given by
\begin{align}
d_{\mu,s}(\bm{r}_i) = \eta_{B_1,\bm{q}=\bm{0}}(\bm{\epsilon}^s_{B_1,\bm{0}})_{\mu},
\end{align}
where $\boldsymbol{\epsilon}^s_{B_1,\bm{0}}$ are $\bm{\epsilon}^1_{B_1,\bm{0}} = (1,-1,0)/\sqrt{8}$, $\bm{\epsilon}^2_{B_1,\bm{0}} = (-1,1,0)/\sqrt{8}$, $\bm{\epsilon}^3_{B_1,\bm{0}} = (1,1,0)/\sqrt{8}$, and $\bm{\epsilon}^4_{B_1,\bm{0}} = (-1,-1,0)/\sqrt{8}$. We then obtain $\eta_{B_1,\bm{0}}$ as
\begin{align}
\eta_{B_1,\bm{0}} =&\sum_{\mu,s}(\boldsymbol{\epsilon}^s_{B_1,\bm{0}})_{\mu}d_{\mu,s}\notag\\
=&\sum_{s=1}^2\frac{(-1)^{s+1}}{\sqrt{8}}\left(
d_{x,s}-d_{y,s}+d_{x,s+2}+d_{y,s+2}
\right)
.\label{eq:disp_D2d}
\end{align}
Here, we omit the ordering-vector subscript in $d_{\mu,s,\bm{q}=\bm{0}}$ for simplicity. In this transition, the RH and LH structures are distinguished by the sign of $\eta_{B_1,\bm{0}}$: $\eta_{B_1,\bm{0}}>0$ ($<0$) corresponds to the RH (LH) structure. Since the $B_1$ irrep breaks all mirror and inversion symmetries of the point group D$_{2\text{d}}$, Eq.~\eqref{eq:disp_D2d} can be identified with $G_u^{(1)}(\bm{\eta})$. Thus, the order parameter of chirality coincides with that of the structural phase transition itself.

For higher-order terms, only the odd-order terms $G_{u}^{(2\ell+1)}$ ($\ell = 0,1,2,\cdots)$ are allowed since reversing the sign of $\eta_{B_1,\bm{0}}$ interchanges the RH and LH structures, implying that $G_{u}(\bm{\eta}) = -G_{u}(-\bm{\eta})$. Accordingly, $G_{u}(\bm{\eta})$ should be written as 
\begin{align}
G_{u}(\bm{\eta}) =C^{(1)}\eta_{B_1,\bm{0}} + C^{(3)}\eta^3_{B_1,\bm{0}} + C^{(5)}\eta^5_{B_1,\bm{0}}  + \cdots.
\label{eq:Gu_D2d}
\end{align}
A recent study has reported such a linear dependence of $G_u$ on $\eta_{B_1,\bm{0}}$ for the electron chirality defined via the Dirac field in the D$_{2\text{d}}$-to-D$_2$ transition~\cite{Miki_chiral}. This agreement indicates that the symmetry-based scaling derived here is shared by independent microscopic measures of chirality, supporting the universality of the present framework.

\begin{comment}
\red{
It should be noted that an AtC transition at $\bm{q} = 0$ does not necessarily lead to finite $G_{0,u}^{(1)}$. As in the case of photo-induced chirality in BPO$_4$~\cite{photo_induced_chirality}, where the doubly degenerate $E$ mode at $\bm{q} = 0$ given by $(x,y)$ induces the $G_{0,u}^{(1)}$ or $B$ mode described by $x^2 - y^2$, $G_{0,u}$ can be a secondary order parameter when the displacive mode belongs to a doubly or triply degenerate mode. In such cases, the leading term of $G_{0,u}$ is usually a second- or higher-order term of $\bm{\eta}$.
}	
\end{comment}

\subsubsection{\label{subsec:D6h}Vector chirality form of $G_u$}
\begin{figure}
\includegraphics[width=1.0\linewidth]{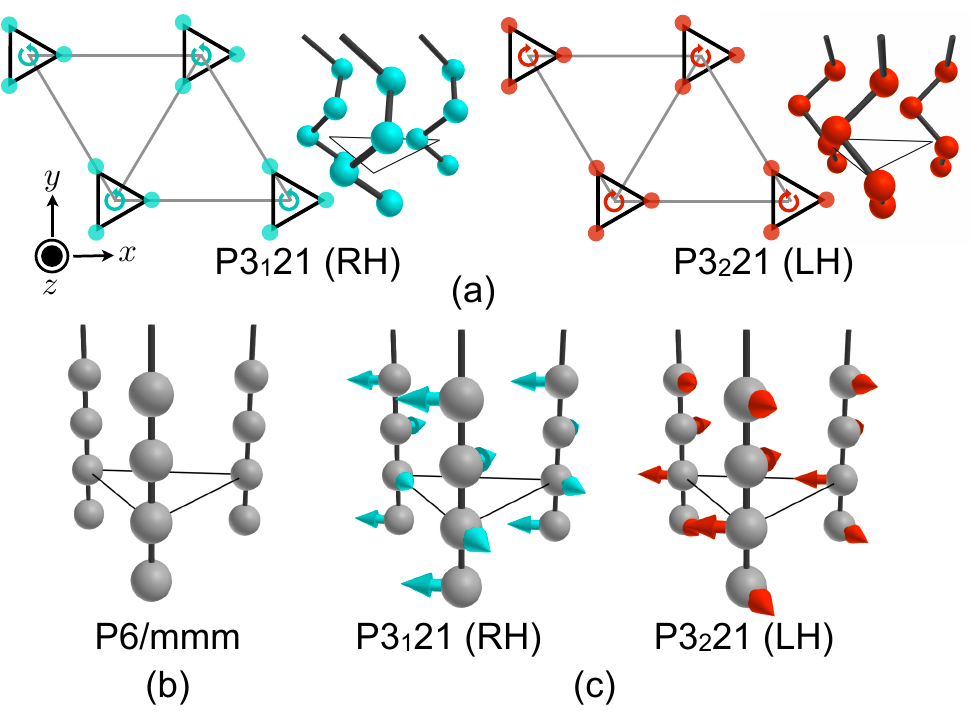}
\caption{\label{fig:D3_image}(a) Schematic illustration of the threefold helical chains belonging to the space groups P$3_121$ (blue) and P$3_221$ (orange).
(b) Crystal structure of the achiral phase of the AtC transition, which belongs to the space group P6/mmm.
(c) Displacement patterns that generate the P$3_121$ and P$3_221$ structures from the achiral phase.
}
\end{figure}

We discuss the formation of a helical structure, a prototypical example of chiral crystals. In particular, we focus on the threefold helical structures belonging to the space groups P$3_121$ (No.~152) and P$3_221$ (No.~154), both of which have the point group D$_3$. Such structures are realized in various systems, including $\alpha$-quartz~\cite{Bragg_quartz} and Te~\cite{Te_Hirayama,CIM_nature,CIM}. In these materials, threefold helices running along the $z$ direction are arranged on a triangular lattice, as schematically shown in Fig.~\ref{fig:D3_image}(a), where the blue and orange helices correspond to the structures with space groups P$3_121$ and P$3_221$, respectively. We hereafter refer to the P$3_121$ (P$3_221$) structure as RH (LH).

There are several possible high-temperature structures ($T>T_c$) that can transform into such helical structures via an AtC transition. As a simple reference configuration, we consider one-dimensional chains aligned along the $z$ axis, forming a triangular lattice in the $xy$ plane, as shown in Fig.~\ref{fig:D3_image}(b). This structure belongs to the space group P6/mmm (No.~191) and the point group D$_{6\text{h}}$.

The displacement patterns generating the helical structures are illustrated in Fig.~\ref{fig:D3_image}(c), where the blue and orange arrows denote the displacements for the RH and LH structures, respectively. These patterns are constructed from a doubly degenerate in-plane eigenmode $(\epsilon_{E_x,\bm{Q}},\epsilon_{E_y,\bm{Q}})$, which transforms as a polar vector $(x,y)$,  at the ordering vector $\bm{Q} = (0,0,1/3)$ (in units of the primitive reciprocal lattice vectors) defined by $\bm{G}_1 = \frac{2\pi}{a}(1,\frac{1}{\sqrt{3}},0)$, $\bm{G}_2 = \frac{2\pi}{a}(0,\frac{2}{\sqrt{3}},0)$, and $\bm{G}_3 = \frac{2\pi}{c}(0,0,1)$, where $a$ and $c$ are the in-plane and out-of-plane lattice constants, respectively.
Thus, the displacements ${\bm d}(\bm{r}_i)$ are described by
\begin{align}
d_{\mu}(\bm{r}_i) = \sum_{\gamma = E_x,E_y}\eta_{\gamma,\bm{Q}}(\epsilon_{\gamma,\bm{Q}})^*_\mu e^{i\bm{Q}\cdot\bm{r}_i}+\text{c.c.}
% + \eta_{\gamma,-\bm{Q}}(\epsilon_{\gamma,-\bm{Q}})^*_\mu e^{-i\bm{Q}\cdot\bm{r}_i}.
\end{align}
Here, we omit the sublattice index $s$ from the displacement $\bm{d}_{s}(\bm{r}_i)$ and the eigenmode $(\boldsymbol{\epsilon}^s_{\gamma,\bm{Q}})_{\mu} (\mu=x,y,z)$ since the primitive unit cell contains a single site.
$\bm{\epsilon}_{\gamma,\bm{Q}}$ is given by $\bm{\epsilon}_{E_x,\bm{Q}} = (1,0,0)$ and $\bm{\epsilon}_{E_y,\bm{Q}} = (0,1,0)$. In this transition, the coefficients $\bm{\eta}_{\bm{Q}} = (\eta_{E_x,\bm{Q}},\eta_{E_y,\bm{Q}}) = \bm{\eta}^{*}_{-\bm{Q}}$ determine the handedness of the chiral structure and are given by
\begin{align}
\bm{\eta}^{\pm}_{\bm{Q}} = \frac{\eta}{\sqrt{2}} e^{i\theta_{\bm{Q}}}(1,\pm i)
%,\quad\bm{\eta}^{-}_{\bm{Q}} =  \frac{\eta}{\sqrt{2}}e^{i\theta_{\bm{Q}}}(1,-i),
\label{eq:helical_dr}
\end{align}
for the RH and LH structures, respectively, where the scalar $\eta$, distinct from the order-parameter set $\bm{\eta}$, is a real number representing the amplitude and the sign, while $\theta_{\bm{Q}}$ ($0\leq\theta_{\bm{Q}}<\pi$) specifies the direction of the displacement. Then, Eq.~\eqref{eq:Psi_d_FT} leads to
\begin{align}
\eta_{E_\mu,\bm{Q}}
=
d_{\mu,\bm{Q}},
\end{align}
which shows that $\eta_{E_\mu,\bm{Q}}$ and $d_{\mu,\bm{Q}}$ are equivalent in the absence of multiple sublattice degrees of freedom.

We now construct $G_u(\bm{\eta})$ for this AtC transition. The first order term vanishes due to a finite ordering vector $\bm{Q}\ne\bm{0}$. Since this transition is driven by the multidimensional irrep and the little group at $\bm{Q}$ is identical to the point group C$_{6\text{v}}$, the second-order term $G_{u}^{(2)}$ must be finite. Noting that $\bm{\eta}_{\bm{Q}}^{\pm}$ is an eigenstate of rotational operations, it follows from Eq.~\eqref{eq:G0_second_gen} that the second-order term is given by
\begin{align}
G_{u}^{(2)}(\bm{\eta}) &\propto (\bm{\eta}^+_{\bm{Q}}\cdot\bm{\eta}^+_{-\bm{Q}}) - (\bm{\eta}^-_{\bm{Q}}\cdot\bm{\eta}^-_{-\bm{Q}})\notag\\
&= i\,(\eta_{x,\bm{Q}}\eta_{y,-\bm{Q}}-\eta_{y,\bm{Q}}\eta_{x,-\bm{Q}}),
\label{eq:Gu_sec}
\end{align}
where we have simplified the expression by denoting the irrep $E_x\rightarrow x$ and $E_y\rightarrow y$. 
%This has the same antisymmetric bilinear structure as the form $\eta_1\xi_2-\eta_2\xi_1$ listed in Table~\ref{tab:multi_copy_G0}, although there is no bilinear entry for D$_{6\rm{h}}$ at $\bm{q}=\bm{0}$ because of the horizontal-mirror parity. By contrast, for finite $\bm{Q}$, the antisymmetric product of the order parameters at $\bm{Q}$ and $-\bm{Q}$ can be odd under the relevant mirror operation.

We note that reversing the displacements, $\bm{\eta}\rightarrow -\bm{\eta}$, does not interchange the RH and LH structures since this operation is identical to the twofold rotation about the $z$ axis, which is the symmetry operation of the achiral structure. Accordingly, $G_u(\bm{\eta})$ must be invariant under $\bm{\eta}\rightarrow -\bm{\eta}$, i.e., $G_u(\bm{\eta})=G_u(-\bm{\eta})$. Therefore, $G_u(\bm{\eta})$ is an even function of the displacement, and only even-order contributions $G_u^{(2\ell)}(\bm{\eta})$ ($\ell=1,2,\cdots$) are allowed. Thus, $G_u(\bm{\eta})$ is expressed as
\begin{align}
G_u(\bm{\eta}) 
= C^{(2)}i\,(\eta_{x,\bm{Q}}\eta_{y,-\bm{Q}}-\eta_{y,\bm{Q}}\eta_{x,-\bm{Q}}) 
+ \mathcal{O}(d^4),
\label{eq:Gu_D6h}
\end{align}
where $C^{(2)}$ is the coefficient of the second-order term.

Although we have considered threefold helical chains as a specific realization of the chiral phase, Eq.~\eqref{eq:Gu_D6h} applies more generally to transitions at finite ordering vectors $\bm{Q}\not\equiv -\bm{Q}$ in which the structure remains invariant under $\bm{\eta}\rightarrow -\bm{\eta}$, provided that the bilinear combination transforms in the same way as $G_u$ under the symmetry of the parent phase. This form can be rewritten as the $z$ component of the vector chirality for uniaxial chiral systems whose helical axis is parallel to the $z$ direction,
\begin{align}
	G_u(\boldsymbol{\eta})\propto i[\boldsymbol{\eta}_{\bm{Q}}\times \boldsymbol{\eta}_{\bm{-Q}} ]_z+\cdots.
\end{align}
This expression is essentially the same as the chirality order parameter discussed in the triangular XY model~\cite{MiyashitaShiba1984,Lee1984,ObuchiKawamura2012}.

\subsubsection{\label{subsec:rhombo}Mixed form of $G_u$}
\begin{figure}
\includegraphics[width=1.0\linewidth]{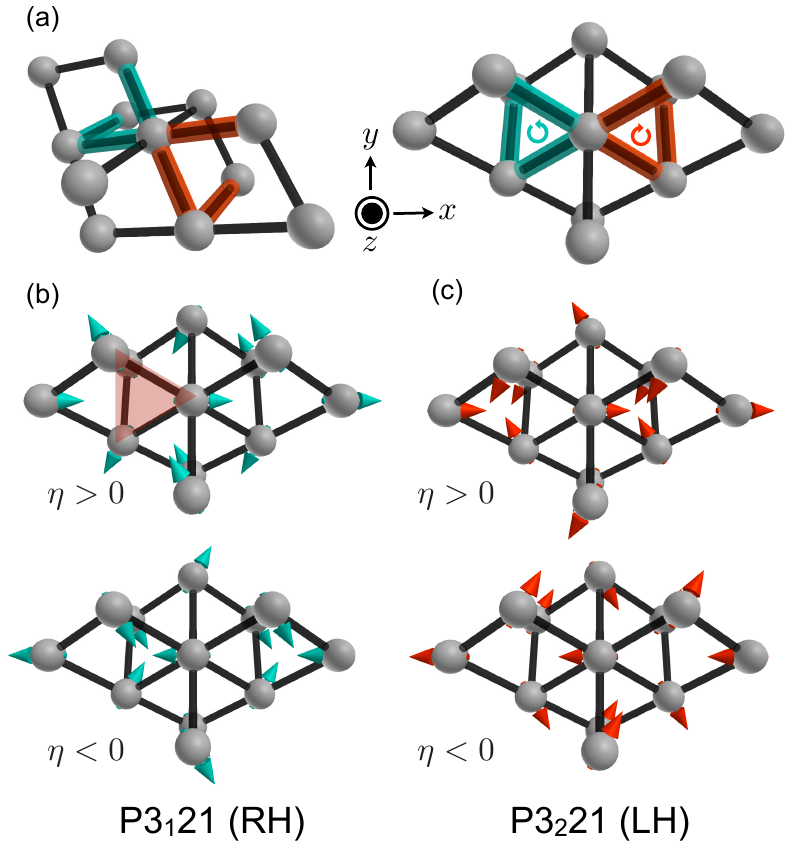}
\caption{\label{fig:D3d_image}(a) Schematic illustration of the nested helices in the achiral phase. The bonds highlighted in blue and orange represent the RH and LH helices, respectively. Displacement configurations that transform the nested helices into (b) the chiral P$3_121$ and (c) the P$3_221$ phases.}
\end{figure}
We next discuss the formation of threefold helices starting from a different parent structure. As discussed by Oiwa and Kusunose in Ref.~\cite{Oiwa_Kusunose_ET}, we consider a structure with nested RH and LH helices as the reference state. 
This structure belongs to the space group R$\bar{3}$m (No.~166) and the point group D$_{3\text{d}}$. 

Figure~\ref{fig:D3d_image}(a) schematically illustrates the nested-helices structure. 
The bonds highlighted in blue and orange form the RH and LH helices, respectively. 
This structure can be transformed into the P$3_121$ and P$3_221$ structures by introducing displacements that render the two helices inequivalent, as shown in Figs.~\ref{fig:D3d_image}(b) and \ref{fig:D3d_image}(c), corresponding to the P$3_121$ and P$3_221$ structures, respectively. 

These displacement patterns closely resemble the 120$^\circ$ spin order on a triangular lattice~\cite{Bernu,Capriotti} if they are mapped onto the triangular lattice.
Accordingly, the displacement configuration can be described by an in-plane eigenmode $(\epsilon_{E_x,\bm{Q}_\text{K}},\epsilon_{E_y,\bm{Q}_\text{K}})$ at the ordering vector $\bm{Q}_\text{K} = \frac{2\pi}{3}(-\sqrt{3},1,0)$. In units of the primitive reciprocal lattice vectors, this is given by $\bm{Q}_\text{K} = \frac{1}{3}(1,1,-2)$, where the reciprocal lattice vectors are defined as $\bm{G}_1 = 2\pi(0,\frac{2}{3a},\frac{1}{3c})$, $\bm{G}_2 = 2\pi(-\frac{1}{\sqrt{3}a},-\frac{1}{3a},\frac{1}{3c})$, and $\bm{G}_3 = 2\pi(\frac{1}{\sqrt{3}a},-\frac{1}{3a},\frac{1}{3c})$, 
where $a$ and $c$ are the in-plane and out-of-plane lattice constants, respectively. 

With this ordering vector, the displacement $\bm{d}_{\mu}(\bm{r}_i)$ is written as
\begin{align}
d_{\mu}(\bm{r}_i) = \sum_{\gamma = E_x,E_y}\eta_{\gamma,\bm{Q}_\text{K}}(\epsilon_{\gamma,\bm{Q}_\text{K}})^*_\mu e^{i\bm{Q}_\text{K}\cdot\bm{r}_i} + \text{c.c.}
\end{align}
As in Sec.~\ref{subsec:D6h}, the eigenmode $\bm{\epsilon}_{\gamma,\bm{Q}_\text{K}} = [(\epsilon_{\gamma,\bm{Q}_\text{K}})_x,(\epsilon_{\gamma,\bm{Q}_\text{K}})_y,(\epsilon_{\gamma,\bm{Q}_\text{K}})_z]$ is given by $\bm{\epsilon}_{E_x,\bm{Q}_\text{K}} = (1,0,0)$ and $\bm{\epsilon}_{E_y,\bm{Q}_\text{K}} = (0,1,0)$. The coefficient $\bm{\eta}_{\bm{Q}_\text{K}} = (\eta_{E_x,\bm{Q}_\text{K}},\eta_{E_y,\bm{Q}_\text{K}})$ is given by $\bm{\eta}^{\pm}_{\bm{Q}_\text{K}} = \eta e^{i\theta_{\bm{Q}_\text{K}}}(1,\pm i)/\sqrt{2}$ and $\bm{\eta}^{+}_{\bm{Q}_\text{K}}$ ($\bm{\eta}^{-}_{\bm{Q}_\text{K}}$) generates the RH (LH) structure. Here, $\theta_{\bm{Q}_\text{K}}$ is not arbitrary; rather, it determines the space group of the resulting chiral structure. The choice $\theta_{\bm{Q}_\text{K}} = \ell\pi/3$ ($\ell = 0,1,2,3,\cdots$) yields the P$3_121$ or P$3_221$ structure, whereas $\theta_{\bm{Q}_\text{K}} \ne \ell\pi/3$ results in the lower-symmetry P$3_1$ or P$3_2$ structure, since the displacement pattern breaks the twofold rotational symmetry about the $x$ axis, $C_2'$.

We now construct $G_u^{(n)}(\bm{\eta})$. 
Because the ordering vector is finite $\bm{Q}_{\text{K}} \neq \bm{0}$, the first-order term $G_u^{(1)}(\bm{\eta})$ is forbidden. Moreover, since the little group at $\bm{Q}_\text{K}$ is the same as the point group C$_{3\text{v}}$ and the eigenmode is multidimensional, the leading contribution arises at the second order. 
However, an important difference from the previous example is that the structures with $\eta>0$ and $\eta<0$ are \textit{inequivalent}. 
As shown in Fig.~\ref{fig:D3d_image}(b), $\eta>0$ expands the triangle highlighted in red, whereas $\eta<0$ shrinks it.  
This asymmetry implies that $|G_u(\bm{\eta})| \neq |G_u(-\bm{\eta})|$, indicating that both even- and odd-order terms appear in the expansion of $G_u(\bm{\eta})$. In the following, we demonstrate that not only the second-order term $G_u^{(2)}(\bm{\eta})$ but also the third-order term $G_u^{(3)}(\bm{\eta})$ arises in the expansion.

The second-order term can be constructed in a similar manner to Eq.~\eqref{eq:Gu_sec} as
\begin{align}
G_u^{(2)}(\bm{\eta}) 
\propto i[\eta_{x,\bm{Q}_\text{K}}\eta_{y,-\bm{Q}_\text{K}}-\eta_{y,\bm{Q}_\text{K}}\eta_{x,-\bm{Q}_\text{K}}],
\label{eq:Gu_sec_Rhombo}
\end{align}
since $\bm{Q}_\text{K} \ne -\bm{Q}_\text{K}+\bm{G}$.
Although this quantity is even under the horizontal mirror operation because $\bm{\eta}_{\bm{Q}_{\text{K}}} \parallel \bm{Q}_{\text{K}}$, there is no such operation in the point group D$_{3\text{d}}$. 
Therefore, Eq.~\eqref{eq:Gu_sec_Rhombo} is regarded as the ET quadrupole in the present system. 

The third-order term $G_u^{(3)}(\bm{\eta})$ is constructed as follows.
$G_u^{(3)}(\bm{\eta})$ must preserve the threefold rotational symmetry while breaking inversion ($I$), threefold rotoinversion ($S_6$), and diagonal mirror ($\sigma_d$) symmetries.
Since $3\bm{Q}_{\text{K}}=\bm{G}$, $G_u^{(3)}(\bm{\eta})$ is uniquely given by
\begin{align}
G_u^{(3)}(\bm{\eta})
\propto&\, \eta^3_{x,\bm{Q}_{\text{K}}} 
- 3 \eta_{x,\bm{Q}_{\text{K}}} \eta^2_{y,\bm{Q}_{\text{K}}} \notag\\
&+\eta^3_{x,-\bm{Q}_{\text{K}}} 
- 3 \eta_{x,-\bm{Q}_{\text{K}}} \eta^2_{y,-\bm{Q}_{\text{K}}}.
\label{eq:Gu_third}
\end{align}
Rewriting Eq.~\eqref{eq:Gu_third} in terms of $\eta$ and $\theta_{\bm{Q}_\text{K}}$ gives
\begin{align}
G_u^{(3)}(\bm{\eta})
\propto \eta^3 \cos(3\theta_{\bm{Q}_\text{K}}),
\end{align}
which is manifestly invariant under threefold rotation
($\theta_{\bm{Q}_\text{K}} \rightarrow \theta_{\bm{Q}_\text{K}} + 2\pi/3$). Here, we have used the relation $\theta_{\bm{Q}_\text{K}} = \theta_{-\bm{Q}_\text{K}}$ arising from $\bm{\eta}_{\bm{Q}_\text{K}} = \bm{\eta}^*_{-\bm{Q}_\text{K}}$.

Thus, for small $|\eta|$, $G_u(\bm{\eta})$ in the present system is expressed as
\begin{align}
G_u(\bm{\eta}) 
\sim& C^{(2)}i[\eta_{x,\bm{Q}_\text{K}}\eta_{y,-\bm{Q}_\text{K}}-\eta_{y,\bm{Q}_\text{K}}\eta_{x,-\bm{Q}_\text{K}}]\notag\\ 
&+ C^{(3)}
\eta^3 \cos(3\theta_{\bm{Q}_\text{K}}),
\label{eq:Gu_D3d}
\end{align}
where $C^{(2)}$ and $C^{(3)}$ are the coefficients of the second- and third-order terms.

As mentioned earlier in this section, the presence of both even- and odd-order terms in $G_{0,u}(\boldsymbol{\eta})$ implies that the structures with $\boldsymbol{\eta}$ and $-\bm{\eta}$ are energetically inequivalent. Therefore, one of the two structures is always chosen at the AtC transition. We note that this third-order contribution vanishes for $\theta_{\bm{Q}_\text{K}}=\pi\ell/3+\pi/6$ $(\ell=0,1,\cdots,5)$. For these special phases, reversing the displacement, $\boldsymbol{\eta}\to-\boldsymbol{\eta}$, is equivalent to a $C_2'$ operation about an axis perpendicular to the displacement direction in the achiral phase. These configurations belong to the lower-symmetry P$3_1$ (P$3_2$) structures for $\theta_{\bm{Q}_\text{K}} \ne \ell\pi/3$ rather than P$3_121$ (P$3_221$).

\subsubsection{\label{subsec:fcc}Cubic achiral-to-chiral transitions with triple-$\boldsymbol{q}$ order}
\begin{figure}
\includegraphics[width=1.0\linewidth]{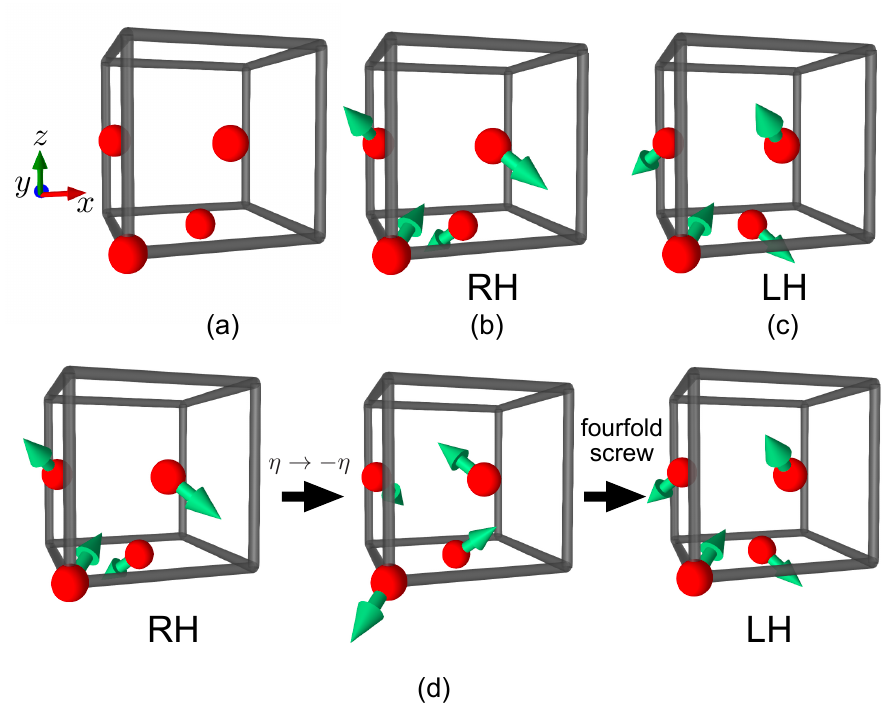}
\caption{\label{fig:fcc_image}(a) Crystal structure of the fcc lattice in the achiral Fm$\bar{3}$m phase. Displacement configurations that induce (b) the RH and (c) the LH chiral P$2_1$3 structures, respectively.
(d) Schematic illustration of the relationship between the RH and LH structures.}
\end{figure}

%In cubic systems, wave vectors in the first BZ satisfying $\bm{q}_1+\bm{q}_2+\bm{q}_3=\bm{G}$ include the M points $(1/2,1/2,0)$ in the simple-cubic (sc) lattice, the X points $(1/2,0,0)$ in the fcc  lattice, and the H points $(1/2,0,0)$ in the body-centered cubic (bcc) lattice. These points lie on the BZ boundary and antisymmetric combinations therefore vanish. An exception arises for lattices with point group T$_\text{d}$, where the little group at these points is D$_{2\text{d}}$. In this case, an $x^2-y^2$-type irrep, constructed from the product of doubly degenerate modes $\sim (x,y)$, can transform as $G_u$, and a finite $G_0^{(2)}(\bm{\eta})$ can be obtained by summing over the three $G_u^{(2)}$ for the ordering vectors $\bm{q}_1$, $\bm{q}_2$, and $\bm{q}_3$. 
Here, we consider an example in which the second-order term vanishes, and thus, a higher-order term becomes the leading term. As a representative example, we consider a triple-$\bm{q}$ order at the X points of the fcc lattice, which belongs to the space group Fm$\bar{3}$m (No.~225) with point group O$_\text{h}$. As we have discussed in Sec.~\ref{subsec:conditions}, the little group at the X points of the space group Fm$\bar{3}$m is identical to the point group D$_{4\text{h}}$, and thus, the second-order term is prohibited. As a result, the third-order term is expected to become the leading contribution for this triple-$\bm{q}$ order.

The resulting structure belongs to the space group P$2_1$3 (No.~198) with point group T, which hosts B20 compounds such as MnSi~\cite{MnSi}, FeGe~\cite{FeGe}, and CoSi~\cite{CoSi}. This space group has been extensively studied and is known to host unconventional band degeneracies, including a sixfold Dirac point~\cite{Barry_science}. It is therefore of particular interest to analyze this structure from the viewpoint of chirality.

The fcc lattice and the displacements leading to the chiral P$2_1$3 phase are illustrated in Figs.~\ref{fig:fcc_image}(a) and \ref{fig:fcc_image}(b), respectively. The displacement pattern shown in Fig.~\ref{fig:fcc_image}(b) is described as follows.
 %As the high-temperature achiral phase, we consider the simple fcc lattice shown in Fig.~\ref{fig:fcc_image}(a), which belongs to the space group Fm$\bar{3}$m with point group O$_\text{h}$. 
Let the four fcc sites in the cubic unit cell be given by 
$\bm{r}_1 = (0,0,0)$, 
$\bm{r}_2 = (1/2,1/2,0)$, 
$\bm{r}_3 = (0,1/2,1/2)$, and 
$\bm{r}_4 = (1/2,0,1/2)$, where the lattice constant is set to unity.  
When these sites are displaced to $\bm{r}_i + \bm{d}(\bm{r}_i)$ ($i=1,2,3,4$), with 
$\bm{d}(\bm{r}_1) = (d,d,d)$, 
$\bm{d}(\bm{r}_2) = (-d,d,-d)$, 
$\bm{d}(\bm{r}_3) = (-d,-d,d)$, and 
$\bm{d}(\bm{r}_4) = (d,-d,-d)$, 
the lattice transforms into a cubic chiral structure belonging to the space group P$2_1$3 with point group T, as illustrated in Fig.~\ref{fig:fcc_image}(b). The resulting atomic positions $\bm{r}_i + \bm{d}(\bm{r}_i)$ correspond to the Wyckoff position $4a$ of P$2_1$3.

We now show that this AtC transition can actually be driven by a triple-$\bm{q}$ order at the three inequivalent X points in the fcc BZ, $\bm{Q}_\text{X} = (1,0,0)$, $\bm{Q}_\text{Y} = (0,1,0)$, and $\bm{Q}_\text{Z} = (0,0,1)$ in units of $2\pi/a$.
For this triple-$\bm{q}$ order, the displacement at the $i$th site with coordinate $\bm{r}_i$ is given by
\begin{align}
\bm{d}(\bm{r}_i) 
=& \sum_{\bm{q} = \bm{Q}_{\text{X}}, \bm{Q}_{\text{Y}}, \bm{Q}_{\text{Z}}}
\bm{\eta}_{\bm{q}} e^{i\bm{q}\cdot\bm{r}_i}\notag\\
=&
\begin{pmatrix}
\eta_{x,\bm{Q}_\text{X}}&\eta_{x,\bm{Q}_\text{Y}}&\eta_{x,\bm{Q}_\text{Z}}\\
\eta_{y,\bm{Q}_\text{X}}&\eta_{y,\bm{Q}_\text{Y}}&\eta_{y,\bm{Q}_\text{Z}}\\
\eta_{z,\bm{Q}_\text{X}}&\eta_{z,\bm{Q}_\text{Y}}&\eta_{z,\bm{Q}_\text{Z}}
\end{pmatrix}
\begin{pmatrix}
e^{i\bm{Q}_\text{X}\cdot\bm{r}_i}\\e^{i\bm{Q}_\text{Y}\cdot\bm{r}_i}\\e^{i\bm{Q}_\text{Z}\cdot\bm{r}_i}
\end{pmatrix},
\label{eq:disp_tripleQ}
\end{align}
where $\bm{\eta}_{\bm{q}} = (\eta_{x,\bm{q}}, \eta_{y,\bm{q}}, \eta_{z,\bm{q}})$. 
At each X point, the vector $\bm{\eta}_{\bm{q}}$ decomposes into one longitudinal mode and two transverse modes due to the D$_{4\mathrm{h}}$ symmetry. The longitudinal mode belongs to the $A_{2\rm{u}}$ irrep, while the twofold-degenerate transverse modes belong to the $E_{\rm u}$ irrep of D$_{4\text{h}}$.

The triple-$\bm{q}$ order of the longitudinal mode leads to a uniform expansion or contraction of the tetrahedron formed by the sites $\bm{r}_i$ ($i=1,2,3,4$). In contrast, the triple-$\bm{q}$ order of the transverse modes induces the AtC transition shown in Fig.~\ref{fig:fcc_image}(b). Retaining only the transverse modes, Eq.~\eqref{eq:disp_tripleQ} reduces to
\begin{align}
\bm{d}(\bm{r}_i)
=&
\begin{pmatrix}
0&\eta_{x,\bm{Q}_\text{Y}}&\eta_{x,\bm{Q}_\text{Z}}\\
\eta_{y,\bm{Q}_\text{X}}&0&\eta_{y,\bm{Q}_\text{Z}}\\
\eta_{z,\bm{Q}_\text{X}}&\eta_{z,\bm{Q}_\text{Y}}&0
\end{pmatrix}
\begin{pmatrix}
e^{i\bm{Q}_\text{X}\cdot\bm{r}_i}\\e^{i\bm{Q}_\text{Y}\cdot\bm{r}_i}\\e^{i\bm{Q}_\text{Z}\cdot\bm{r}_i}
\end{pmatrix}.
\label{eq:disp_fcc}
\end{align}
If the transition occurs with 
$\eta_{y,\bm{Q}_{\text{X}}} = \eta_{z,\bm{Q}_{\text{Y}}} = \eta_{x,\bm{Q}_{\text{Z}}} = \eta$ while all other components vanish, Eq.~\eqref{eq:disp_fcc} reduces to
\begin{align}
\bm{d}(\bm{r}_i)
=&\eta
\begin{pmatrix}
0&0&1\\
1&0&0\\
0&1&0
\end{pmatrix}
\begin{pmatrix}
e^{i\bm{Q}_\text{X}\cdot\bm{r}_i}\\e^{i\bm{Q}_\text{Y}\cdot\bm{r}_i}\\e^{i\bm{Q}_\text{Z}\cdot\bm{r}_i}
\end{pmatrix}.
\label{eq:fcc_RH}
\end{align}
Here, $\eta$ is real since $\bm{Q}_{\text{X,Y,Z}}\equiv-\bm{Q}_{\text{X,Y,Z}}$. This yields 
$\bm{d}(\bm{r}_1) = (\eta,\eta,\eta)$, 
$\bm{d}(\bm{r}_2) = (-\eta,\eta,-\eta)$, 
$\bm{d}(\bm{r}_3) = (-\eta,-\eta,\eta)$, and 
$\bm{d}(\bm{r}_4) = (\eta,-\eta,-\eta)$, reproducing the structure in Fig.~\ref{fig:fcc_image}(b). 

By contrast, choosing the complementary set of transverse modes, 
$\eta_{z,\bm{Q}_{\text{X}}} = \eta_{x,\bm{Q}_{\text{Y}}} = \eta_{y,\bm{Q}_{\text{Z}}} = \eta$, yields
\begin{align}
\bm{d}(\bm{r}_i)
=&\eta
\begin{pmatrix}
0&1&0\\
0&0&1\\
1&0&0
\end{pmatrix}
\begin{pmatrix}
e^{i\bm{Q}_\text{X}\cdot\bm{r}_i}\\e^{i\bm{Q}_\text{Y}\cdot\bm{r}_i}\\e^{i\bm{Q}_\text{Z}\cdot\bm{r}_i}
\end{pmatrix}.
\label{eq:fcc_LH}
\end{align}
The resulting structure is the mirror image of that in Eq.~(\ref{eq:fcc_RH}), as shown in Fig.~\ref{fig:fcc_image}(c). We therefore refer to the structures in Fig.~\ref{fig:fcc_image}(b) [Eq.~(\ref{eq:fcc_RH})] and Fig.~\ref{fig:fcc_image}(c) [Eq.~(\ref{eq:fcc_LH})] as the RH and LH structures, respectively.

Moreover, the RH and LH structures are interchanged by $\bm{\eta} \rightarrow -\bm{\eta}$ combined with a fourfold screw operation about the $z$ axis, $\lbrace C_4^{-1}|\bm{\tau}\rbrace$ with $\bm{\tau} = (0,0,1/2)$, about the center of the tetrahedron, as illustrated in Fig.~\ref{fig:fcc_image}(d). This symmetry relation implies that $G_0(\bm{\eta}) = -G_0(-\bm{\eta})$, so that only odd-order contributions are allowed. However, the first-order term $G_0^{(1)}(\bm{\eta})$ is forbidden because $\delta_{\bm{Q}_{\text{X}},\bm{G}}=0$. Therefore, the third-order term $G_0^{(3)}(\bm{\eta})$ provides the leading contribution, since $\bm{Q}_{\text{X}}+\bm{Q}_{\text{Y}}+\bm{Q}_{\text{Z}}=\bm{G}$.

We now construct $G_0^{(3)}(\bm{\eta})$. The third-order contribution must be invariant under all rotational operations, including the fourfold rotation about $[001]$ and the threefold rotation about $[111]$, while being odd under all mirror and inversion operations. A quantity satisfying these constraints can be constructed as
\begin{align}
G_0^{(3)}(\bm{\eta}) 
\propto \eta_{y,\bm{Q}_{\text{X}}}
 \eta_{z,\bm{Q}_{\text{Y}}}
  \eta_{x,\bm{Q}_{\text{Z}}}
- \eta_{z,\bm{Q}_{\text{X}}}
  \eta_{x,\bm{Q}_{\text{Y}}}
  \eta_{y,\bm{Q}_{\text{Z}}}.
\label{eq:G0_fcc}
\end{align}
If the transverse modes given by Eq.~(\ref{eq:fcc_RH}), 
$\eta_{y,\bm{Q}_{\text{X}}} 
= \eta_{z,\bm{Q}_{\text{Y}}} 
= \eta_{x,\bm{Q}_{\text{Z}}} = \eta$, 
condense to form the RH structure, Eq.~\eqref{eq:G0_fcc} yields $G_0^{(3)}(\bm{\eta})=\eta^3$. In contrast, for the complementary set of transverse modes 
$\eta_{z,\bm{Q}_{\text{X}}} 
= \eta_{x,\bm{Q}_{\text{Y}}} 
= \eta_{y,\bm{Q}_{\text{Z}}} = \eta$ [Eq.~(\ref{eq:fcc_LH})], 
which induces the LH structure, one obtains $G_0^{(3)}(\bm{\eta})=-\eta^3$. This confirms that $G_0^{(3)}(\bm{\eta})$ defined in Eq.~\eqref{eq:G0_fcc} reflects the chirality of the structure and serves as the ET monopole, i.e., the order parameter of chirality for this cubic-to-cubic AtC transition.

Moreover, the sign choices of $\eta_{\mu,\bm{q}}$ correspond to the triple-$\bm{q}$ domains. In addition to the choice $(\eta_{y,\bm{Q}_X},\eta_{z,\bm{Q}_Y},\eta_{x,\bm{Q}_Z})=(\eta,\eta,\eta)$, other sign choices $(\eta,-\eta,-\eta)$, $(-\eta,\eta,-\eta)$, and $(-\eta,-\eta,\eta)$ give an identical structure up to a simple translation, with $G_0^{(3)}(\boldsymbol{\eta})=\eta^3$ kept invariant. As noted above, the choice $(\eta_{y,\bm{Q}_X},\eta_{z,\bm{Q}_Y},\eta_{x,\bm{Q}_Z})=(-\eta,-\eta,-\eta)$ gives opposite $G_0^{(3)}(\boldsymbol{\eta})=-\eta^3$. This also means 
that if one changes the sign of one of the three components as $(-\eta,\eta,\eta)$, $(\eta,-\eta,\eta)$, or $(\eta,\eta,-\eta)$ from $(\eta,\eta,\eta)$, the resulting structure is identical to that with $\boldsymbol{\eta}\to -\boldsymbol{\eta}$, i.e.,
equivalent to the other choice
$(\eta_{z,\bm{Q}_\text{X}},\eta_{x,\bm{Q}_\text{Y}},\eta_{y,\bm{Q}_\text{Z}})=(\eta,\eta,\eta)$ in Eq.~(\ref{eq:fcc_LH}). Such a multi-domain structure is expected to be encoded in the universality class as discussed in triple-$\bm{q}$ orders in quadrupolar triangular systems~\cite{Hattori2024}, where the critical Ashkin-Teller universality class emerges. Thus, examining the statistical aspect of this cubic AtC transition is also an interesting future problem. 

%Furthermore, this result suggests that the sixth-order term in the free energy, $\mathcal{F}^{(6)}$, plays an essential role in stabilizing this triple-$\bm{q}$ order, since the lowest-order contribution of $G_u(\bm{\eta})$ to the free energy is given by $\mathcal{F}^{(6)} \propto [G_0^{(3)}(\bm{\eta})]^2$.

\section{\label{sec:numerical}Chirality-induced phenomena}
So far, we have derived the ET monopole $G_0$ and quadrupole $G_u$ in terms of the displacive modes for several structural phase transitions. These results are based on the symmetry arguments for each AtC phase transition. In this section, we clarify how they influence physical observables. 
To this end, we calculate the displacement dependence of chiral-phonon splitting (CPS) and current-induced magnetization (CIM) for the model systems introduced in Sec.~\ref{sec:toroidals}. We then compare these results with the symmetry-derived displacement dependence of $G_{0,u}$.

%So far we have derived the ET monopole $G_0$ and quadrupole $G_u$ in terms of the displacive modes of the several structural phase transitions. However, it is still unclear how they influence physical observables. Therefore, in this section, we numerically calculate the displacement dependence of chirality induced phenomena such as the CPS and the current-induced magnetization of the model systems we have considered in Sec.~\ref{sec:toroidals} to compare the displacement dependence of such quantities and that of $G_{0,u}$ we have already derived. 

\subsection{\label{subsec:phonon}Chiral-phonon splitting}
We investigate changes in phonon dispersions induced by AtC structural phase transitions, with a particular focus on the displacement dependence of CPS. We begin with a brief overview of chiral phonons. Chiral phonons are modes that carry finite angular momentum $\bm{L}$ along their propagation direction $\bm{k}$, where the angular momentum of the phonon labeled by the wave vector $\bm{k}$ and the band index $n$ is defined as~\cite{Zhu_ChiralPhonon},
\begin{align}
L_\mu(\bm{k},n) =u^\dagger_{n,\bm{k}}\mathscr{L}_\mu u_{n,\bm{k}},\label{eq:phonon_angular_def}
\end{align}
where $u_{n,\bm{k}}$ is the eigenvector for the $n$th phonon branch at $\bm{k}$ and $\mathscr{L}_\mu$ is given by the tensor product of the $N_s\times N_s$ identity matrix $I_{N_s\times N_s}$ in the sublattice sector and the $\mu$ component of the generator of SO$(3)$ rotations for the displacements. The matrix elements are $(\mathscr{L}_\mu)_{\nu\lambda} \equiv I_{N_s\times N_s}\otimes (-i\varepsilon_{\mu\nu\lambda})$ ($\mu,\nu,\lambda = x,y,z$).  Using $\bm{L}$ and $\bm{k}$, one can construct an electric toroidal monopole $G_0$ and a quadrupole $G_u$ as $G_0 = \bm{k}\cdot\bm{L}$ and $G_u = 2k_zL_z - k_xL_x - k_yL_y$. Here, $\bm{L}$ is an axial vector, whereas $\bm{k}$ is the phonon wave vector and thus a polar vector. For $G_u$, the $z$ axis is taken to be the $c$ axis in noncubic systems. When such symmetry-allowed couplings appear in the dynamical matrix, phonon modes with opposite angular momenta can split at a given $\bm{k}$, leading to CPS.

%Here, we investigate changes in phonon dispersions due to the structural AtC transitions, particularly we focus on displacement dependences of CPS. 
%First, let us briefly review the chiral phonons. The chiral phonons are phonons which possess finite angular momenta $\bm{L}$ along their propagating direction $\bm{k}$. With $\bm{L}$ and $\bm{k}$, the ET monopole $G_0$ and the quadrupole $G_u$ can be constructed as $G_0 = \bm{k}\cdot\bm{L}$ and $G_u = 2k_zL_z -k_xL_x -k_yL_y$ since $\bm{L}$ is an axial vector while $\bm{k}$ is the wavenumber of phonons, and thus, a polar vector. Therefore, at a certain $\bm{k}$, the energy of the phonon with $\bm{L}>0$ and $\bm{L}<0$ splits, which yields the CPS.
\subsubsection{\label{subsubsec:Model_Method}Model and Method}
To examine CPS, we construct a simple spring model for phonons within the harmonic approximation. The potential energy of the lattice system is written as
\begin{align}
V =  \sum_{i,j,s,s',\mu,\nu}
u^\mu_{i,s}
\mathcal{D}_{\mu,s;\nu,s'}(i,j)
u^\nu_{j,s'},
\label{eq:Hami_phonon}
\end{align}
where $i,j=1,2,\cdots,N$ are the unit-cell indices and $s,s'=1,2,\cdots,N_s$ are the sublattice indices, with $N$ the total number of unit cells and $N_s$ the number of sublattices per unit cell. $u^\mu_{i,s}$ is the $\mu$ component of the lattice displacement from the equilibrium position at site $(i,s)$, and $\mathcal{D}_{\mu,s;\nu,s'}(i,j)$ is the dynamical matrix, which encodes the crystal structure and interatomic forces.

In the spring model, the dynamical matrix is constructed by projecting relative displacements onto the bond direction, yielding
\begin{align}
&\sum_{i,j,s,s',\mu,\nu}u^\mu_{i,s}
\mathcal{D}_{\mu,s;\nu,s'}(i,j)
u^\nu_{j,s'}\notag\\
&\hspace{8mm}=\sum_{i,j,s,s',\mu,\nu}k_{i,s;j,s'}
\big[
(\bm{u}_{i,s} - \bm{u}_{j,s'})
\cdot
\hat{R}_{i,s;j,s'}
\big]^2,\\
&\hat{R}_{i,s;j,s'}=\frac{\bm{r}_{i,s}-\bm{r}_{j,s'}}{|\bm{r}_{i,s}-\bm{r}_{j,s'}|},\label{eq:def_unitR}
\end{align}
where $\bm{r}_{i,s}$ is the position of the $s$th sublattice in the $i$th unit cell, $\hat{R}_{i,s;j,s'}$ is the unit vector along the bond direction, and $k_{i,s;j,s'}$ is the corresponding spring constant. For convenience, we use the vector notation $\bm{u}_{i,s} = (u^x_{i,s},u^y_{i,s},u^z_{i,s})$.
We model the spring constant as $k_{i,s;j,s'} \equiv \exp[-(|\bm{R}_{i,s;j,s'}|/R_{\text{nn}}-1)]$, assuming a Born--Mayer-type potential~\cite{BornMayer1932}, where $R_\text{nn}$ is the nearest-neighbor distance. 
Here, we adopt the Born--Mayer form assuming short-range forces dominate. As long as the spring constant is isotropic, its functional form affects only quantitative details, while the symmetry-governed qualitative behavior remains unchanged.
In this way, structural transformations are incorporated into the dynamical matrix through $k_{i,s;j,s'}$ and $\hat{R}_{i,s;j,s'}$.

In our calculations, we use the enlarged unit cell of the chiral phase for both chiral and achiral phases, since some transitions occur at finite ordering vectors. We fix the displacement pattern and parametrize it by the amplitude and sign of the coefficient $\eta_{\gamma,\bm{q}}$. Hereafter, we denote $\eta \equiv \eta_{\gamma,\bm{q}}$ for simplicity. For periodic systems, $\mathcal{D}_{\mu,s;\nu,s'}(i,j)$ is diagonalized by Fourier transformation, yielding $\mathcal{D}_{\mu,s;\nu,s'}(\bm{k})$, from which the phonon dispersion $\omega_{\bm{k},n}$ of the $n$th branch and the corresponding eigenvector $u_{n,\bm{k}}$ are obtained.

In an achiral structure, some phonon branches are degenerate at high-symmetry points or along high-symmetry lines in the BZ, i.e., $\omega_{\bm{k},\ell}=\omega_{\bm{k},\ell'}$. Such degeneracies can be lifted by an AtC transition away from symmetry-enforced degeneracy points, leading to the emergence of chiral phonons. We evaluate CPS by calculating the splitting $|\omega_{\bm{k},\ell'}-\omega_{\bm{k},\ell}| \equiv |\Delta\omega^{(\ell,\ell')}_{\bm{k}}|$ as a function of the displacement amplitude $\eta$.

In the following sections, we present results for CPS in the model systems introduced in Secs.~\ref{subsubsec:D2d}, \ref{subsec:D6h}, \ref{subsec:rhombo}, and \ref{subsec:fcc}. We evaluate $|\Delta\omega^{(\ell,\ell')}_{\bm{k}}|$ at $\bm{k}=(0,0,k)$ for noncubic systems and at $\bm{k}=(k,k,k)$ for cubic systems. This choice is motivated by symmetry: in noncubic systems, a finite $G_u$ induces couplings of the form $k_zL_z$ or $2k_zL_z - k_xL_x - k_yL_y$, whereas in cubic systems, $G_0$ induces $k_xL_x + k_yL_y + k_zL_z$ due to the threefold rotational symmetry about the $[111]$ direction.

\subsubsection{\label{sec:CPS_D2d}$\text{D}_{2\text{d}}$ to $\text{D}_2$}
\begin{figure}[t]
\includegraphics[width=1.0\linewidth]{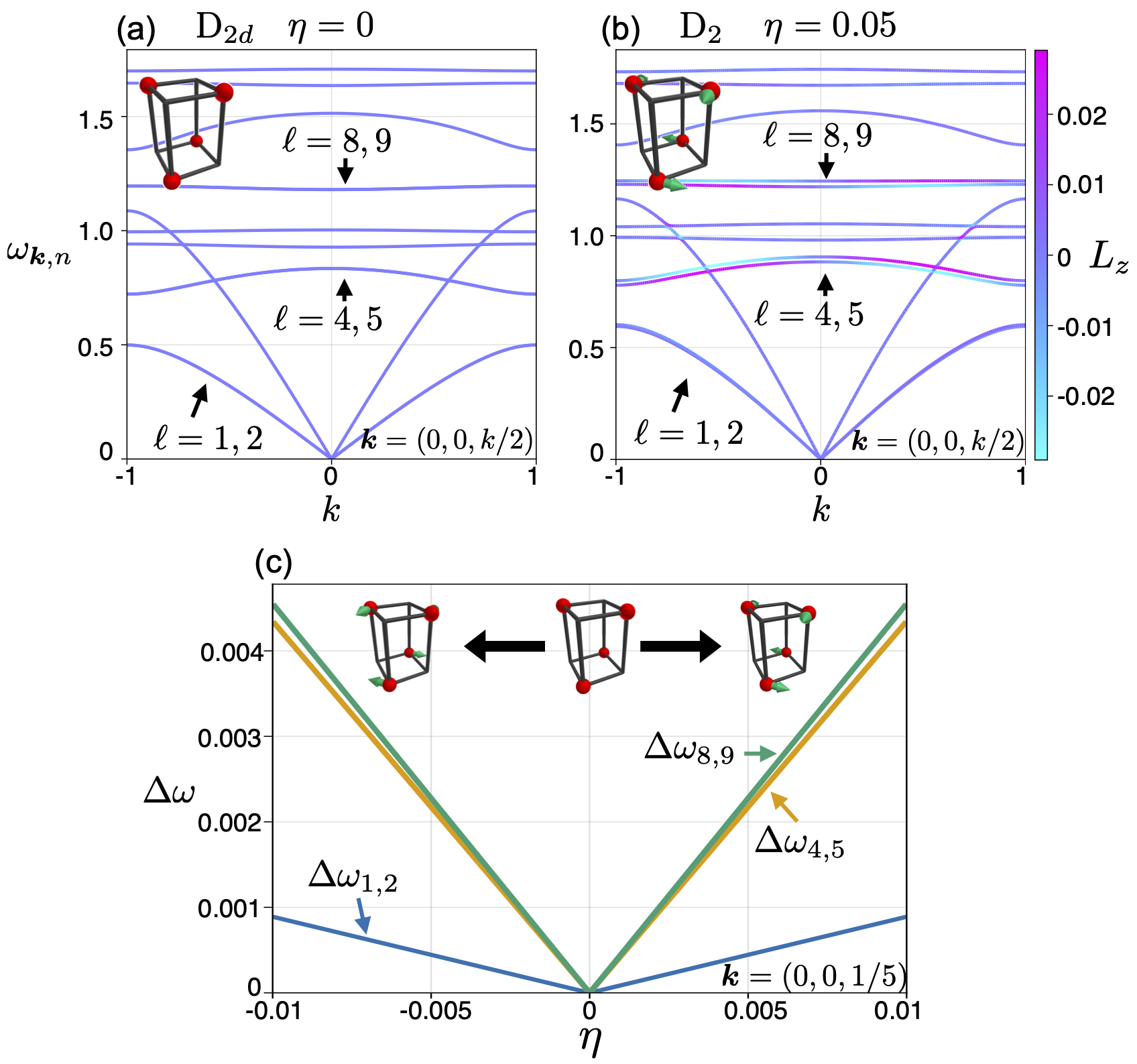}
\caption{\label{fig:CPS_D2d} 
Phonon dispersions of (a) the D$_{2\text{d}}$ and (b) the D$_2$
structures along $\bm{k} = (0,0,k/2)$ ($-1 \leq k \leq 1$).
The color represents the phonon angular momentum.
The branches exhibiting the splitting, $\ell = 1,2$, $\ell = 4,5$,
and $\ell = 8,9$, are indicated by arrows.
(c) The $\eta$ dependence of the splittings
$\Delta\omega_{1,2}(\eta)$, $\Delta\omega_{4,5}(\eta)$,
and $\Delta\omega_{8,9}(\eta)$ for $-0.01 \leq \eta \leq 0.01$ at $\bm{k} = (0,0,1/5)$.
}
\end{figure}

We first consider the transition from the D$_{2\text{d}}$ to D$_2$ structure discussed in Sec.~\ref{subsubsec:D2d}. The phonon dispersions along $\bm{k}=(0,0,k)$ ($-1/2 \leq k \leq 1/2$) for the achiral D$_{2\text{d}}$ ($\eta=0$) and chiral D$_2$ ($\eta=0.05$) phases are shown in Figs.~\ref{fig:CPS_D2d}(a) and \ref{fig:CPS_D2d}(b), respectively. The color of the bands represents the phonon angular momentum along the $z$ axis $L_z(\bm{k},n)$ at each wave vector $\bm{k}$ for the $n$th band. 

In the achiral phase, $L_z(\bm{k},n)=0$ for all $\bm{k}$ and $n$. In contrast, in the chiral phase, some branches acquire finite $L_z(\bm{k},n)$, as shown in Fig.~\ref{fig:CPS_D2d}(b). Despite the emergence of finite angular momentum, time-reversal symmetry (TRS) is preserved, and $L_z(\bm{k},n)$ satisfies $L_z(\bm{k},n) = -L_z(-\bm{k},n)$. Some branches that are twofold degenerate in the achiral phase exhibit splitting due to the symmetry lowering from D$_{2\text{d}}$ to D$_2$. These branches are labeled as $\ell=1,2$, $\ell=4,5$, and $\ell=8,9$ in Figs.~\ref{fig:CPS_D2d}(a) and \ref{fig:CPS_D2d}(b).

We next examine how the splittings of the branches $\ell=1,2$, $\ell=4,5$, and $\ell=8,9$ evolve as the AtC transition progresses. We calculate $|\Delta \omega^{(\ell,\ell')}_{\bm{k}}|$ for these pairs while varying $\eta=\eta_{B_1,\bm{0}}$ in Eq.~(\ref{eq:disp_D2d}). Fixing $\bm{k}=(0,0,1/5)$, we denote the $\eta$ dependence of the splitting as $\Delta \omega_{\ell,\ell'}(\eta)$ for simplicity. The results are shown in Fig.~\ref{fig:CPS_D2d}(c), where the blue, yellow, and green lines represent $\Delta \omega_{1,2}(\eta)$, $\Delta \omega_{4,5}(\eta)$, and $\Delta \omega_{8,9}(\eta)$, respectively. As expected from the expression for $G_u$ in Eq.~\eqref{eq:Gu_D2d}, the splittings grow linearly with $\eta$, i.e., $\Delta \omega_{\ell,\ell'}(\eta) \propto \eta\sim G_u^{(1)}$.

It should be noted that this splitting is not solely due to the emergence of chirality. In the point group D$_{2\text{d}}$, $x$ and $y$ form the two-dimensional $E$ representation and are thus degenerate. Upon the transition, however, $x$ and $y$ become inequivalent, lifting this degeneracy, since the point group D$_2$ belongs to the orthorhombic system. As a result, the twofold degeneracy is lifted even at the $\Gamma$ point (except for the acoustic modes), as shown in Fig.~\ref{fig:CPS_D2d}(b) for $\ell=4,5$ and $\ell=8,9$.

\subsubsection{\label{sec:CPS_D6h}$\text{P}6/\text{mmm}$ to $\text{P}3_121$ and $\text{P}3_221$}
\begin{figure}[t]
\includegraphics[width=1.0\linewidth]{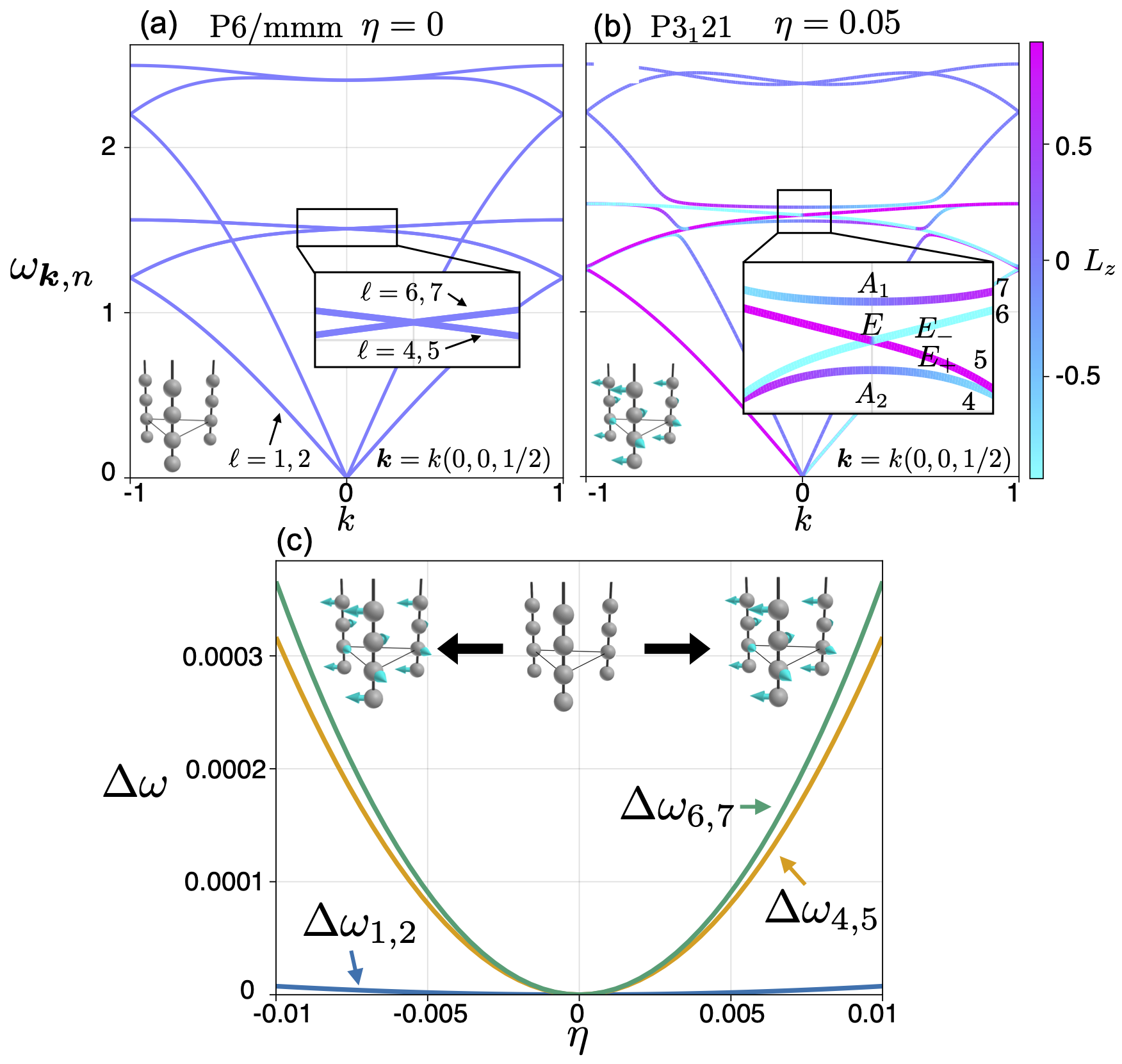}
\caption{\label{fig:CPS_Te} Phonon dispersions of the (a) achiral P6/mmm structure
($\eta = 0$) and (b) chiral P$3_121$ ($\eta = 0.05$) structures
along $\bm{k} = (0,0,k/2)$ ($-1 \leq k \leq 1$).
The insets in (a) and (b) show the fourfold degeneracy of the optical modes
at $\bm{k} = 0$ and its splitting into the corresponding irreps, respectively. (c) The $\eta$ dependence of the splittings
$\Delta\omega_{1,2}(\eta)$, $\Delta\omega_{4,5}(\eta)$,
and $\Delta\omega_{6,7}(\eta)$ for $-0.01 \leq \eta \leq 0.01$.
The splittings are measured at $\bm{k} = (0,0,1/5)$.
}
\end{figure}

We now evaluate $|\Delta \omega^{(\ell,\ell')}_{\bm{k}}|$ for the AtC transition from the P6/mmm to the P$3_121$ (P$3_221$) structure discussed in Sec.~\ref{subsec:D6h}. The phonon dispersions for the achiral P6/mmm and chiral P$3_121$ structures along $\bm{k}=(0,0,k/2)$ ($-1 \leq k \leq 1$) are shown in Figs.~\ref{fig:CPS_Te}(a) and \ref{fig:CPS_Te}(b), respectively. Here, we represent  $\bm{k}$ in the reciprocal lattice vectors for P$3_121$ or P$3_221$ for both structures. 

In the achiral phase, all phonon branches carry zero angular momentum, $L_z(\bm{k},n)=0$, reflecting the presence of mirror symmetry, which forbids chiral phonons. A fourfold degeneracy appears among the optical modes at the $\Gamma$ point as shown in the inset of Fig.~\ref{fig:CPS_Te}(a); however, this degeneracy is not symmetry-protected but instead arises from band folding due to the enlarged unit cell.

The situation changes qualitatively in the chiral phase. Once inversion and mirror symmetries are broken, several branches acquire finite $L_z(\bm{k},n)$, signaling the emergence of chiral phonons. Simultaneously, the fourfold degeneracy is lifted into a $1+2+1$ structure corresponding to the $A_2$, $E$, and $A_1$ irreps of D$_3$, in order of increasing energy [Fig.~\ref{fig:CPS_Te}(b)].

A key feature emerges away from the $\Gamma$ point: the two-dimensional $E$ representation splits into two nondegenerate modes, $E_+\sim x+iy$ and $E_-\sim x-iy$, which carry opposite angular momenta, as highlighted in the inset of Fig.~\ref{fig:CPS_Te}(b). This splitting is not limited to optical modes but also occurs in the transverse acoustic (TA) branches, reflecting their common symmetry origin.

To quantify this effect, we evaluate $|\Delta \omega^{(\ell,\ell')}_{\bm{k}}|$ at $\bm{k}=(0,0,1/5)$. The splittings of the TA and optical modes correspond to $\Delta \omega_{1,2}$ and $\Delta \omega_{4,5}, \Delta \omega_{6,7}$, respectively. Although one might expect that the CPS arises between $E_+$ and $E_-$ for the optical modes and $\Delta\omega_{5,6}(\eta)$ should be evaluated, these two branches are already split in the achiral phase. Therefore, the splitting induced by the transition is better captured by $\Delta \omega_{4,5}, \Delta \omega_{6,7}$ rather than $\Delta\omega_{5,6}(\eta)$. The results are shown in Fig.~\ref{fig:CPS_Te}(c), where all splittings exhibit a quadratic dependence on the displacement amplitude, $\Delta \omega_{\ell,\ell'}(\eta)\propto \eta^2$. Here, $\eta$ is defined in Eq.~(\ref{eq:helical_dr}). This behavior is consistent with the symmetry analysis in Sec.~\ref{subsec:D6h}, where the leading contribution to $G_u$ arises at second order, $G_u^{(2)} \sim \eta^2$.

\subsubsection{\label{sec:CPS_D3d}$\text{R}\bar{3}\text{m}$ to $\text{P}3_121$ and $\text{P}3_221$}
\begin{figure}[t]
\includegraphics[width=1.0\linewidth]{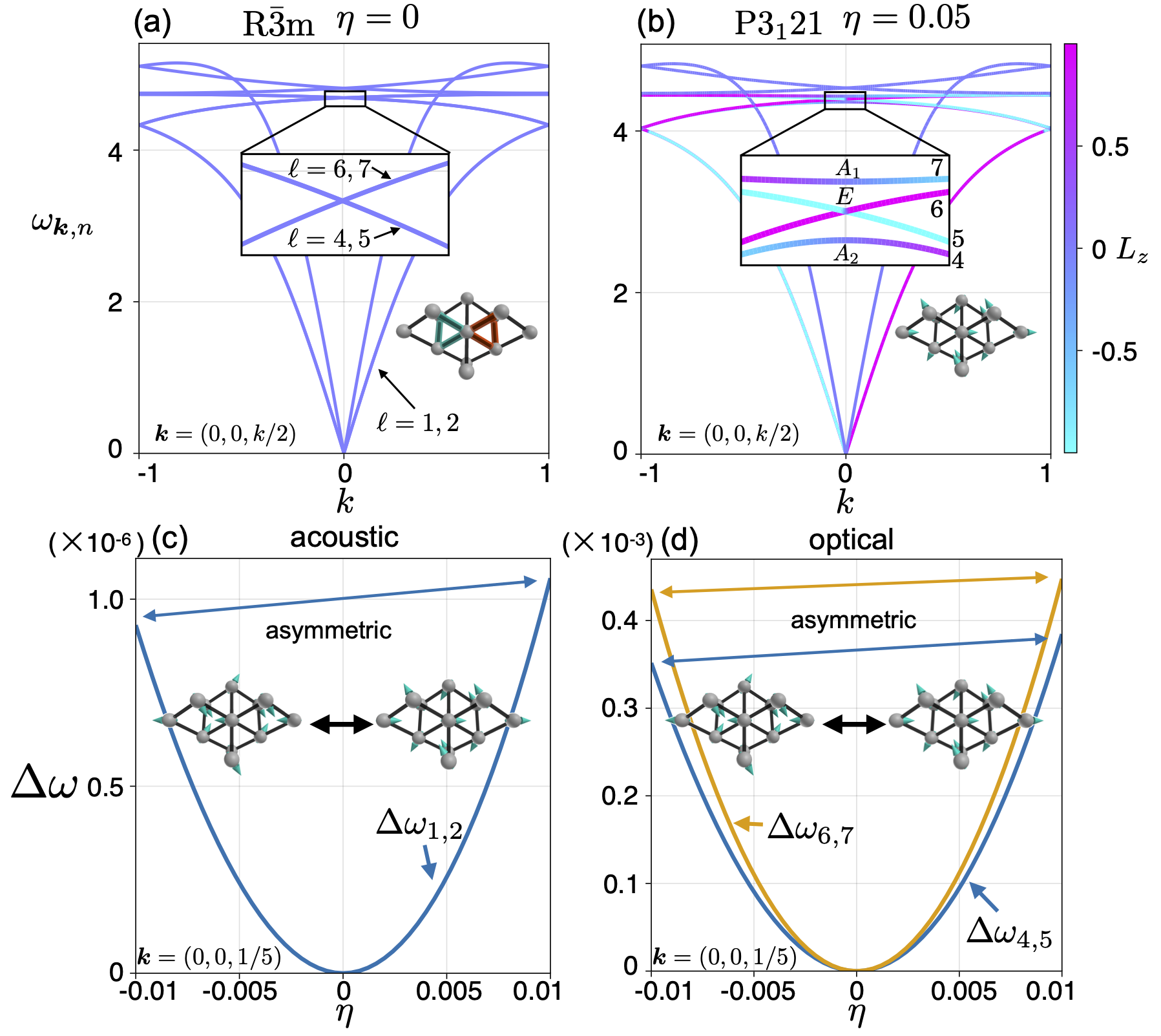}
\caption{\label{fig:CPS_rhombo} 
Phonon dispersions of (a) the achiral R$\bar{3}$m
($\eta = 0$) and (b) the chiral P$3_121$ ($\eta = 0.05$) structures
along $\bm{k} = (0,0,k/2)$ ($-1 \leq k \leq 1$).
The insets in (a) and (b) show the fourfold-degenerate branches and their splittings with corresponding irreps, respectively. The $\eta$ dependence of the splittings of (c) the TA modes
$\Delta\omega_{1,2}(\eta)$ and (d) the optical modes
$\Delta\omega_{4,5}(\eta)$ and $\Delta\omega_{6,7}(\eta)$
for $-0.01 \leq \eta \leq 0.01$.
The splittings are measured at $\bm{k} = (0,0,1/5)$.}
\end{figure}

Next, we examine the CPS in the transition from the achiral R$\bar{3}$m structure to the chiral P$3_121$ (P$3_221$) structure discussed in Sec.~\ref{subsec:rhombo}. The phonon dispersions of the R$\bar{3}$m and P$3_121$ structures along $\bm{k}=k(0,0,1/2)$ ($-1 \leq k \leq 1$) are shown in Figs.~\ref{fig:CPS_rhombo}(a) and \ref{fig:CPS_rhombo}(b), respectively. We set $\eta=0.05$ for the chiral phase. Again, the color denotes $L_z(\bm{k},n)$. As in the previous two examples, $L_z(\bm{k},n) = 0$ in the achiral phase, whereas some branches exhibit $L_z(\bm{k},n) \neq 0$ in the chiral phase. We again use an enlarged unit cell, which leads to a fourfold degeneracy at the $\Gamma$ point [inset of Fig.~\ref{fig:CPS_rhombo}(a)]. This degeneracy is reduced to $A_1 \oplus E \oplus A_2$ upon the AtC transition for the same reason as in the previous case, as shown in the inset of Fig.~\ref{fig:CPS_rhombo}(b).

As the displacement parameter $\eta$ is varied, qualitatively distinct behavior emerges in $|\Delta \omega^{(\ell,\ell')}_{\bm{k}}|$, as shown in Figs.~\ref{fig:CPS_rhombo}(c) and \ref{fig:CPS_rhombo}(d), where $|\Delta \omega^{(\ell,\ell')}_{\bm{k}}|$ is evaluated at $\bm{k}=(0,0,1/5)$ and $\eta$ is defined in Eq.~(\ref{eq:helical_dr}). The branches that exhibit splitting are $\ell=1,2$, $\ell=4,5$, and $\ell=6,7$, where $\ell=1,2$ correspond to the TA modes, while the others are optical modes. See Fig.~\ref{fig:CPS_rhombo}(b). One immediately notices that $\Delta \omega_{\ell,\ell'}(\eta)$ is asymmetric with respect to $\eta$ for both the TA modes in Fig.~\ref{fig:CPS_rhombo}(c) and the optical branches in Fig.~\ref{fig:CPS_rhombo}(d). This asymmetry arises from the structural inequivalence between the chiral structures with $\eta>0$ and $\eta<0$, as discussed in Sec.~\ref{subsec:rhombo}. 

We further analyze the results and find that $\Delta \omega_{\ell,\ell'}(\eta)$ can be expressed as $\Delta \omega_{\ell,\ell'}(\eta) \sim A_{\bm{k}} \eta^2 + B_{\bm{k}} \eta^3$ for $|\eta| \ll 1$, where $A_{\bm{k}}$ and $B_{\bm{k}}$ are $\bm{k}$-dependent coefficients of the second- and third-order contributions in $\eta$, respectively. This $\eta$ dependence is consistent with $G_u(\bm{\eta})$ derived in Sec.~\ref{subsec:rhombo}. The appearance of the $\eta^3$ term is in stark contrast to its absence in Fig.~\ref{fig:CPS_Te} in Sec.~\ref{sec:CPS_D6h}, {\it even though the space group of the symmetry-broken phase is the same}. This demonstrates the importance of the supergroup symmetry at high temperature.

\subsubsection{\label{sec:CPS_Oh}$\text{Fm}\bar{3}\text{m}$ to $\text{P}2_13$}
\begin{figure}[t]
\includegraphics[width=1.0\linewidth]{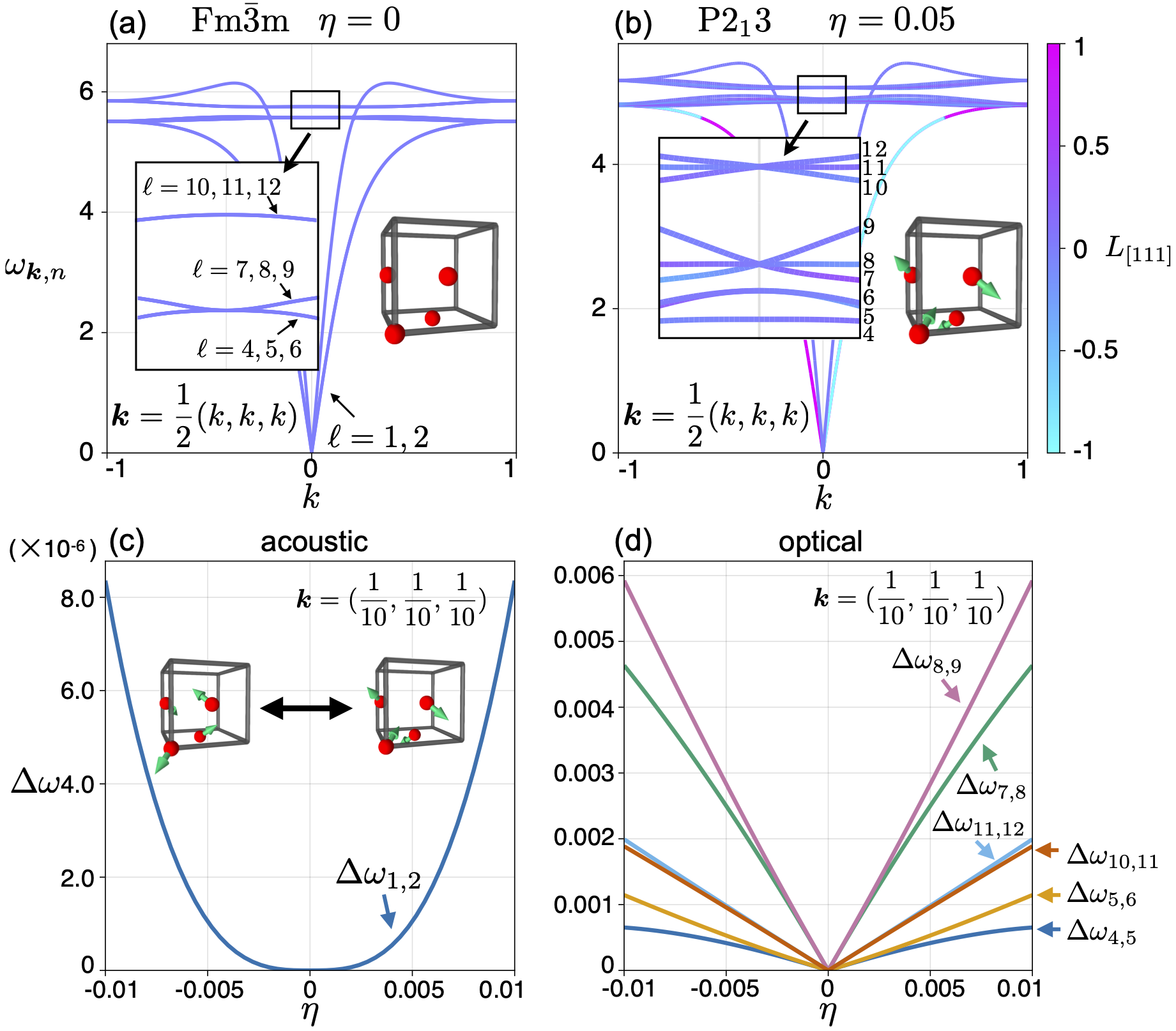}
\caption{\label{fig:CPS_Oh}Phonon dispersions of (a) the achiral Fm$\bar{3}$m
($\eta = 0$) and (b) the chiral P$2_1$3 ($\eta = 0.05$) structures
along $\bm{k} = \frac{1}{2}(k,k,k)$ ($-1 \leq k \leq 1$).
The inset in (a) shows the sixfold and threefold degeneracies
at $\bm{k} = 0$, and that in (b) shows the splittings of the
degenerate modes.
The $\eta$ dependence of the splittings of (c) the TA modes
$\Delta\omega_{1,2}(\eta)$ and (d) the optical modes
for $-0.01 \leq \eta \leq 0.01$.
The splittings are measured at
$\bm{k} = \left(\frac{1}{10},\frac{1}{10},\frac{1}{10}\right)$.
}
\end{figure}
Lastly, we investigate $|\Delta \omega^{(\ell,\ell')}_{\bm{k}}|$ for the transition from the Fm$\bar{3}$m to the P$2_1$3 structure. Since this is a cubic-to-cubic AtC transition, we examine $|\Delta \omega^{(\ell,\ell')}_{\bm{k}}|$ along the $[111]$ direction. First, we show the phonon dispersions of both the Fm$\bar{3}$m and P$2_1$3 phases along $\bm{k} = \frac{1}{2}(k,k,k)$ ($-1 \leq k \leq 1$) in Figs.~\ref{fig:CPS_Oh}(a) and \ref{fig:CPS_Oh}(b), respectively, where $\bm{k}$ is represented by the reciprocal lattice vectors in P$2_1$3 for a clear comparison between the two structures. In the P$2_1$3 phase, the dispersion is calculated with $\eta=0.05$, where $\eta$ is defined in Eq.~(\ref{eq:fcc_RH}) or Eq.~(\ref{eq:fcc_LH}). The color represents the phonon angular momentum along the $[111]$ direction, $L_{[111]}(\bm{k},n) = \tfrac{1}{\sqrt{3}}[L_x(\bm{k},n)+L_y(\bm{k},n)+L_z(\bm{k},n)]$, since the CPS is characterized by $\bm{k} \cdot \bm{L}$ and we consider the dispersion along the $[111]$ direction. 

Although one can also consider the splitting along the $[100]$ direction, $G_0$ is not the sole contribution because the $x$, $y$, and $z$ directions become inequivalent at finite $\bm{k}=(k,0,0)$ in the chiral P$2_1$3 phase. This inequivalence also induces the splitting apart from the effect of $G_0$, which complicates the analysis. On the other hand, along the $[111]$ direction, the $x$, $y$, and $z$ directions remain equivalent, and this complication is therefore absent, allowing the splitting to cleanly reflect the contribution of $G_0$. Therefore, we focus on the splitting along the $[111]$ direction.

We also adopt an enlarged unit cell in both the achiral and chiral phases to accommodate the triple-$\bm{q}$ order. As a result, sixfold and threefold degeneracies of optical modes appear at the $\Gamma$ point, and a threefold degeneracy arises at finite $k$ in the achiral phase due to band folding, as shown in the inset of Fig.~\ref{fig:CPS_Oh}(a). In the chiral phase, the sixfold degeneracy splits into $A \oplus E \oplus T$, and the threefold degeneracy at $\bm{k} \neq 0$ is fully lifted. Accompanied by the degeneracy lifting, some branches acquire finite $L_{[111]}(\bm{k},n)$ at $\bm{k} \neq \bm{0}$. The splitting of the optical modes is shown in the inset of Fig.~\ref{fig:CPS_Oh}(b), where the numbers next to the bands denote the band index $\ell$.

The lowest-energy band $\ell=4$ is a non-degenerate mode corresponding to the $A$ irrep, while $\ell=5,6$ form a doubly degenerate $E$ mode at the $\Gamma$ point, which splits at finite $\bm{k}$. In the achiral phase, $\ell=4,5,6$ are triply degenerate, and the AtC transition lifts this degeneracy. Accordingly, to evaluate the splitting, we compute $\Delta \omega_{4,5}(\eta)$ and $\Delta \omega_{5,6}(\eta)$.

The middle bands $\ell=7,8,9$ and the upper bands $\ell=10,11,12$ are triply degenerate at the $\Gamma$ point and thus belong to the $T$ representation. They split into three nondegenerate modes at finite $\bm{k}$. Accordingly, for the middle bands, we compute $\Delta \omega_{7,8}(\eta)$ and $\Delta \omega_{8,9}(\eta)$, and for the upper bands, we compute $\Delta \omega_{10,11}(\eta)$ and $\Delta \omega_{11,12}(\eta)$ to analyze the splitting.

The splittings of the TA and optical modes $\Delta \omega_{\ell,\ell'}(\eta)$ at $\bm{k} = \left(\frac{1}{10},\frac{1}{10},\frac{1}{10}\right)$ exhibit qualitatively different behavior, as shown in Figs.~\ref{fig:CPS_Oh}(c) and \ref{fig:CPS_Oh}(d). For the TA modes, $\Delta \omega_{1,2}(\eta) \propto \eta^3$, which is consistent with the displacement dependence of $G_0$ derived in Sec.~\ref{subsec:fcc}. In contrast, the splitting of the optical modes is found to be proportional to $\eta$. This implies that $G_0$ is not the dominant contribution to the splitting of the optical modes, whereas the acoustic modes split primarily due to $G_0$. 
This is because the ordering vectors satisfy $\bm{Q}_\text{X} + \bm{Q}_\text{Y} + \bm{Q}_\text{Z} = \bm{G}$, which allows the acoustic phonons at the X points in the BZ of the primitive unit cell, which correspond to the optical phonons at $\bm{k}=0$ in the BZ of the enlarged unit cell, to couple linearly in $\eta_{\mu,\bm{Q}}$. This point is discussed in detail in Sec.~\ref{subsec:nonchiral}.

\subsection{\label{subsec:CIM}Current-induced magnetization}
Next, we evaluate the displacement dependence of the CIM in the AtC transitions considered in Sec.~\ref{sec:toroidals}. In the linear-response regime, the CIM is expressed as
$M_{\mu} = \alpha_{\mu\nu} J_{\nu}$,
where $M_{\mu}$ ($J_{\mu}$) is the $\mu$ component of the magnetization $\bm{M}$ (current density $\bm{J}$), and $\alpha_{\mu\nu}$ is the current--magnetization correlation function, i.e., the CIM susceptibility.

In chiral systems, the induced magnetization is parallel to the applied current ($\bm{M} \parallel \bm{J}$), since the couplings $G_0(\bm{J}\!\cdot\!\bm{M})$ and $G_u(2J_z M_z - J_x M_x - J_y M_y)$ are symmetry-allowed invariants~\cite{Hayami2018PRB}. In other words, the diagonal components of $\alpha_{\mu\nu}$ become finite in chiral phases, i.e., $M_{\mu} = \alpha_{\mu\mu} J_{\mu}$. Therefore, the displacement dependence of $G_{0,u}(\eta)$ should be reflected in that of $\alpha_{\mu\mu}$.

It should be noted that the in-plane components $\alpha_{xx}$ and $\alpha_{yy}$ can become finite even in some achiral noncubic systems belonging to the point groups C$_{3\text{v}}$, C$_{4\text{v}}$, C$_{6\text{v}}$, D$_{2\text{d}}$, and S$_4$, whereas the out-of-plane component $\alpha_{zz}$ becomes finite only in chiral systems. Therefore, we evaluate $\alpha_{zz}$ for noncubic systems as a primary indicator of chirality.

\subsubsection{Model}
To incorporate the structural transformation into electronic systems, we consider a $p$-electron tight-binding model as a minimal setting. We focus on local orbital magnetization, since inducing spin magnetization requires spin-orbit coupling (SOC), which would increase the number of parameters without altering the essential physics. The present calculation should therefore be regarded as a minimal orbital realization of the symmetry-allowed CIM tensor, rather than a material-specific prediction of its absolute magnitude.

We define the Hamiltonian of the $p$-electron tight-binding model as
\begin{align}
H = \sum_{i,j} \sum_{s,s'}
t_{i,s;j,s'}
(\bm{p}^\dagger_{i,s} \cdot \hat{R}_{i,s;j,s'})
(\bm{p}_{j,s'} \cdot \hat{R}_{i,s;j,s'})
+ \text{h.c.},
\label{eq:Hami_el}
\end{align}
where 
$\bm{p}^\dagger_{i,s} = (p^\dagger_{x,i,s}, p^\dagger_{y,i,s}, p^\dagger_{z,i,s})$ denotes the creation operators for the $p_\mu$ orbitals at the $s$th sublattice in the $i$th unit cell. 
$\hat{R}_{i,s;j,s'}$ denotes the unit vector along the bond direction, defined in Eq.~(\ref{eq:def_unitR}). 
We assume that the $p$-orbital electronic wave functions are localized at each site and define the hopping parameter as
$t_{i,s;j,s'} = \exp\!\left[-\left(|\bm{R}_{i,s;j,s'}|/R_{\text{nn}} - 1\right)\right]$, 
where $R_{\text{nn}}$ is the nearest-neighbor distance in the achiral phase, similarly to the approximation used in Sec.~\ref{subsubsec:Model_Method}. 
The exponential form reflects the decay of overlap between localized orbitals with distance; as in Sec.~\ref{subsubsec:Model_Method}, its specific form affects only quantitative details, not the symmetry-governed behavior.
Throughout this section, the unit of energy in this parameterization is the nearest-neighbor hopping $t_{\rm nn}=1$.

In this model, we consider only the $\sigma$-bond hopping~\cite{slater_koster}, which typically provides the dominant contribution because of its large overlap integral. Other hopping parameters such as $\pi$ bonds may lead to quantitative differences but do not qualitatively alter our main results regarding the displacement dependence of the CIM. Retaining only the $\sigma$-bond term is therefore sufficient for our purpose. The structural transformation is incorporated through changes in $\hat{R}_{i,s;j,s'}$ and $t_{i,s;j,s'}$, which are ultimately reflected in the band dispersion. It should also be noted that we use the enlarged unit cell of the chiral phase for both the chiral and achiral phases, since the transition occurs at finite ordering wave vectors and enlarges the unit cell in the models discussed in Secs.~\ref{subsec:D6h}, \ref{subsec:rhombo}, and \ref{subsec:fcc}.

\subsubsection{Method}
The CIM susceptibility is obtained using the Kubo formula~\cite{Kubo_murakami}, given by
\begin{align}
\alpha_{\mu\nu}
=& \frac{1}{\Omega}
\sum_{n,m,\bm{k}}
\mel{\bm{k},n}{\mathcal{M}_{\mu}}{\bm{k},m}
\mel{\bm{k},m}{\mathcal{J}_{\nu}}{\bm{k},n}
\nonumber \\
&\times
\Bigg[
\frac{-i\left\{ f(\varepsilon_{\bm{k},n}) - f(\varepsilon_{\bm{k},m}) \right\}}
{(\varepsilon_{\bm{k},n}-\varepsilon_{\bm{k},m})^2+\delta^2}
-
\frac{f'(\varepsilon_{\bm{k},m})}{\delta}
\delta_{\varepsilon_{\bm{k},n},\varepsilon_{\bm{k},m}}
\Bigg],
\label{eq:CIM}
\end{align}
where $\mathcal{M}_{\mu}$ and 
$\mathcal{J}_{\mu} \equiv -\partial \mathcal{H}(\bm{k})/\partial k_{\mu}$ 
are the $\mu$ ($\mu = x,y,z$) components of the magnetization and current operators, respectively. 
$\ket{\bm{k},n}$ is the eigenstate of the Hamiltonian in Eq.~\eqref{eq:Hami_el} with energy $\varepsilon_{\bm{k},n}$. 
$f(\varepsilon)$ denotes the Fermi--Dirac distribution function $f(\varepsilon)=1/\{\exp[(\varepsilon-\mu_c)/T]+1\}$, where $\mu_c$ is the chemical potential for the $p$ electrons, and $f'(\varepsilon)=\partial f(\varepsilon)/\partial \varepsilon$. $\Omega$ is the volume of the system, and $\delta$ is regarded as a temperature-independent scattering rate due to nonmagnetic impurities, where we focus on the low-temperature response. Ignoring the vertex corrections associated with impurity scattering is an approximation, but it is sufficient for capturing the qualitative aspects of the response while maintaining simplicity~\cite{Kubo_murakami}. Since our model incorporates only the orbital degrees of freedom and does not include spin degrees of freedom, the induced magnetization arises solely from the orbital angular momentum, and $\mathcal{M}_{\mu}$ is given by $\mathcal{M}_{\mu} = \mathscr{L}_\mu$. Here, $\mathscr{L}_{\mu}$ denotes the local angular momentum of the $p$ electrons. The contribution from the orbital motion of itinerant electrons to the magnetization~\cite{orbital_magne} is neglected, as it is not expected to qualitatively modify the present results.

When computing $\alpha_{\mu\nu}$, we fix the chemical potential $\mu_c$ and the temperature $T$ while varying $\eta$ for simplicity. In reality, both quantities change during the structural phase transition. However, $f(\varepsilon)$ and $f'(\varepsilon)$ are not expected to change significantly near the transition temperature $T_c$, where our expressions for $G_{0,u}$ are valid. Therefore, fixing $\mu_c$ and $T$ does not qualitatively affect the results. The data with the electron filling fixed are discussed in Sec.~\ref{discussion-Tdep} for examining the temperature dependence of CIM.

It should be noted that the first and second terms in Eq.~\eqref{eq:CIM} have different transformation properties under time-reversal operation. The first term changes sign under time reversal and thus vanishes in time-reversal-symmetric systems. On the other hand, the second term remains invariant under time reversal. Therefore, only the second term contributes to $\alpha_{\mu\nu}$ in our calculation, since the Hamiltonian in Eq.~\eqref{eq:Hami_el} preserves time-reversal symmetry.

In the following sections, we present the displacement dependence of $\alpha_{\mu\nu}$ for the model systems introduced in Sec.~\ref{sec:toroidals}, based on Eq.~\eqref{eq:CIM}. The quantity $\alpha_{\mu\nu}$ is calculated for several system sizes $L_\text{size} = 10,\,20,\,40,\,60$, and $80$ with volume $\Omega=L_\text{size}^3$, and we perform a linear extrapolation in $1/L_\text{size}$ to obtain the thermodynamic-limit value.

\subsubsection{$\text{D}_{2\text{d}}$ to $\text{D}_2$}
\begin{figure}[t]
\includegraphics[width=1.0\linewidth]{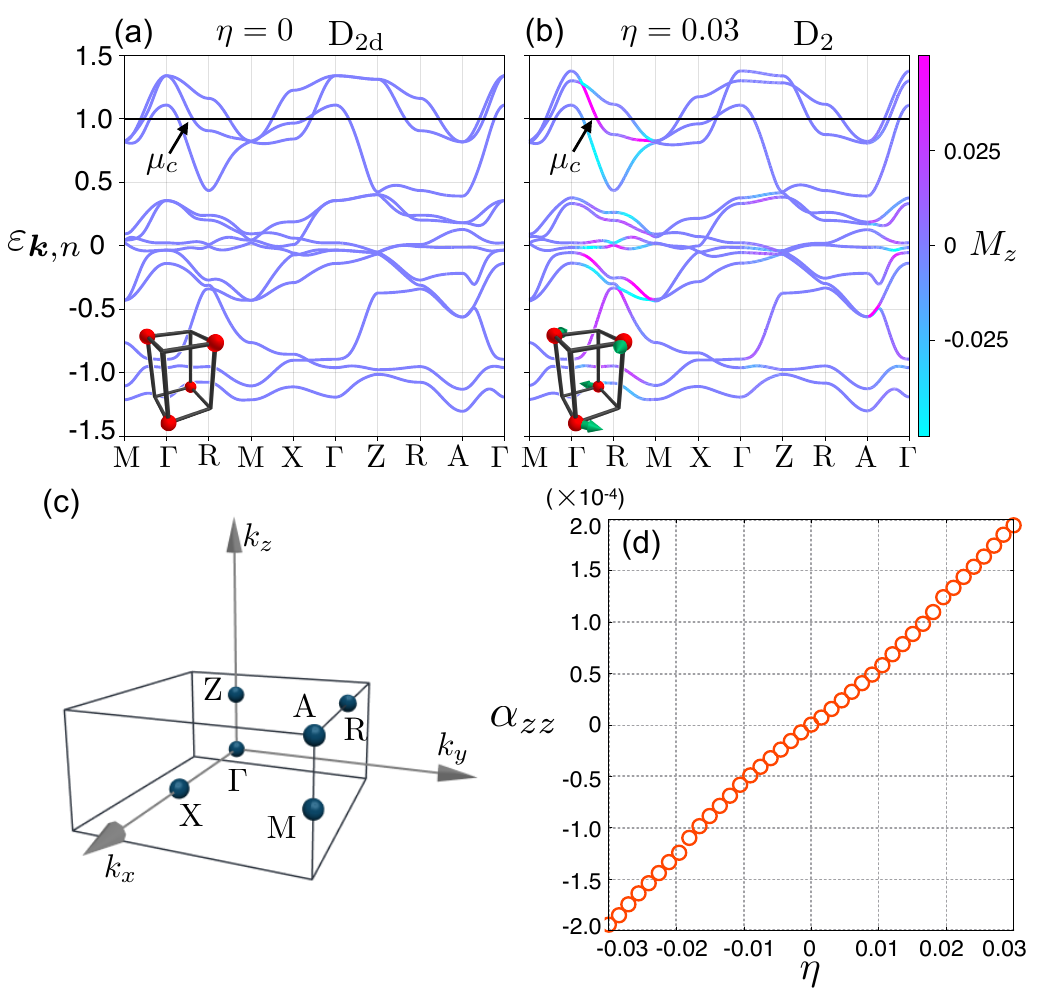}
\caption{\label{fig:CIM_D2d} Electronic band dispersions of (a) the achiral D$_{2\text{d}}$ and
(b) the chiral D$_2$ structures along the high-symmetry lines of the
tetragonal BZ.
The color represents the orbital angular momentum and the horizontal
black line indicates the chemical potential $\mu_c = 1.0$.
(c) The first BZ of a tetragonal lattice and its high-symmetry
$\bm{k}$ points.
(d) Displacement dependence of $\alpha_{zz}$ calculated at $T = 0.1$
and $\mu_c = 1.0$, showing a linear dependence on $\eta$ characteristic
of the electric toroidal quadrupole $G_u$.
}
\end{figure}

We first discuss the AtC transition from the D$_{2\text{d}}$ structure to the chiral D$_2$ structure. As discussed in Sec.~\ref{sec:CPS_D2d}, the ET quadrupole satisfies $G_u \propto \eta$, and the CPS correspondingly exhibits the linear dependence $\Delta \omega_{\ell,\ell'}(\eta) \propto \eta$. We show below that the CIM exhibits the same symmetry-dictated displacement dependence.

Finite orbital magnetization $M_z = \ev{\mathcal{M}_z}{\bm{k},n}$ emerges only in the chiral D$_2$ phase, reflecting the activation of the ET quadrupole $G_u$. 
Figures~\ref{fig:CIM_D2d}(a) and \ref{fig:CIM_D2d}(b) show the corresponding electronic band dispersion for the achiral D$_{2\text{d}}$ and chiral D$_2$ structures, respectively, along the high-symmetry points of the tetragonal BZ given in Fig.~\ref{fig:CIM_D2d}(c).
%The corresponding electronic band dispersions for the achiral D$_{2\text{d}}$ and chiral D$_2$ structures are shown in Figs.~\ref{fig:CIM_D2d}(a) and \ref{fig:CIM_D2d}(b), respectively. 
In the chiral phase, we set $\eta=0.03$ as a representative parameter. As seen in Fig.~\ref{fig:CIM_D2d}(b), several Bloch states acquire finite orbital magnetization, which contributes to a finite $\alpha_{zz}$.

The displacement dependence of $\alpha_{zz}$ for $T=0.1$ and $\mu_c = 1.0$ is shown in Fig.~\ref{fig:CIM_D2d}(d). Near $\eta=0$, $\alpha_{zz}$ exhibits a linear dependence on $\eta$, consistent with the symmetry analysis for $G_u$. Although the linear behavior is modified at larger $\eta$, the relation $\alpha_{zz}(\eta)=-\alpha_{zz}(-\eta)$ remains satisfied over the entire range, indicating that the sign of the CIM is switched by reversing the chirality. These results consistently reflect the odd-parity nature of $G_u$ derived in Sec.~\ref{subsec:D2d}. 

\subsubsection{$\text{P}6/\text{mmm}$ to $\text{P}3_121$}
\begin{figure}[t]
\includegraphics[width=1.0\linewidth]{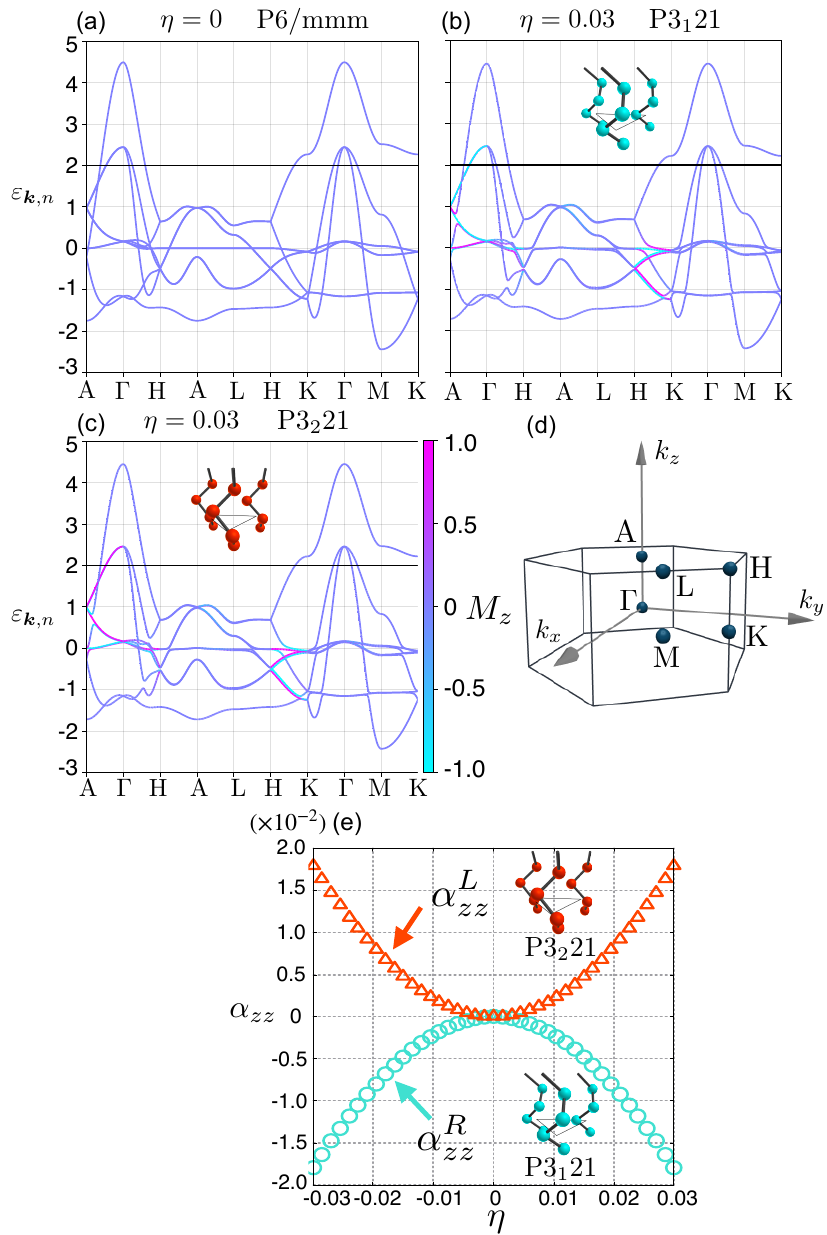}
\caption{\label{fig:CIM_Te} Electronic band dispersions of (a) the achiral P6/mmm, (b) the chiral
P$3_121$, and (c) the chiral P$3_221$ structures.
For the two chiral structures, the displacement parameter $\eta$ is
set to $\eta = 0.03$.
The color represents the orbital angular momentum $M_z$ along the $z$ axis, and the horizontal black line indicates the chemical potential
$\mu_c$.
(d) The first BZ of a hexagonal lattice with the high-symmetry
$\bm{k}$ points.
(e) The $\eta$ dependence of $\alpha_{zz}$ of the P$3_121$ structure
$\alpha_{zz}^R$ (blue circles) and the P$3_221$ structure
$\alpha_{zz}^L$ (orange triangles), demonstrating opposite signs for
opposite chirality and a dominant quadratic dependence on $\eta$.
}
\end{figure}

We next discuss the CIM for the system analyzed in Sec.~\ref{subsec:D6h}, namely the transition from the achiral P6/mmm structure to the chiral P$3_121$ and P$3_221$ structures. In contrast to the previous D$_{2\text{d}} \to$ D$_2$ transition, the leading-order ET quadrupole satisfies $G_u \sim \eta^2$, implying that the CIM should also exhibit a dominant quadratic dependence on $\eta$.

In the chiral phases, several bands acquire finite orbital magnetization $M_z$, particularly along the $\Gamma$--A line, reflecting the symmetry-allowed coupling $G_u \sim k_z M_z$. The corresponding band dispersions for the P6/mmm, P$3_121$, and P$3_221$ structures are shown in Figs.~\ref{fig:CIM_Te}(a), \ref{fig:CIM_Te}(b), and \ref{fig:CIM_Te}(c), respectively. 
The high-symmetry $\bm{k}$ points are illustrated in Fig.~\ref{fig:CIM_Te}(d).
Analogously to the previous example, we set $\eta=0.03$ in the chiral phases, where the definition of $\eta$ is given in Eq.~(\ref{eq:helical_dr}). Importantly, the sign of $M_z$ for a Bloch state $\ket{\bm{k},n}$ is reversed between the P$3_121$ and P$3_221$ structures, reflecting their opposite chirality.

The displacement dependence of the CIM is shown in Fig.~\ref{fig:CIM_Te}(e) for $T = 0.1$ and $\mu_c = 2.0$. The CIM $\alpha_{zz}^R$ for the RH structure and $\alpha_{zz}^L$ for the LH structure exhibit opposite signs, while their absolute values are proportional to $\eta^2$, consistent with the expression for $G_u$ derived in Sec.~\ref{subsec:D6h}. The absence of odd-order contributions follows from the equivalence between the structures with $\eta$ and $-\eta$, leading to the relations $\alpha_{zz}^{R}(\eta)=\alpha_{zz}^R(-\eta)=-\alpha_{zz}^{L}(\eta)=-\alpha_{zz}^L(-\eta)$. Thus, the CIM directly reflects the even-parity displacement dependence imposed by the symmetry of the parent phase.

\subsubsection{$\text{R}\bar{3}\text{m}$ to $\text{P}3_121$}
\begin{figure}[t]
\includegraphics[width=1.0\linewidth]{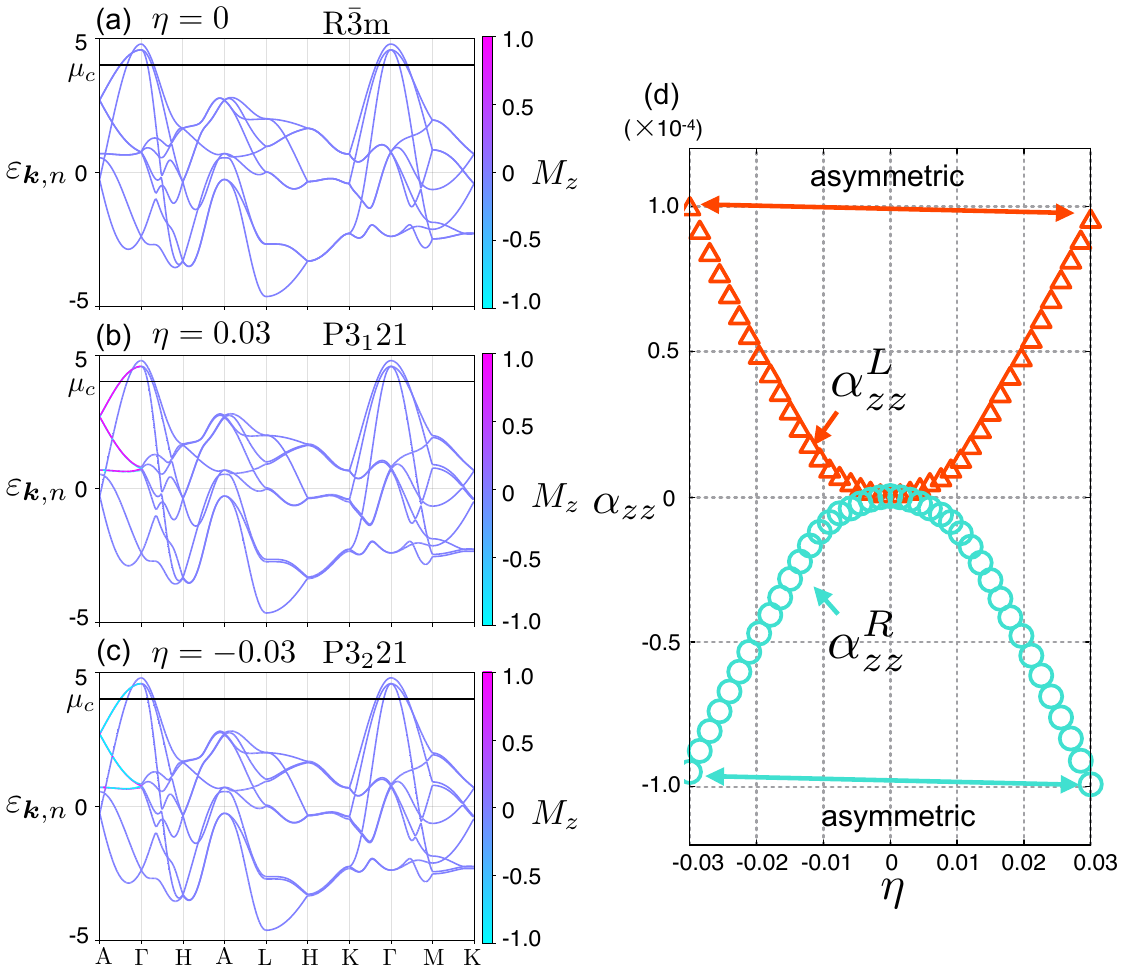}
\caption{\label{fig:CIM_rhombo}Electronic band dispersions of (a) the achiral R$\bar{3}$m, (b) the chiral
P$3_121$, and (c) the chiral P$3_221$ structures.
For the P$3_121$ structure, the displacement parameter $\eta$ is set
to $\eta = 0.03$, while for the P$3_221$ structure $\eta = -0.03$.
The color represents the orbital angular momentum $M_z$ along the $z$ axis, and the horizontal black line indicates the chemical potential
$\mu_c$.
(d) The $\eta$ dependence of $\alpha_{zz}$ of the P$3_121$ structure
$\alpha_{zz}^R$ (blue circles) and the P$3_221$ structure
$\alpha_{zz}^L$ (orange triangles), exhibiting both quadratic and cubic
contributions to the CIM response.
}
\end{figure}

Next, we discuss the system introduced in Sec.~\ref{subsec:rhombo}, where the low-temperature chiral phase is again described by the P$3_121$ or P$3_221$ structure, while the high-temperature phase belongs to the R$\bar{3}$m structure. Although the low-temperature chiral structures are identical to those in the previous section, the different symmetry of the parent phase qualitatively changes the displacement dependence of the CIM.

The electronic band dispersions of the R$\bar{3}$m, P$3_121$, and P$3_221$ structures are shown in Figs.~\ref{fig:CIM_rhombo}(a), \ref{fig:CIM_rhombo}(b), and \ref{fig:CIM_rhombo}(c), respectively. Analogously to the previous examples, the color denotes the orbital magnetization $M_z$. Recalling that the mirror image of the P$3_121$ structure with $\eta>0$ corresponds to the P$3_221$ structure with $\eta<0$ as shown in Figs.~\ref{fig:D3d_image}(b) and \ref{fig:D3d_image}(c), we set $\eta=0.03$ and $\eta=-0.03$ for the P$3_121$ and P$3_221$ structures, respectively, such that the two structures become mirror images of each other. Several bands along the $\Gamma$--A line acquire finite orbital magnetization in the chiral phases, and the sign of $M_z$ is reversed between the two opposite chiralities.

The displacement dependence of $\alpha_{zz}$ is shown in Fig.~\ref{fig:CIM_rhombo}(d) for $T=0.1$ and $\mu_c=4.0$. Near $\eta=0$, the quadratic contribution $\alpha_{zz}^{R,L} \propto \eta^2$ is dominant. However, with increasing $|\eta|$, the response becomes asymmetric with respect to $\eta$, indicating the emergence of odd-order contributions. Indeed, the numerical results are well reproduced by $\alpha_{zz}^{R} \sim A \eta^2 + B \eta^3$ and $\alpha_{zz}^{L} \sim -A \eta^2 + B \eta^3$, where $A \sim -0.21$ and $B \sim -0.14$. This behavior is fully consistent with the symmetry analysis of $G_u$ in Sec.~\ref{subsec:rhombo}, where both quadratic and cubic terms are allowed. These results demonstrate that the displacement dependence of the CIM is governed not only by the symmetry of the chiral phase itself but also by that of the parent phase.

\subsubsection{$\text{Fm}\bar{3}\text{m}$ to $\text{P}2_13$}
\begin{figure}[t]
\includegraphics[width=1.0\linewidth]{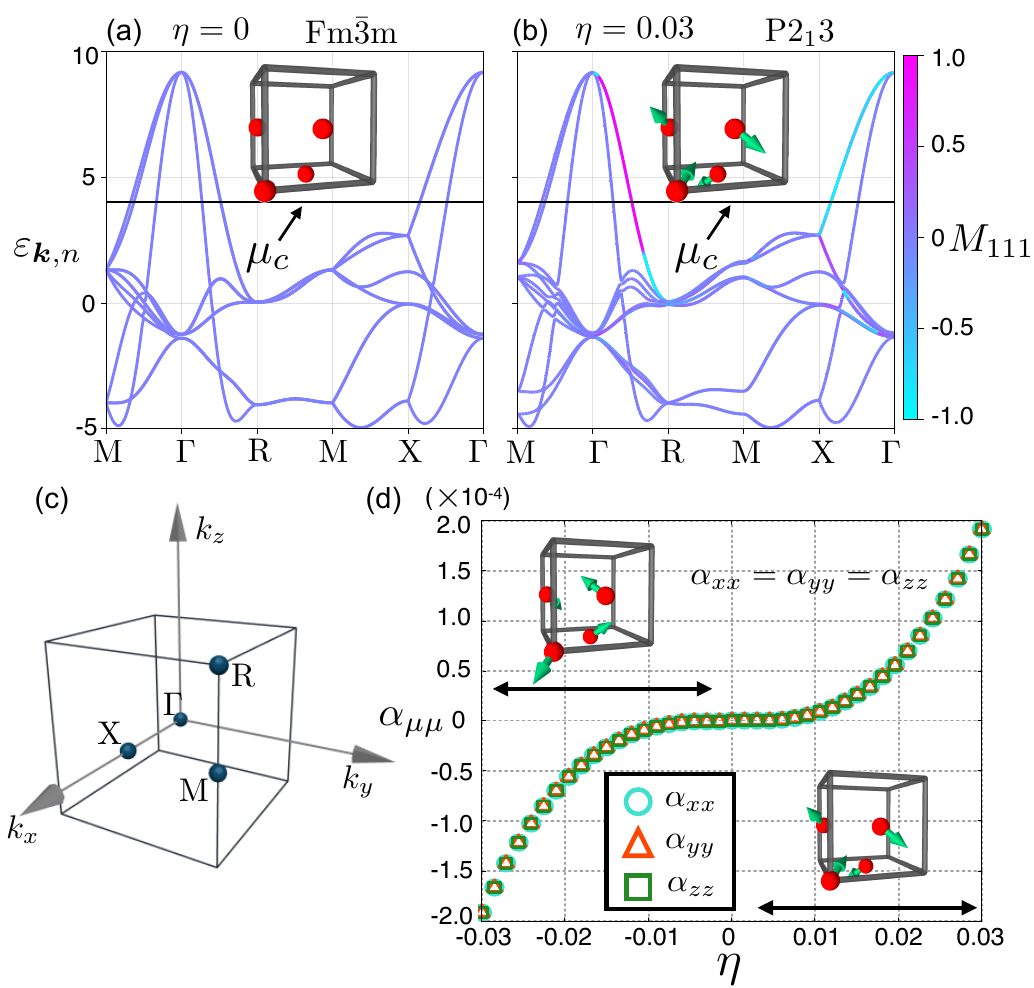}
\caption{\label{fig:CIM_B20} 
Electronic band dispersions of the electrons of (a) the achiral
Fm$\bar{3}$m and (b) the chiral P$2_1$3 structures along the
high-symmetry path of the simple cubic BZ.
The color denotes the orbital angular momentum along the $[111]$
direction $M_{111} = (M_x + M_y + M_z)/\sqrt{3}$, and the horizontal
black lines indicate the chemical potential $\mu_c$.
(c) The first BZ of a simple cubic lattice and its
high-symmetry $\bm{k}$ points.
(d) The $\eta$ dependence of $\alpha_{\mu\mu}$ ($\mu = x,y,z$) for
$T = 0.1$ and $\mu_c = 4.0$, exhibiting the cubic dependence.}
\end{figure}

Lastly, we discuss the CIM for the cubic-to-cubic transition from the achiral Fm$\bar{3}$m structure to the chiral P$2_1$3 structure. In this case, the ET monopole satisfies $G_0 \propto \eta^3$, implying that lower-order contributions to the CIM are forbidden by the high symmetry of the cubic parent phase.

The electronic band dispersions of the achiral and chiral structures are shown in Figs.~\ref{fig:CIM_B20}(a) and \ref{fig:CIM_B20}(b), respectively. For the chiral structure, we set $\eta=0.03$. The color denotes the orbital magnetization along the $[111]$ direction, $M_{111} = (M_x + M_y + M_z)/\sqrt{3}$, reflecting the cubic symmetry of the transition. In the chiral phase, several bands acquire finite $M_{111}$, particularly along the $\Gamma$--R and $\Gamma$--X paths, whereas the orbital magnetization vanishes identically in the achiral phase. For the high-symmetry points in the first BZ, see Fig.~\ref{fig:CIM_B20}(c). 

The displacement dependence of $\alpha_{xx}(=\alpha_{yy}=\alpha_{zz})$ for $T=0.1$ and $\mu_c=4.0$ is shown in Fig.~\ref{fig:CIM_B20}(d). The CIM exhibits the cubic dependence $\alpha_{\mu\mu} \propto \eta^3$, which is consistent with the symmetry-derived form $G_0\propto \eta^3$ obtained in Sec.~\ref{subsec:fcc}. Furthermore, owing to the equivalence between the structures with $\eta$ and $-\eta$, even-order contributions are forbidden and thus $\alpha_{\mu\mu}(\eta)=-\alpha_{\mu\mu}(-\eta)$. These results demonstrate that the CIM in this cubic chiral system is governed by the isotropic monopolar response associated with the ET monopole.

\section{\label{sec:discussion}Discussion}
So far, we have shown that $G_{0,u}(\bm{\eta})$ can be probed through the displacement dependence of the phonon band splittings $|\Delta \omega^{(\ell,\ell')}_{\bm{k}}|$ in lattice systems and through that of the CIM in electronic systems. In this section, we focus on changes in the phonon energies and eigenmodes induced by AtC transitions, as these are closely related to the structural transformation of the system. We discuss the microscopic mechanism by which $G_{0,u}(\bm{\eta})$ contributes to the CPS and leads to the emergence of chiral phonons.

\subsection{\label{sec:mechanism}Mechanism of the CPS}
To encode structural chirality in phonons, the phonon degrees of freedom must couple to $G_{0,u}$. Because the Hamiltonian must remain invariant under any symmetry operations in the achiral phase, such coupling can occur only through an ET-multipole coupling of the form $g_{0,u} G_{0,u}$, where $g_{0,u}$ is an ET monopole or quadrupole constructed from the phonon degrees of freedom.

Within the harmonic approximation, the potential energy $V$ in Eq.~\eqref{eq:Hami_phonon} can be decomposed into the achiral part $V_0$ and a coupling term $V'$ as
\begin{align}
V = V_0 + V'.
\end{align}
Here, we use the eigenmodes of the achiral phase as the basis of the dynamical matrix. In this basis, the dynamical matrix for $V_0$ is diagonal, and $V_0$ is written as
\begin{align}
V_0 = \sum_{\bm{k},n}
\left[\omega^{(0)}_{n,\bm{k}}\right]^2
u_{n,\bm{k}} \,
u_{n,-\bm{k}},
\end{align}
where $\omega^{(0)}_{n,\bm{k}}$ is the eigenfrequency of the $n$th eigenmode $u_{n,\bm{k}}$ in the achiral phase, with the atomic masses appropriately normalized. Furthermore, $V'$ can be decomposed into the coupling with the ET monopole (quadrupole), $V'_{c}$, and other contributions, $V'_{\text{other}}$. In the following, we focus on $V'_{c}$, as it is expected to provide the dominant contribution to the CPS.

We first consider the CPS in noncubic systems. In such systems, the CPS appears along $\bm{k}=(0,0,k)$ due to the coupling $k_z L_z$. Therefore, we restrict our discussion to phonons with wave vector $\bm{k}=(0,0,k)$. In noncubic achiral systems with $C_n$ ($n \ge 3$) symmetry about the $z$ axis, phonon eigenmodes at finite $\bm{k}$ are at most doubly degenerate. From a group-theoretical perspective, such doubly degenerate modes can be represented by $(u_{x,\bm{k}}, u_{y,\bm{k}})$ or $(u_{x^2-y^2,\bm{k}}, u_{2xy,\bm{k}})$, where the former transforms as a polar vector $(x,y)$, while the latter transforms as $(x^2 - y^2, 2xy)$. Except for systems with sixfold rotational symmetry, it is sufficient to consider $(u_{x,\bm{k}}, u_{y,\bm{k}})$, since they belong to the same irrep.

For the present case with $\bm{k}$ along the $z$ axis, one can construct $g_u$ from the doubly degenerate modes as antisymmetric products:
\begin{align}
g_u^{(1)}(\bm{k}) &= i \left( u_{x,\bm{k}} u_{y,-\bm{k}} - u_{y,\bm{k}} u_{x,-\bm{k}} \right)\label{eq:gu1}, \\
g_u^{(2)}(\bm{k}) &= i \left( u_{x^2-y^2,\bm{k}} u_{2xy,-\bm{k}} - u_{2xy,\bm{k}} u_{x^2-y^2,-\bm{k}} \right).
\end{align}
Accordingly, $G_u(\bm{\eta})$ enters $V'_{c}$ as the ET-multipole coupling:
\begin{align}
V'_{c} = \sum_{\bm{k}} \lambda_{\bm{k}} G_u(\bm{\eta})  g_u(\bm{k}), 
\label{eq:phonon_disp}
\end{align}
where $g_u$ denotes either $g_u^{(1)}$ or $g_u^{(2)}$, the sum is taken over $\bm{k}=(0,0,k)$, and $\lambda_{\bm{k}}$ is the coupling constant.

Denoting the phonon basis as $u_{x,\bm{k}},\,u_{x^2-y^2,\bm{k}} = (1,0)^{\text{T}}$ and $u_{y,\bm{k}},\,u_{2xy,\bm{k}} = (0,1)^{\text{T}}$, where $\text{T}$ denotes the transpose, the dynamical matrix of $V'_{c}$ is written as
\begin{align}
\mathcal{D}_c'(\bm{k}) = G_u(\bm{\eta})
\begin{pmatrix}
0 & i \lambda_{\bm{k}} \\
- i \lambda_{\bm{k}} & 0
\end{pmatrix}.
\label{eq:dyna}
\end{align}
The dynamical matrix for $V_0 + V'_c$ in the degenerate subspace is then given by
\begin{align}
\mathcal{D}(\bm{k}) =
\begin{pmatrix}
\big[\omega^{(0)}_{n,\bm{k}}\big]^2 & i \lambda_{\bm{k}} G_u(\bm{\eta}) \\
- i \lambda_{\bm{k}} G_u(\bm{\eta}) & \big[\omega^{(0)}_{n,\bm{k}}\big]^2
\end{pmatrix}.
\end{align}
Diagonalizing this matrix, one obtains the phonon dispersions in the chiral phase, $\omega_{\pm,\bm{k}}$, as
\begin{align}
\omega_{\pm,\bm{k}}
\sim
\omega^{(0)}_{n,\bm{k}}
\pm
\frac{\lambda_{\bm{k}} G_u(\bm{\eta})}{2 \, \omega^{(0)}_{n,\bm{k}}},
\end{align}
for $|G_u(\bm{\eta})| / \omega^{(0)}_{n,\bm{k}} \ll 1$. As a result, the splitting $\Delta \omega_{\bm{k}} \equiv |\omega_{+,\bm{k}} - \omega_{-,\bm{k}}|$ is proportional to $G_u(\bm{\eta})$, i.e., $\Delta \omega_{\bm{k}} \propto G_u(\bm{\eta})$.

Moreover, the matrix form $\mathcal{D}_c'$ for Eq.~\eqref{eq:dyna} in this restricted basis can be rewritten as
\begin{align}
\mathcal{D}_c' = \sum_{\bm{k}} \lambda_{\bm{k}} \, G_u(\bm{\eta}) \, \mathscr{L}_z,
\end{align}
where $\mathscr{L}_z$ is the phonon angular momentum operator along the $z$ axis in the degenerate subspace, defined as
\begin{align}
\mathscr{L}_z =
\begin{pmatrix}
0 & i \\
- i & 0
\end{pmatrix}.
\end{align}
This form indicates that the eigenmodes $u_{\pm,\bm{k}}$ with energies $\omega_{\pm,\bm{k}}$, given by $u_{\pm,\bm{k}} = (1, \pm i)^\text{T} e^{i\phi}/\sqrt{2}$ with an arbitrary phase $\phi$, are also eigenstates of $\mathscr{L}_z$ with eigenvalues $\pm 1$. Therefore, these eigenmodes can be regarded as chiral phonons.

When the AtC transition breaks the equivalence between the $x$ and $y$ axes, as in the transition from the D$_{2\text{d}}$ to the D$_2$ structure, one must consider additional forms of $g_u$:
\begin{align}
g_u^{\prime(1)}(\bm{k}) =&\, u_{x,\bm{k}}u_{x,-\bm{k}} - u_{y,\bm{k}}u_{y,-\bm{k}},\label{eq:gu_prime1}\\
g_u^{\prime(2)}(\bm{k}) =&\, u_{x^2-y^2,\bm{k}}u_{x^2-y^2,-\bm{k}} - u_{2xy,\bm{k}}u_{2xy,-\bm{k}}. 
\label{eq:gu_prime2}
\end{align}
Including $g_u^{\prime(p)}$ ($p=1,2$), the matrix form of $\mathcal{D}_c'$ becomes
\begin{align}
\mathcal{D}_c^{\prime} = \sum_{\bm{k}}\Big[\lambda_{\bm{k}}\mathscr{L}_z + \lambda^{\prime}_{\bm{k}}\mathscr{Q}_{v}\Big]G_u(\bm{\eta}),
\label{eq:ortho_dyna}
\end{align}
where $\lambda^{\prime}_{\bm{k}}$ is the coupling constant between $g_u^{\prime}(\bm{k})=g_u^{\prime(1)}(\bm{k})$ or $g_u^{\prime(2)}(\bm{k})$ and $G_u(\bm{\eta})$, and $\mathscr{Q}_{v}$ is defined as
\begin{align}
\mathscr{Q}_{v} =
\begin{pmatrix}
1 & 0 \\
0 & -1
\end{pmatrix}.
\end{align}
The eigenvalues $\Lambda_{\pm,\bm{k}}$ of Eq.~\eqref{eq:ortho_dyna} are given by
\begin{align}
\Lambda_{\pm,\bm{k}}=\pm G_u(\bm{\eta})\sqrt{|\lambda_{\bm{k}}|^2+|\lambda^\prime_{\bm{k}}|^2}.
\end{align}
This implies that the splitting of the phonon frequencies remains proportional to $G_u(\bm{\eta})$. However, the eigenmodes $u_{\pm,\bm{k}}$ corresponding to $\Lambda_{\pm,\bm{k}}$ are modified by a finite $\lambda^{\prime}_{\bm{k}}$ as
\begin{align}
u_{\pm,\bm{k}} 
&=e^{i\phi} \left(
	\cos\frac{\theta_{\pm,\bm{k}}}{2},\ 
	\pm i\sin\frac{\theta_{\pm,\bm{k}}}{2}
\right)^{\rm T},\label{eq:chiral_mode_D2d}\\
\tan\frac{\theta_{\pm,\bm{k}}}{2}&\equiv \sqrt{1+\left|\frac{\lambda^\prime_{\bm{k}}}{\lambda_{\bm{k}}}\right|^2}\mp\frac{\lambda^\prime_{\bm{k}}}{\lambda_{\bm{k}}},\quad (0\le \theta_{\pm,\bm{k}}\le \pi),\label{eq:tan}
\end{align}
where $\phi$ is an arbitrary phase factor. Equation~\eqref{eq:chiral_mode_D2d} is no longer an eigenstate of $\mathscr{L}_z$, resulting in elliptically polarized phonons. The expectation value of $\mathscr{L}_z$ for Eq.~\eqref{eq:chiral_mode_D2d} is obtained as
\begin{align}
L_z^{\pm} = u_{\pm,\bm{k}}^\dag \mathscr{L}_z u_{\pm,\bm{k}}
=\pm \sin\theta_{\pm,\bm{k}}.
\label{eq:ortho_Lz}
\end{align}
For $\lambda^{\prime}_{\bm{k}} = 0$, $\theta_{\pm,\bm{k}}=\pi/2$ and $L_z^\pm = \pm 1$, while a finite $\lambda^{\prime}_{\bm{k}}$ leads to $|L^\pm_z| \leq 1$ due to the elliptic polarization. This result also indicates that the angular momentum of phonons may be suppressed in the presence of anisotropy. Anisotropic impurities play such a role in a chiral crystal by breaking the local rotational symmetry. Therefore, it is important to minimize such impurities when synthesizing chiral crystals in order to utilize the angular momentum.

For acoustic phonons, one finds that $\lambda^\prime_{\bm{k}}/\lambda_{\bm{k}}$ is at least proportional to $1/k$ for $\bm{k}=(0,0,k)$ in the $k \to 0$ limit, due to the requirement of finite positive sound velocities. We first consider the denominator $\lambda_{\bm{k}}$. This term is responsible for chiral phonons, and it has been pointed out that the sound velocities of chiral phonons in acoustic branches must be identical to avoid nonanalyticity~\cite{tsunetsugu_kusunose}. If $\lambda_{\bm{k}} \propto k^2$, then $\omega_{\pm,\bm{k}}$ acquires a term linear in $k$, leading to $\omega_{\pm,\bm{k}} \sim \omega_{\bm{k}} \pm c k$, where $c$ is a constant. This implies that the sound velocities of the two chiral phonons become different, which is prohibited. Therefore, $\lambda_{\bm{k}}$ must start at least from order $k^3$ in the $k \to 0$ limit. In contrast, $\lambda^{\prime}_{\bm{k}}$ breaks the equivalence between the $x$ and $y$ directions, allowing the two TA modes to have different sound velocities. Therefore, $\lambda^{\prime}_{\bm{k}}$ can start from order $k^2$. As a result, $L_z$ in the $k \to 0$ limit behaves as $L_z \propto k$. For further details, see Appendix~\ref{apdx:D2}.

For cubic systems, a similar analysis can be performed. Instead of $G_0(\bm{\eta})$ coupling to a phonon mode at a single $\bm{k} = (k,k,k)$, it couples to phonon modes at symmetry-equivalent $\bm{k}$ points along the $[111]$ direction, $\bm{k}_m = \hat{C}_4^m \bm{k}$ ($m=0,1,2,3$), where $\hat{C}_4$ is the fourfold rotation operator about the $[001]$ axis, ensuring cubic symmetry. Therefore, $g_0$ is written as
\begin{align}
g_0(\bm{k})= i\sum_{m=0,1,2,3}\left(u_{x,\bm{k}_m}u_{y,-\bm{k}_m} - u_{y,\bm{k}_m}u_{x,-\bm{k}_m}\right), \label{g0_cubic}
\end{align}
where, for each mode $(u_{x,\bm{k}_m}, u_{y,\bm{k}_m})$, the $z$ axis is taken to be parallel to $\bm{k}_m$. Accordingly, $x$ and $y$ do not refer to the global Cartesian directions but to the local coordinates in the plane perpendicular to each $\bm{k}_m$. Any doubly degenerate modes at $\bm{k} = (k,k,k)$ can be represented by $(u_{x,\bm{k}}, u_{y,\bm{k}})$, since there is no $C_6$ symmetry about the $[111]$ axis in cubic systems.

Then, $G_0(\bm{\eta})$ enters $V_c'$ as
\begin{align}
V_c' = \sum_{\bm{k}} \lambda_{\bm{k}} g_0(\bm{k}) G_0(\bm{\eta}), \label{eq:Vcprime}
\end{align}
where $\lambda_{\bm{k}}=-\lambda_{-\bm{k}}$, i.e., it is odd with respect to $\bm{k}$. By choosing $u_{x,\bm{k}_m}=(1,0)^{\mathrm{T}}$ and $u_{y,\bm{k}_m}=(0,1)^{\mathrm{T}}$, one obtains the matrix form of Eq.~(\ref{eq:Vcprime})
\begin{align}
\mathcal{D}_c'=\sum_{\bm{k}}\lambda_{\bm{k}} G_0(\bm{\eta})\mathscr{L}_z,\label{eq:cubic_dyna}
\end{align}
which is identical to Eq.~\eqref{eq:dyna} except for the replacement $G_u(\bm{\eta}) \rightarrow G_0(\bm{\eta})$. Therefore, the splitting is given by $\Delta\omega_{\bm{k}} \propto G_0(\bm{\eta})$, and the eigenmodes at $\bm{k}_m$, $u_{\pm,\bm{k}_m}$, are simultaneous eigenstates of $\mathscr{L}_z$. Here, $\mathscr{L}_z$ should be interpreted as the angular momentum operator along $\bm{k}_m$, since the $z$ axis is taken to be parallel to $\bm{k}_m$.
%In summary, we have discussed the mechanism of the CPS. We have shown that the antisymmetric product of doubly degenerate modes at $\bm{k}=(0,0,k)$ in noncubic systems and at $\bm{k}=(k,k,k)$ in cubic systems couples to $G_{0,u}(\bm{\eta})$. This coupling leads to the splitting $\Delta\omega_{\bm{k}} \propto G_{0,u}(\bm{\eta})$ and contributes to the emergence of chiral phonons.

 \subsection{\label{subsec:nonchiral}Nonchiral contribution to optical modes}
In Sec.~\ref{sec:CPS_Oh}, we have shown that the splitting of the triply degenerate optical modes at finite $\bm{k}\parallel [111]$ in the transition from the Fm$\bar{3}$m to the P2$_1$3 structure is not proportional to $G_0(\bm{\eta})$ but instead scales linearly with the displacements. Here, we discuss the origin of this splitting. In Sec.~\ref{sec:mechanism}, we have considered mainly the ET-multipole coupling $V'_c$ between phonon modes and $G_{0,u}$. However, there is an additional term $V'_{\text{other}}$ that couples phonons to other displacement-constructed quantities beyond $G_{0,u}$.

For this splitting, $V'_{\text{other}}$ is expected to provide the dominant contribution, since its displacement dependence does not follow that of $G_0$. Below, we analyze $V'_{\text{other}}$ to first order in the displacements.

Since this transition is driven by transverse displacive modes $\bm{\eta}_{\bm{Q}}$ ($\bm{\eta}_{\bm{Q}}\perp \bm{Q}$, $\bm{Q} = \bm{Q}_{\text{X}}, \bm{Q}_{\text{Y}}, \bm{Q}_{\text{Z}}$) at the X points, $V'_{\text{other}}$ within the harmonic approximation is written as
\begin{align}
V'_{\text{other}} &= \sum_{\bm{k},\bm{p},\bm{Q}}\sum_{\mu=x,y,z}\sum_{n,n'}C^{\bm{k},\bm{p},\bm{Q}}_{nn'\mu}\,u_{n,\bm{k}}u_{n',\bm{p}}\eta_{\mu,\bm{Q}}\,\delta_{\bm{k} + \bm{p} + \bm{Q},\bm{G}},\label{eq:D_other}
\end{align}
where $\bm{G}$ is a reciprocal lattice vector. 
%This contrasts with the ET-multipole coupling in Eq.~(\ref{eq:phonon_disp}).

Apart from the trivial symmetry under the interchange $\bm{k}\leftrightarrow \bm{p}$ and $n\leftrightarrow n'$, i.e., $C^{\bm{k},\bm{p},\bm{Q}}_{nn'\mu}=C^{\bm{p},\bm{k},\bm{Q}}_{n'n\mu}$, the coupling constant $C^{\bm{k},\bm{p},\bm{Q}}_{nn'\mu}$ satisfies the following symmetry constraints:
\begin{align}
    &C^{\bm{k},\bm{p},\bm{Q}}_{nn'\mu}=[C^{-\bm{k},-\bm{p},\bm{Q}}_{nn'\mu}]^*\qquad (\rm{TRS\ or\ Hermitian}),\label{eq:constraint_TRS}\\
    &C^{\bm{k},\bm{p},\bm{Q}}_{nn'\mu}=-C^{-\bm{k},-\bm{p},\bm{Q}}_{nn'\mu}\qquad (\rm{spatial\ inversion}),\label{eq:constraint_I}\\ 	
    &C^{\bm{k},\bm{p},\bm{Q}}_{nn'\mu}=-[C^{\bm{k},\bm{p},\bm{Q}}_{nn'\mu}]^*\qquad {\rm from}\ {\rm Eqs.}\ (\ref{eq:constraint_TRS}) 	 \ {\rm and}\  (\ref{eq:constraint_I}).	\label{eq:constraint_TRS_I}
\end{align}
To derive Eq.~(\ref{eq:constraint_I}), we use the transformation properties of the phonons and the displacement under spatial inversion: $u_{n,\bm{k}}\to -u_{n,-\bm{k}}$ and $\eta_{\mu,\bm{Q}}\to -\eta_{\mu,-\bm{Q}} \equiv -\eta_{\mu,\bm{Q}}$. From Eq.~(\ref{eq:constraint_TRS_I}), one finds that $C^{\bm{k},\bm{p},\bm{Q}}_{nn'\mu}$ is purely imaginary. Here, we use the primitive unit cell rather than the enlarged cubic unit cell so that the coupling to the displacive modes $\eta_{\mu,\bm{Q}}$ is explicit.

The three triply degenerate optical modes in Fig.~\ref{fig:CPS_Oh}(a) originate from the phonon modes $u_{n,\bm{k}+\bm{Q}}$ with $\bm{Q}=\bm{Q}_{\rm X}$, $\bm{Q}_{\rm Y}$, and $\bm{Q}_{\rm Z}$ in the primitive BZ with $\bm{k} \parallel [111]$. This degeneracy is enforced by the $C_3$ rotation about the $[111]$ axis and its equivalents. Thus, the eigenvalues of $u_{n,\bm{k}+\bm{Q}_{\text{X}}}$, $u_{n,\bm{k}+\bm{Q}_{\text{Y}}}$, and $u_{n,\bm{k}+\bm{Q}_{\text{Z}}}$ are degenerate:
\begin{align}
\omega^{(0)}_{n,\bm{k}+\bm{Q}_{\text{X}}}
=
\omega^{(0)}_{n,\bm{k}+\bm{Q}_{\text{Y}}}
=
\omega^{(0)}_{n,\bm{k}+\bm{Q}_{\text{Z}}}.
\end{align}

A coupling among these modes arises upon setting $\bm{k}\to \bm{k}+\bm{Q}_{\rm X}$, $\bm{p}\to -\bm{k}+\bm{Q}_{\rm Y}$, and $\bm{Q}\to \bm{Q}_{\rm Z}$ in Eq.~\eqref{eq:D_other}, as well as similar combinations, where the Kronecker delta $\delta_{\bm{k} + \bm{p} + \bm{Q},\bm{G}}$ is satisfied by the Umklapp condition $\bm{Q}_{\rm X}+\bm{Q}_{\rm Y}+\bm{Q}_{\rm Z}=\bm{G}$. The resulting coupling takes the form,
\begin{align}
&\sum_{\bm{k},\mu,n,n'}C^{\bm{k}-\bm{Q}_{\text{X}},-\bm{k}+\bm{Q}_{\text{Y}},\bm{Q}_{\text{Z}}}_{nn'\mu}[u_{n,\bm{k}-\bm{Q}_{\text{X}}}u_{n',-\bm{k}+\bm{Q}_{\text{Y}}}\notag\\
&\hspace{2cm}-
u_{n,-\bm{k}+\bm{Q}_{\text{X}}}u_{n',\bm{k}-\bm{Q}_{\text{Y}}}
]
\eta_{\mu,\bm{Q}_{\text{Z}}}+{\rm cyclic},
\label{eq:couple_tripleQ}
\end{align}
where $C^{\bm{k}-\bm{Q}_{\text{X}},-\bm{k}+\bm{Q}_{\text{Y}},\bm{Q}_{\text{Z}}}_{nn'\mu}$ is purely imaginary. Here, ``cyclic'' denotes a cyclic permutation with respect to $\bm{Q}$. The sign convention for $\pm\bm{Q}$ is chosen such that the wave vector remains inside the first BZ, with $\bm{k}=k(1,1,1)$ and small $k>0$.

%Thus, for the triple-$\bm{q}$ order, the leading effects arise from 
We note that the coefficient $C^{\bm{k}-\bm{Q}_{\text{X}},-\bm{k}+\bm{Q}_{\text{Y}},\bm{Q}_{\text{Z}}}_{nn'\mu}$ is odd in $\bm{k}$ since 
\begin{align}
	C^{\bm{k}-\bm{Q}_{\text{X}},-\bm{k}+\bm{Q}_{\text{Y}},\bm{Q}_{\text{Z}}}_{nn'\mu}
	&=
-C^{-\bm{k}+\bm{Q}_{\text{X}},\bm{k}-\bm{Q}_{\text{Y}},\bm{Q}_{\text{Z}}}_{nn'\mu}\notag\\
&=
-C^{-\bm{k}-\bm{Q}_{\text{X}},\bm{k}+\bm{Q}_{\text{Y}},\bm{Q}_{\text{Z}}}_{nn'\mu}.
\end{align}
Here, we use $\bm{Q}_{\rm X}\equiv -\bm{Q}_{\rm X}$ and $\bm{Q}_{\rm Y}\equiv -\bm{Q}_{\rm Y}$.

%An important point is that $\bm{k}+\bm{Q}_{\text{X}}$, $\bm{k}+\bm{Q}_{\text{Y}}$, and $\bm{k}+\bm{Q}_{\text{Z}}$ are related by a threefold rotation about the $[111]$ direction. When we adopt the convention that the index $n$ labels the eigenmodes in ascending order of their eigenvalues, the three 

In the enlarged cubic unit cell, the BZ is folded and $\bm{Q}_{\text{X}}$, $\bm{Q}_{\text{Y}}$, $\bm{Q}_{\text{Z}}$ are mapped onto the $\Gamma$ point. Denoting the common frequency by $\omega'_{n,\bm{k}}$,
\begin{align}
\omega^{(0)}_{n,\bm{k}+\bm{Q}_{\text{X}}}
=
\omega^{(0)}_{n,\bm{k}+\bm{Q}_{\text{Y}}}
=
\omega^{(0)}_{n,\bm{k}+\bm{Q}_{\text{Z}}}
=
\omega'_{n,\bm{k}}.
\end{align}
To linear order in $\eta_{\mu,\bm{Q}}$ and at each $\bm{k}$ with $|\bm{k}|>0$, the dynamical matrix for the degenerate manifold $\lbrace u_{n,\bm{k}+\bm{Q}_\text{X}},u_{n,\bm{k}+\bm{Q}_\text{Y}},u_{n,\bm{k}+\bm{Q}_\text{Z}}\rbrace$ is 
\begin{align}
	\begin{pmatrix}
		(\omega'_{n,\bm{k}})^2 & ig_{n,\bm k} & ig_{n,\bm k}\\
		-ig_{n,\bm k} & (\omega'_{n,\bm{k}})^2 & ig_{n,\bm k} \\
		-ig_{n,\bm k} & -ig_{n,\bm k} &(\omega'_{n,\bm{k}})^2 
	\end{pmatrix},
\end{align}
where $ig_{n,\bm{k}}$ $=$ $C^{\bm{k}-\bm{Q}_{\text{X}},-\bm{k}+\bm{Q}_{\text{Y}},\bm{Q}_{\text{Z}}}_{nnx}\eta_{x,\bm{Q}_{\rm Z}}$ $=$ $C^{\bm{k}-\bm{Q}_{\text{Y}},-\bm{k}+\bm{Q}_{\text{Z}},\bm{Q}_{\text{X}}}_{nny}\eta_{y,\bm{Q}_{\rm X}}$ $=$ $C^{\bm{k}-\bm{Q}_{\text{Z}},-\bm{k}+\bm{Q}_{\text{X}},\bm{Q}_{\text{Y}}}_{nnz}\eta_{z,\bm{Q}_{\rm Y}}$. The resulting phonon frequencies are 
\begin{align}
	\omega'_{n,\bm{k}},\quad \omega'_{n,\bm{k}}\pm \frac{\sqrt{3}}{2} \frac{|g_{n,\bm{k}}|}{\omega'_{n,\bm{k}}},
\end{align}
which indicates that the splittings are equidistant within the degenerate manifold, as found in Fig.~\ref{fig:CPS_Oh}(d) for small $\eta$.

Other possible terms couple phonons near the $\Gamma$ point to those near $\bm{Q} = \bm{Q}_\text{X}$, $\bm{Q}_\text{Y}$, and $\bm{Q}_\text{Z}$, corresponding to $\bm{p}\to-\bm{k}+\bm{Q}$ in Eq.~\eqref{eq:D_other}. However, since the phonon frequencies at these distinct $\bm{k}$ points are generically nondegenerate, such couplings contribute to the splitting only at the second order in perturbation theory and are suppressed by the energy denominator. In contrast, the coupling of the triply degenerate modes in Eq.~\eqref{eq:couple_tripleQ} contributes to the splitting at the first order in perturbation theory, making it the dominant contribution.

\subsection{Temperature dependence of chirality-induced phenomena}\label{discussion-Tdep}

We discuss the temperature dependence of chirality-sensitive responses across the AtC phase transition. 
We emphasize that experiments should not only detect whether chirality is finite or zero, but also examine the temperature dependence of chirality-induced phenomena such as CPS and CIM. Such scaling behavior provides information about how structural chirality develops from the underlying displacement order parameter. In this section, we focus on CIM as a representative response. Assuming the mean-field temperature dependence $\eta\equiv |\boldsymbol{\eta}|\sim\sqrt{T_c-T}$, we present the corresponding temperature dependence of CIM for representative examples discussed in this paper. This perspective provides a way to extract the symmetry-controlled functional form of chirality from experimentally measurable responses.

In second-order phase transitions, Landau theory predicts a mean-field temperature dependence $\eta\sim \sqrt{T_c-T}$ below the structural transition temperature $T_c$. This dependence is generally modified by fluctuation effects~\cite{G-Furuya}. However, the mean-field result provides a useful starting point for understanding the underlying physics.
Moreover, structural phase transitions are typically accompanied by long-range elastic interactions, which suppress critical fluctuations and extend the validity of the mean-field description over a wide temperature range.
Structural phase transitions can also be first order. Strictly speaking, our formulation does not directly apply to this case; however, the expressions of $G_{0,u}$ still provide important information, such as even/odd/hybrid properties with respect to $\bm{\eta}$. When the discontinuity at the first-order transition is small, one can discuss the temperature dependence qualitatively by analogy with second-order transitions.

\begin{figure}
\includegraphics[width=1.0\linewidth]{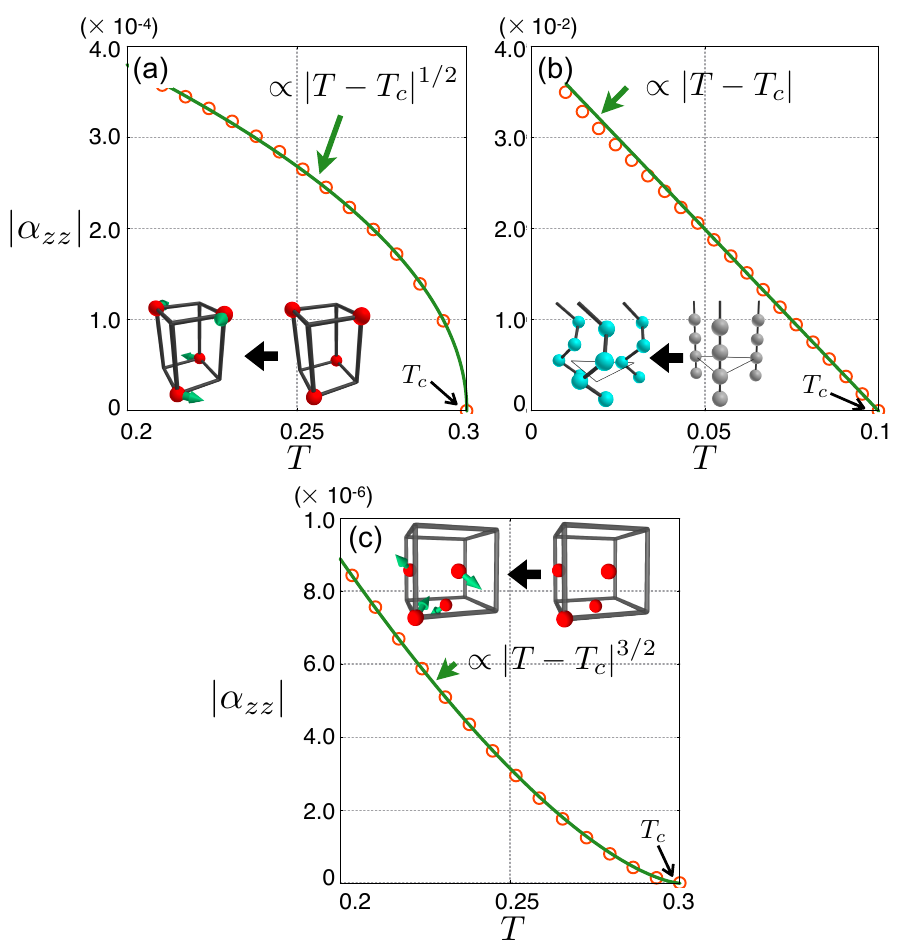}
\caption{\label{fig:CIM_temp}
Temperature dependence of $|\alpha_{zz}|$ for (a) $\text{D}_{2\text{d}}\to \text{D}_2$, (b) $\text{P}6/\text{mmm}\to \text{P}3_121$, and (c) $\text{Fm}\bar{3}\text{m}\to \text{P}2_13$. The green lines are $\propto |T-T_c|^{n/2}$, where $n$ corresponds to the leading order of $G_{0,u}(\bm{\eta})$ obtained in
Sec.~\ref{sec:toroidals}: (a) $n=1$ and $T_c=0.3$, (b) $n=2$ and $T_c=0.1$, and (c) $n=3$ and $T_c=0.3$. The electron filling $n_f$ is set to (a) $n_f=11/12$, (b) $n_f=8/9$, and (c) $n_f=11/12$. 
}
\end{figure}

Figures~\ref{fig:CIM_temp}(a)--\ref{fig:CIM_temp}(c) show the temperature
dependence of $|\alpha_{zz}|$ for the transitions
$\text{D}_{2\text{d}} \rightarrow \text{D}_2$,
$\text{P}6/\text{mmm} \rightarrow \text{P}3_121$, and
$\text{Fm}\bar{3}\text{m} \rightarrow \text{P}2_13$, respectively.
The same scaling argument applies to CPS when the phonon splitting is governed by the ET-multipole coupling. The orange circles represent the numerical results and the green lines
are fit by $\propto |T-T_c|^{n/2}$,
where $n$ is the leading order of $G_{0,u}(\bm{\eta})\propto \eta^n$ obtained in
Sec.~\ref{sec:toroidals}: (a) $n=1$, (b) $n=2$, and (c) $n=3$.
Here, unlike the analyses in Sec.~\ref{subsec:CIM}, we fix the electron filling $n_f$ by varying the chemical potential $\mu_c$ for each 
$T$: $n_f = 11/12$ for (a) and (c), and
$n_f = 8/9$ for (b), which leads to the  values of $\mu_c$ similar to those
in Sec.~\ref{subsec:CIM}.
One can see in Fig.~\ref{fig:CIM_temp} that the temperature
dependences of $\alpha_{zz}$ follow 
the power law of $G_{0,u}\propto \eta^n\propto (T_c-T)^{n/2}$ for each system. Although we do not show the data, there are AtC transitions with $G_{0,u}\propto \eta^n\propto|T_c-T|^{\frac{n}{2}}$ ($n\geq 4$). For example, a triple-$\bm{q}$ order at the W point in the fcc lattice shows $G_0\propto\eta^6$ as demonstrated in Appendix~\ref{subsec:others}.

From an experimental viewpoint, the key test is to compare chirality-sensitive responses with the structural order parameter itself. A useful strategy is to determine the displacement amplitude $\eta(T)$ by diffraction or structural refinement and to examine whether CPS, CIM, or other responses, denoted collectively by $R(\eta)$, scale as $R(\eta)\propto G_{0,u}(\boldsymbol{\eta})\sim \eta^n$ across the same AtC transition. When the response is governed by the ET-multipole coupling, this scaling is fixed by symmetry and is largely independent of microscopic details. The same analysis can be extended to nonlinear optical and transport responses, such as second-harmonic generation (SHG) and nonlinear Hall effects, once the relevant symmetry-selected tensor components are identified. SHG is sensitive to inversion- or mirror-symmetry breaking, as demonstrated for odd-parity multipolar order in Cd$_2$Re$_2$O$_7$~\cite{Harter2017,DiMatteo2017}, whose ET-multipole character has been discussed theoretically~\cite{Hayami2019}. Nonlinear Hall responses probe inversion-broken electronic structures through mechanisms such as the Berry-curvature dipole~\cite{Sodemann2015}. These responses are not purely chiral probes in general, because nonchiral noncentrosymmetric distortions, electronic-structure effects, and scattering processes can also contribute. Nevertheless, the chirality-controlled tensor components should inherit the leading order of the corresponding $G_{0,u}(\boldsymbol{\eta})$. In phonon measurements, one should similarly distinguish ET-multipole-controlled CPS from generic symmetry-lowering splittings; acoustic branches and the circular-polarization character of the eigenmodes provide cleaner indicators of the chirality-induced contribution.

\section{\label{sec:conclusion}Conclusion}
We have investigated structural achiral-to-chiral (AtC) phase transitions driven by atomic displacements and formulated the electric toroidal (ET) monopole $G_0$ and quadrupole $G_u$ as functions of the displacement order parameters. This formulation provides a symmetry-based way to quantify structural chirality beyond a binary classification into achiral and chiral phases. We have shown that the leading order of $G_{0,u}(\bm{\eta})$ depends on the symmetry of the parent structure and on the character of the structural order parameter.
Specifically, for uniform transitions at $\bm{q}=\bm{0}$, we have classified the leading-order expressions of $G_{0,u}(\bm{\eta})$ for all achiral point groups. For nonuniform transitions at $\bm{q}\ne\bm{0}$, we have derived the conditions for the existence of the second-order term $G_{0,u}^{(2)}$ according to the symmetry of the little group at the ordering vector, including the $\bm{k}$-arm chirality realized when the little group contains only proper operations. When these conditions are not satisfied, third- or higher-order terms become the leading contribution; we have identified the minimal order allowed by the Umklapp condition and explicitly constructed the corresponding forms for representative multiple-$\bm{q}$ orders.
As a result, different AtC transitions are distinguished not only by their low-temperature chiral space groups but also by the functional form of chirality in terms of the displacement order parameter.

We have demonstrated that this displacement dependence is directly reflected in chirality-induced responses. In the lattice sector, the chiral-phonon splitting (CPS) follows the leading behavior of $G_{0,u}(\bm{\eta})$ when the splitting is governed by the coupling between phonon degrees of freedom and the ET monopole or quadrupole. In the electronic sector, the current-induced magnetization (CIM) exhibits the same symmetry-dictated dependence on the displacement amplitude. These results establish CPS and CIM as complementary probes of structural chirality and show that their scaling behavior can reveal the symmetry character of the underlying AtC transition.

We have also clarified the microscopic mechanism of CPS. For degenerate phonon modes, the coupling between $G_{0,u}(\bm{\eta})$ and the ET monopole or quadrupole constructed from phonon degrees of freedom lifts the degeneracy and produces chiral phonon eigenmodes. This mechanism leads to a CPS proportional to $G_{0,u}(\bm{\eta})$. We have further shown that additional nonchiral couplings can contribute to the splitting of optical modes even across an AtC transition, particularly in folded band structures, whereas the acoustic-branch splitting provides a cleaner probe of the ET-multipole contribution. Anisotropic terms can also make the phonon polarization elliptical and suppress the phonon angular momentum.

Finally, since the displacement order parameter evolves with temperature across a structural phase transition, the functional form of $G_{0,u}(\bm{\eta})$ determines the temperature dependence of chirality-induced phenomena. Thus, measuring not only whether CPS or CIM is finite but also how it changes with temperature, preferably in comparison with the independently determined $\eta(T)$, provides a route to identifying the symmetry-controlled order-parameter dependence of chirality. This perspective offers a refined classification of chiral structural phase transitions based on the leading order and parity of $G_{0,u}(\bm{\eta})$, and it may be useful for interpreting and designing chiral materials with controllable lattice and electronic responses.

\appendix
\section{\label{subsec:others}$G_{0,u}$ in more general systems}

In the main text, we have mainly considered minimal models with simple sublattice structures. Here,
we consider explicit expressions for $G_{0,u}$ in more complex systems with multiple sublattice degrees of freedom. In real materials, a primitive unit cell typically contains multiple sites. It is therefore important to derive $G_{0,u}$ for such systems.

The first two examples are simplified models of the materials GaNb$_4$Se$_8$ and K$_3$NiO$_2$, which exhibit structural AtC transitions~\cite{Kitou_chiral,KNO_experi}. Then, we examine AtC transitions from a diamond structure to two chiral structures with distinct space groups. These transitions are driven by two doubly degenerate eigenmodes originating from the nonsymmorphic nature of the diamond structure. Next, we show an AtC transition in the bcc lattice, where the third-order term $G_{0,u}^{(3)}$ constructed from a one-dimensional irrep serves as the leading contribution. Lastly, we demonstrate that an even higher, sixth-order term $G_{0,u}^{(6)}$ arises as the leading term in an AtC transition in the fcc lattice.

\subsection{\label{subsubsec:Breathing}Breathing pyrochlore lattice}
\begin{figure}
\includegraphics[width=1.0\linewidth]{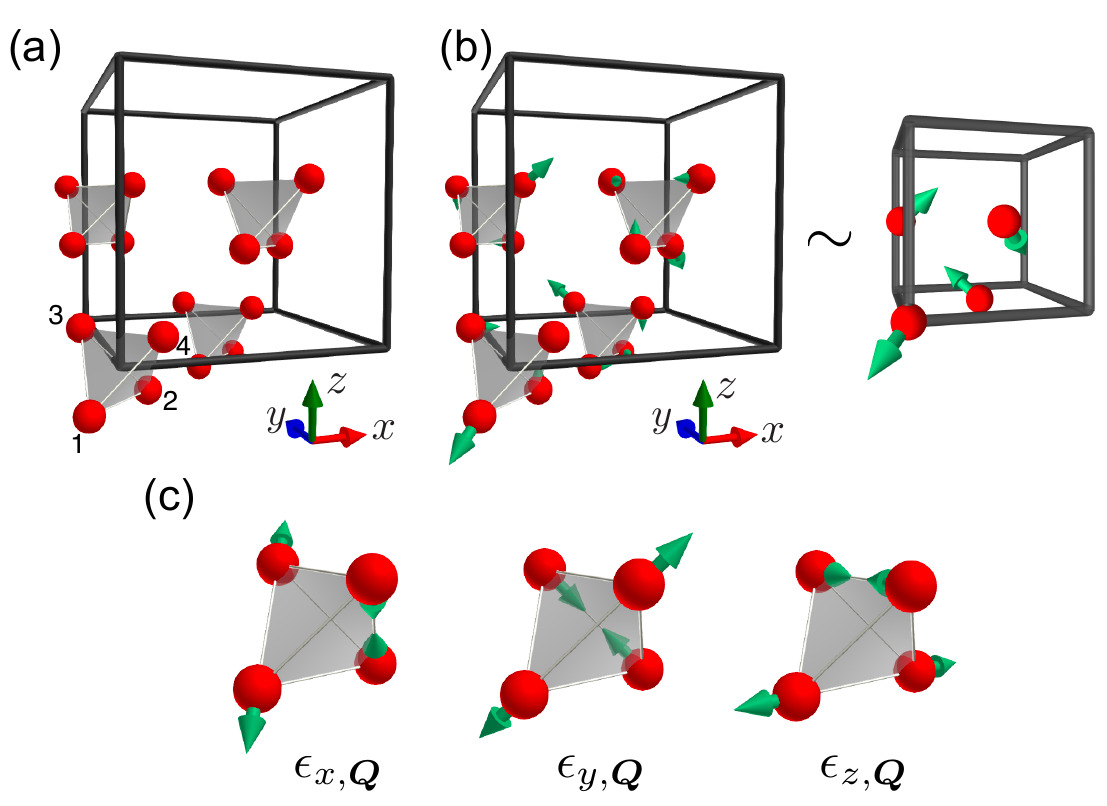}
\caption{\label{fig:Bpyro_image}(a) Schematic illustration of the breathing pyrochlore lattice. The numbers indicate the sublattice indices. (b) Displacement pattern that induces the AtC transition. (c) Schematic illustration of the $\epsilon_{x,\bm{Q}}$, $\epsilon_{y,\bm{Q}}$, and $\epsilon_{z,\bm{Q}}$ modes of the tetrahedron.}
\end{figure}

As a first example, we consider a transition from the breathing pyrochlore lattice with space group F$\bar{4}3$m (No.~216) to the P$2_1$3 (No.~198) structure. The breathing pyrochlore lattice can be viewed as an fcc lattice of tetrahedra, as shown in Fig.~\ref{fig:Bpyro_image}(a). This transition has been experimentally observed in the spinel compound GaNb$_4$Se$_8$~\cite{Kitou_chiral}, where the Nb atoms form tetrahedra and a chiral charge order drives a rearrangement of the Nb tetrahedral clusters from a tetramer to a trimer--monomer configuration. As a result, the four tetrahedra in the cubic unit cell are distorted along the four equivalent $[111]$ directions, leading to a finite dipole moment for each tetrahedron.

By regarding each distorted tetrahedron as an effective electric-dipole degree of freedom,  the system can be mapped onto the fcc model discussed in Sec.~\ref{subsec:fcc}, as illustrated in Fig.~\ref{fig:Bpyro_image}(b). This mapping suggests that the transition can be described as a triple-$\bm{q}$ order at the X points, $\bm{Q}_{\text{X}} = (1,0,0)$, $\bm{Q}_{\text{Y}} = (0,1,0)$, and $\bm{Q}_{\text{Z}} = (0,0,1)$ in units of $2\pi/a$. As in Sec.~\ref{subsec:fcc}, transverse modes at each X point drive this transition. 
For the present case, the transverse modes at $\bm{Q}_{\rm X}$, $(\epsilon_{y,\bm{Q}_\text{X}},\epsilon_{z,\bm{Q}_\text{X}})$, are given by 
\begin{align}
	\bm{\epsilon}_{y,\bm{Q}_{\rm X}}^{1}&=\frac{1}{\sqrt{8}}(-1,0,-1),\quad 
	\bm{\epsilon}_{z,\bm{Q}_{\rm X}}^{1}=\frac{1}{\sqrt{8}}(-1,-1,0),\label{eq:ep_X_1}\\
	\bm{\epsilon}_{y,\bm{Q}_{\rm X}}^{2}&=\frac{1}{\sqrt{8}}(-1,0,1),\quad 
	\bm{\epsilon}_{z,\bm{Q}_{\rm X}}^{2}=\frac{1}{\sqrt{8}}(1,1,0),\\
	\bm{\epsilon}_{y,\bm{Q}_{\rm X}}^{3}&=\frac{1}{\sqrt{8}}(1,0,-1),\quad 
	\bm{\epsilon}_{z,\bm{Q}_{\rm X}}^{3}=\frac{1}{\sqrt{8}}(1,-1,0),\\
	\bm{\epsilon}_{y,\bm{Q}_{\rm X}}^{4}&=\frac{1}{\sqrt{8}}(1,0,1),\quad 
	\bm{\epsilon}_{z,\bm{Q}_{\rm X}}^{4}=\frac{1}{\sqrt{8}}(-1,1,0).\label{eq:ep_X_4}
\end{align}
Here, we have used the notation $\bm{\epsilon}_{\gamma,\bm{Q}_{\rm X}}^{s}=[(\bm{\epsilon}^s_{\gamma,\bm{Q}_\text{X}})_{x},(\bm{\epsilon}^s_{\gamma,\bm{Q}_\text{X}})_{y},(\bm{\epsilon}^s_{\gamma,\bm{Q}_\text{X}})_{z}]$ and $s$ is the sublattice index shown in Fig.~\ref{fig:Bpyro_image}(a). Similarly, the transverse modes at $\bm{Q}_{\rm Y,Z}$ are obtained from Eqs.~(\ref{eq:ep_X_1})--(\ref{eq:ep_X_4}) by the threefold rotation about the $[111]$ direction.  We denote the coefficients of the corresponding two-dimensional irreps as $\bm{\eta}_{\bm{Q}_{\text{X}}} = (\eta_{y,\bm{Q}_{\text{X}}},\eta_{z,\bm{Q}_{\text{X}}})$, $\bm{\eta}_{\bm{Q}_{\text{Y}}} = (\eta_{z,\bm{Q}_{\text{Y}}},\eta_{x,\bm{Q}_{\text{Y}}})$, and $\bm{\eta}_{\bm{Q}_{\text{Z}}} = (\eta_{x,\bm{Q}_{\text{Z}}},\eta_{y,\bm{Q}_{\text{Z}}})$. These transform as the $\mu$ components of a polar vector, i.e., dipole moments. Thus,
$G_0(\bm{\eta})$ can be constructed in a manner similar to that in Sec.~\ref{subsec:fcc} as
\begin{align}
G^{(3)}_0(\bm{\eta}) \propto 
\eta_{y,\bm{Q}_{\text{X}}}
\eta_{z,\bm{Q}_{\text{Y}}}
\eta_{x,\bm{Q}_{\text{Z}}}
-
\eta_{z,\bm{Q}_{\text{X}}}
\eta_{x,\bm{Q}_{\text{Y}}}
\eta_{y,\bm{Q}_{\text{Z}}}.
\end{align}
However, there is an important difference between the present system and the fcc lattice; the breathing pyrochlore lattice belongs to the space group F$\bar{4}3$m rather than Fm$\bar{3}$m. Consequently, the little group at the X points is D$_{2\text{d}}$. Within this little group, for example, the $x^2 - y^2$-type irrep at $\bm{Q}_{\text{Z}}$ transforms as $G_u$, as discussed in Sec.~\ref{subsec:fcc}. Accordingly, a second-order contribution $G^{(2)}_0(\bm{\eta})$ can be constructed by summing $G_u$ over all X points as
\begin{align}
G^{(2)}_0(\bm{\eta}) \propto 
\eta^2_{y,\bm{Q}_{\text{X}}} - \eta^2_{z,\bm{Q}_{\text{X}}}
+ \text{cyclic},
\end{align}
where ``cyclic'' denotes the cyclic permutation $x \rightarrow y \rightarrow z \rightarrow x$ and $\bm{Q}_{\text{X}} \rightarrow \bm{Q}_{\text{Y}} \rightarrow \bm{Q}_{\text{Z}} \rightarrow \bm{Q}_{\text{X}}$. The existence of $G_0^{(2)}$ for $\mathscr{G}_{\bm{p}}\cong \text{D}_{2\text{d}}$ is consistent with the condition (i) discussed in Sec.~\ref{subsubsec:nonuniform_condition}.

In summary, near the transition temperature $T_c$, we obtain
\begin{align}
G_0(\bm{\eta})= C^{(2)}G^{(2)}_0(\bm{\eta}) + C^{(3)} G^{(3)}_0(\bm{\eta}),
\end{align}
where $C^{(2)}$ and $C^{(3)}$ are the coefficients of the second- and third-order contributions, respectively. The coexistence of finite $C^{(2)}$ and $C^{(3)}$ reflects the inequivalence of the structures associated with $\bm{\eta}$ and $-\bm{\eta}$. This can be understood by noting that elongated tetrahedra are not equivalent to compressed ones and are not related by mirror symmetry.

The RH and LH domains are constructed in a manner similar to that discussed in Sec.~\ref{subsec:fcc}. An important difference, however, is that arbitrary sign changes $\eta_{\mu,\bm{Q}}\to \pm \eta_{\mu,\bm{Q}}$ are not allowed. For example, RH configurations are given by $\{\eta_{y,\bm{Q}_\text{X}},\eta_{z,\bm{Q}_\text{Y}},\eta_{x,\bm{Q}_\text{Z}}\}=\{1,1,1\},\{-1,-1,1\},\{1,-1,-1\}$, and $\{-1,1,-1\}$. In contrast, the configuration $\{\eta_{y,\bm{Q}_\text{X}},\eta_{z,\bm{Q}_\text{Y}},\eta_{x,\bm{Q}_\text{Z}}\}=\{-1,-1,-1\}$ leads to a qualitatively different tetrahedral deformation and hence to an inequivalent structure.

For comparison, in the fcc case, the equivalence between the structures with $\bm{\eta}$ and $-\bm{\eta}$ implies that configurations $\{\eta_{y,\bm{Q}_\text{X}},\eta_{z,\bm{Q}_\text{Y}},\eta_{x,\bm{Q}_\text{Z}}\}=\{-1,-1,-1\},\{1,1,-1\},\{-1,1,1\}$, and $\{1,-1,1\}$ correspond to LH states that are enantiomorphic to $\{1,1,1\}$. In this sense, the distinction between RH and LH domains in the breathing pyrochlore system is more explicit than in the fcc system. The handedness is uniquely determined by the triple-$\bm{q}$ set $\{\eta_{y,\bm{Q}_\text{X}},\eta_{z,\bm{Q}_\text{Y}},\eta_{x,\bm{Q}_\text{Z}}\}$ or $\{\eta_{z,\bm{Q}_\text{X}},\eta_{x,\bm{Q}_\text{Y}},\eta_{y,\bm{Q}_\text{Z}}\}$.

%Figure~\ref{fig:Bpyro_image}(d) shows the deformation of the tetrahedron for $\bm{\eta}$ and $-\bm{\eta}$, where $\bm{\eta}$ elongates the tetrahedron along the $[111]$ direction while $-\bm{\eta}$ compresses it. Due to this difference, the structure with $\bm{\eta}$ and $-\bm{\eta}$ do not form domains and either of the structure is always chosen. 

\subsection{Rutile structure}\label{app:rutile}

\begin{figure}
\includegraphics[width=1.0\linewidth]{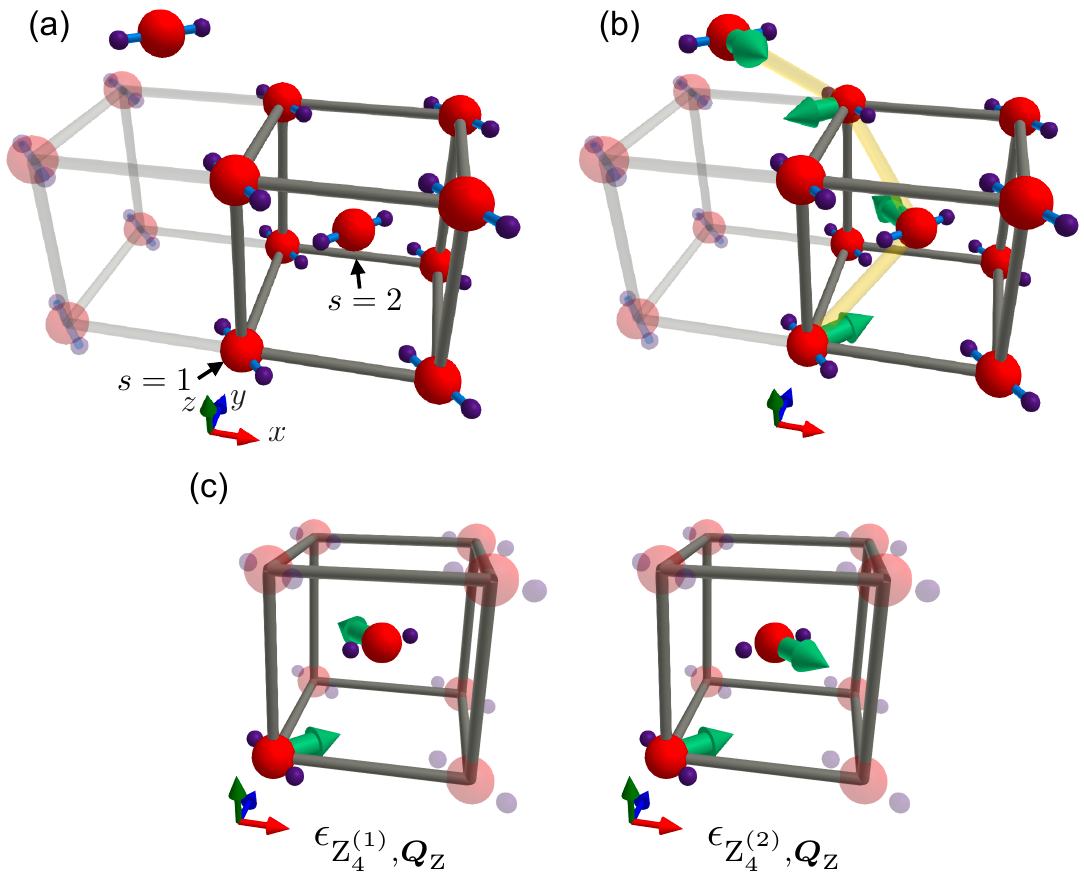}
\caption{\label{fig:rutile_image}
(a) Schematic illustration of the rutile structure in K$_3$NiO$_2$. The red and purple spheres represent K and O atoms. 
(b) Displacement pattern (green arrows) of the AtC transition from the rutile structure (space group P$4_2$/mnm) to the chiral P$4_12_12$ structure. The yellow tube represents one of the fourfold helices emerging in this transition. 
(c) Schematic illustration of the doubly degenerate Z$_4$ mode $(\epsilon^{(1)}_{\text{Z}_4},\epsilon^{(2)}_{\text{Z}_4})$ responsible for the transition.
}
\end{figure}

As a second example, we consider a transition from the rutile structure with space group P$4_2$/mnm (No.~136) to a fourfold helical structure with space group P$4_12_12$ (No.~92) or P$4_32_12$ (No.~96), motivated by the recent discovery of an AtC transition in K$_3$NiO$_2$~\cite{KNO_experi,KNO}. In this compound, the K and O atoms form a rutile structure, as shown in Fig.~\ref{fig:rutile_image}(a), where the red and purple spheres represent the K and O atoms, respectively. The rutile structure contains two sublattices, $s=1,2$, as illustrated in Fig.~\ref{fig:rutile_image}(a). The local environments of the K atoms at $s=1$ and $s=2$ are rotated by $90^\circ$ relative to each other due to the different arrangements of the surrounding O atoms, rendering the system nonsymmorphic.

In K$_3$NiO$_2$, the displacement-driven structural phase transition takes place at $T = 400\text{--}423$ K~\cite{KNO_experi}. In the low-temperature phase, the K atoms move within the $xy$ plane, as illustrated in Fig.~\ref{fig:rutile_image}(b). The resulting displacements lower the symmetry from the rutile structure to the P$4_12_12$ or P$4_32_12$ structure. The displacement pattern that yields the P$4_12_12$ structure is schematically illustrated by green arrows in Fig.~\ref{fig:rutile_image}(b), where the yellow tube represents one of the fourfold helices emerging in this transition. First-principles calculations indicate that this transition is driven by an in-plane eigenmode, referred to as the Z$_4$ mode at the Z point $\bm{Q}_{\text{Z}} = (0,0,1/2)$ of the tetragonal BZ~\cite{KNO}. The Z$_4$ mode is doubly degenerate, and we denote it as $(\epsilon_{\text{Z}^{(1)}_4,\bm{Q}_\text{Z}},\epsilon_{\text{Z}^{(2)}_4,\bm{Q}_\text{Z}})$, where $\text{Z}^{(m)}_4$ ($m=1,2$) represents the $m$th component of the Z$_4$ mode. 
The eigenmodes $\bm{\epsilon}^s_{\text{Z}_4^{(m)},\bm{Q}_\text{Z}} = [(\epsilon_{\text{Z}_4^{(m)},\bm{Q}_\text{Z}})_{x,s},(\epsilon_{\text{Z}_4^{(m)},\bm{Q}_\text{Z}})_{y,s},(\epsilon_{\text{Z}_4^{(m)},\bm{Q}_\text{Z}})_{z,s}]$ are described by
\begin{align}
	\bm{\epsilon}^{1}_{\text{Z}_4^{(1)},\bm{Q}_{\rm Z}}&= \frac{1}{2}(1,1,0),\quad 
	\bm{\epsilon}^{1}_{\text{Z}_4^{(2)},\bm{Q}_{\rm Z}}= \frac{1}{2}(-1,1,0),\\
	\bm{\epsilon}^{2}_{\text{Z}_4^{(1)},\bm{Q}_{\rm Z}}&= \frac{1}{2}(1,1,0),\quad 
	\bm{\epsilon}^{2}_{\text{Z}_4^{(2)},\bm{Q}_{\rm Z}}= \frac{1}{2}(1,-1,0),
\end{align}
which are schematically shown in Fig.~\ref{fig:rutile_image}(c).
Here, we label the two displacive modes by $m=1,2$ instead of $x$ and $y$, since they cannot be classified by irreps of the point group owing to the nonsymmorphic nature of the system and $\eta_{\text{Z}_4^{(m)},\bm{Q}_\text{Z}}$ is related to the Fourier modes ${\bm d}_{\mu,s,\bm{Q}_Z}$ by Eq.~(\ref{eq:Psi_d_FT}).

Similarly to the example in Sec.~\ref{subsec:D6h}, the coefficient of the ordering mode determines the space group of the chiral phase. Since $\bm{Q}_\text{Z}$ is at the BZ boundary, $\eta_{\text{Z}_4^{(m)},\bm{Q}_\text{Z}}$ is real and $\eta_{\text{Z}_4^{(1)},\bm{Q}_\text{Z}}=\eta,\,\eta_{\text{Z}_4^{(2)},\bm{Q}_\text{Z}}=0$ ($\eta\in\mathbb{R}$) yields the P$4_12_12$ structure while its counterpart $\eta_{\text{Z}_4^{(1)},\bm{Q}_\text{Z}}=0,\,\eta_{\text{Z}_4^{(2)},\bm{Q}_\text{Z}}=\eta$ yields the P$4_32_12$ structure.

The modes $\epsilon_{\text{Z}_4^{(1)},\bm{Q}_{\text{Z}}}$ and $\epsilon_{\text{Z}_4^{(2)},\bm{Q}_{\text{Z}}}$ are eigenstates of the fourfold screw operation about the $z$ axis and the twofold screw operations about the $x$ and $y$ axes, while they are interchanged by glide or inversion operations. The detailed symmetry properties of the Z$_4$ mode are summarized in Appendix~\ref{apdx:irrep}.

Since the first-order term is forbidden due to $\bm{Q}_\text{Z}\ne0$ and the characters for the improper operations of the Z$_4$ mode are zero as shown in Table~\ref{tab:rutile}, $G_u(\bm{\eta})$ should start at least from the second order. 
Taking the transformation properties of  $\epsilon_{\text{Z}_4^{(m)},\bm{Q}_{\text{Z}}}$ into account, the second-order contribution $G_u^{(2)}(\bm{\eta})$ is given by
\begin{align}
G_u^{(2)}(\bm{\eta}) \propto (\eta_{\text{Z}_4^{(1)},\bm{Q}_{\text{Z}}})^2 - (\eta_{\text{Z}_4^{(2)},\bm{Q}_{\text{Z}}})^2\label{eq:Gu_rutile}.
\end{align}
This accords with Eq.~\eqref{eq:G0_second_gen}.
One can easily see that Eq.~\eqref{eq:Gu_rutile} is even under the rotation or screw operations since screw or rotational operations transform the coefficient $\eta_{\text{Z}_4^{(m)},\bm{Q}_\text{Z}}\rightarrow\pm\eta_{\text{Z}_4^{(m)},\bm{Q}_\text{Z}}$, whereas mirror and inversion operations lead to $\eta_{\text{Z}_4^{(m)},\bm{Q}_\text{Z}}\rightarrow\pm\eta_{\text{Z}_4^{(m')},\bm{Q}_\text{Z}}$ $(m\ne m')$, which changes the sign of $G^{(2)}_u(\bm{\eta})$.

In addition, odd-order contributions are forbidden since $(2n+1)\bm{Q}_{\text{Z}} \neq \bm{G}$ for $n=0,1,2,\cdots$, where $\bm{G}$ is a reciprocal lattice vector, which suggests the equivalence of the structure with $\bm{\eta}$ and $-\bm{\eta}$. It should also be emphasized that this transition occurs owing to the nonsymmorphic nature of the rutile structure. If the system is symmorphic, eigenmodes at the Z point belong to either the $E_{\rm g}$ or $E_{\rm u}$ representation of the point group D$_{4\text{h}}$, and it is impossible to construct $G_u$, which belongs to the $A_{1\rm{u}}$ representation, solely from eigenmodes belonging to the $E_{\rm g}$ or $E_{\rm u}$ representations. This suggests that nonsymmorphic systems provide additional opportunities for realizing AtC transitions.

\subsection{Diamond structure}\label{app:diamond}
\begin{figure}
\includegraphics[width=1.0\linewidth]{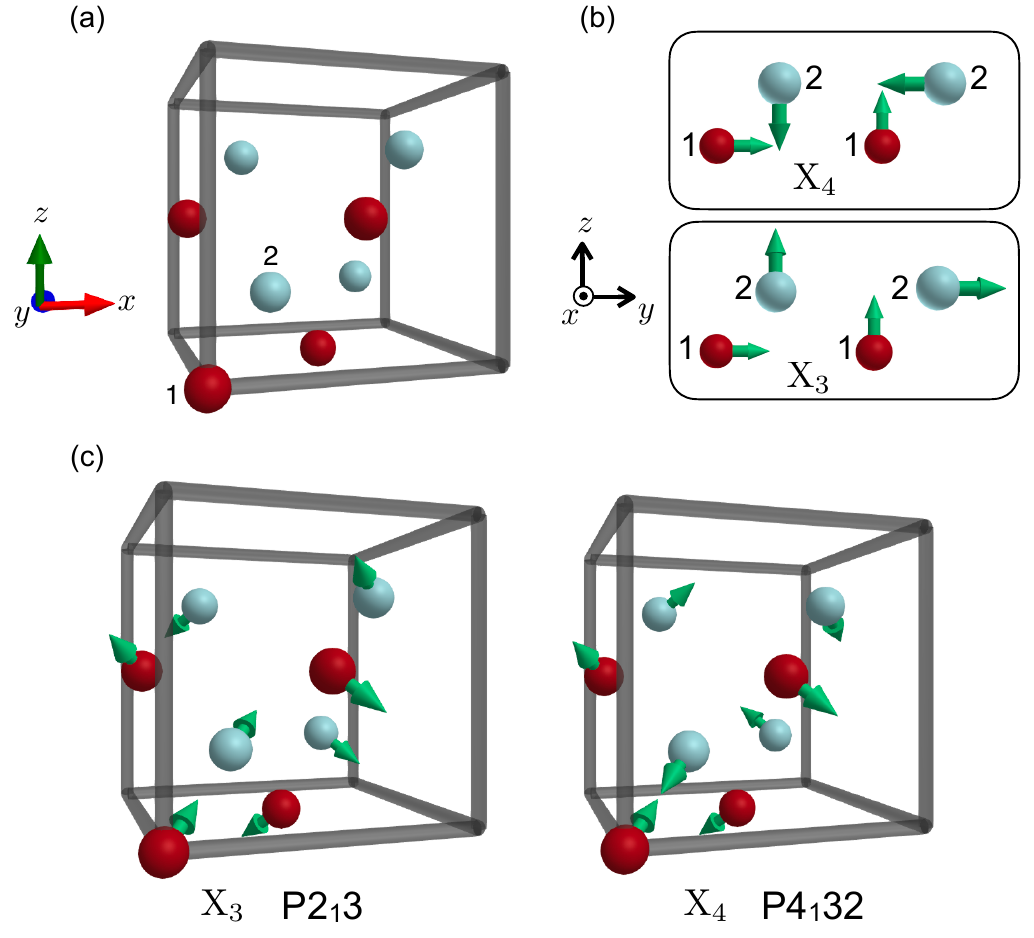}
\caption{\label{fig:diamond_image}
(a) Cubic unit cell of the diamond structure. The red and blue spheres represent sublattices 1 and 2, respectively. 
(b) Schematic illustration of the X$_3$ and X$_4$ modes at $\bm{Q}_{\text{X}} = (1,0,0)$ in units of $2\pi/a$.
(c) Displacement configurations leading to the P$2_1$3 and P$4_132$ structures.
}
\end{figure}
As a nonsymmorphic example of an AtC transition to cubic chiral space groups, we consider AtC transitions from the diamond structure to two cubic chiral structures. The diamond structure consists of two sublattices, $s=1,2$, each forming an fcc lattice, as illustrated in Fig.~\ref{fig:diamond_image}(a), where the red and blue spheres represent $s=1$ and $s=2$, respectively. As in the cubic-to-cubic transitions in Secs.~\ref{subsec:fcc} and \ref{subsubsec:Breathing}, we consider a triple-$\bm{q}$ order at the three X points. There exist two doubly degenerate eigenmodes at the X points, X$_3$ and X$_4$,  that induce chiral structures, which we denote as $(\epsilon_{\text{X}_\ell^{(1)},\bm{Q}},\epsilon_{\text{X}_\ell^{(2)},\bm{Q}})$, where the subscript $\text{X}_\ell^{(m)}$ represents the $m$th ($m=1,2$) component of the X$_\ell$ ($\ell=3,4$) mode. The eigenmodes $\bm{\epsilon}^s_{\text{X}^{(m)}_3,\bm{Q}_\text{X}}$ for X$_3$ at $\bm{Q}=\bm{Q}_{\rm X}$, 
% =[( \epsilon^s_{\text{X}^{(m)}_3,\bm{Q}_\text{X}})_{x},( \epsilon^s_{\text{X}^{(m)}_3,\bm{Q}_\text{X}})_{y},( \epsilon^s_{\text{X}^{(m)}_3,\bm{Q}_\text{X}})_{z}]$ 
are given by 
\begin{align}
	\bm{\epsilon}^1_{\text{X}^{(1)}_3,\bm{Q}_\text{X}} &=\frac{1}{\sqrt{2}}(0,1,0),\quad 
	\bm{\epsilon}^2_{\text{X}^{(1)}_3,\bm{Q}_\text{X}} =\frac{1}{\sqrt{2}}(0,0,1),\\	
	\bm{\epsilon}^1_{\text{X}^{(2)}_3,\bm{Q}_\text{X}} &=\frac{1}{\sqrt{2}}(0,0,1),\quad 
	\bm{\epsilon}^2_{\text{X}^{(2)}_3,\bm{Q}_\text{X}} =\frac{1}{\sqrt{2}}(0,1,0),
\end{align}
while those for X$_4$ at $\bm{Q}=\bm{Q}_{\rm X}$ are 
\begin{align}
	\bm{\epsilon}^1_{\text{X}^{(1)}_4,\bm{Q}_\text{X}} &=\frac{1}{\sqrt{2}}(0,1,0),\quad 
	\bm{\epsilon}^2_{\text{X}^{(1)}_4,\bm{Q}_\text{X}} =\frac{1}{\sqrt{2}}(0,0,-1),\\	
	\bm{\epsilon}^1_{\text{X}^{(2)}_4,\bm{Q}_\text{X}} &=\frac{1}{\sqrt{2}}(0,0,1),\quad 
	\bm{\epsilon}^2_{\text{X}^{(2)}_4,\bm{Q}_\text{X}} =\frac{1}{\sqrt{2}}(0,-1,0),\end{align}
as shown in Fig.~\ref{fig:diamond_image}(b).

We introduce corresponding coefficients in vector form as 
$\bm{\eta}_{\text{X}_\ell,\bm{Q}} = (\eta_{\text{X}^{(1)}_\ell,\bm{Q}},\eta_{\text{X}^{(2)}_\ell,\bm{Q}})$ ($\ell=3,4$). The modes at different X points, $\bm{\eta}_{\text{X}_\ell,\bm{Q}_\text{X}}$, $\bm{\eta}_{\text{X}_\ell,\bm{Q}_\text{Y}}$, and $\bm{\eta}_{\text{X}_\ell,\bm{Q}_\text{Z}}$, are related by the threefold rotation about the $[111]$ direction, $\hat{C}_3$, as $\bm{\eta}_{\text{X}_\ell,\bm{Q}_\text{Y}}= \hat{C}_3 \bm{\eta}_{\text{X}_\ell,\bm{Q}_\text{X}}$ and $\bm{\eta}_{\text{X}_\ell,\bm{Q}_\text{Z}} = \hat{C}_3^2 \bm{\eta}_{\text{X}_\ell,\bm{Q}_\text{X}}$.

The condensation of the X$_3$ mode leads to the P$2_1$3 (No.~198) structure, while that of the X$_4$ mode transforms the diamond lattice into the P$4_132$ (No.~213) or P$4_332$ (No.~212) structures, as shown in Fig.~\ref{fig:diamond_image}(c). For the P$2_1$3 structure, the two sublattices move in phase, whereas for the P$4_132$ and P$4_332$ structures they move out of phase. The P$4_132$ (P$4_332$) structure is realized in various materials such as SrSi$_2$~\cite{SrSi2} and $\beta$-Mn~\cite{beta-Mn}. The transition from the Fd$\bar{3}$m (No.~227) to the P$4_132$ (P$4_332$) structures has been analyzed within Landau theory~\cite{Talanov2016}, mainly focusing on the spinel compounds and the hyper-kagome structure. Thus, we focus on constructing $G_0$ in the diamond structure, i.e., the 8$a$ sites in the Fd$\bar{3}$m and 8$c$ sites in the P$4_132$ and P$4_332$ structures.

Similarly to the Z$_4$ mode of the rutile structure, each component of $\bm{\eta}_{\text{X}_\ell,\bm{Q}}$ is an eigenstate of the rotational and screw operations of the little group at the X point, while the two components are interchanged by mirror, glide, and inversion operations, which also means the characters for these improper operations are zero as shown in Table~\ref{tab:diamond} in Appendix~\ref{apdx:irrep}.
Therefore, the second-order contribution $G^{(2)}_0(\bm{\eta})$ can be constructed in a manner similar to that for the rutile structure as
\begin{align}
G^{(2)}_0(\bm{\eta}) \propto \sum_{\bm{Q} = \bm{Q}_{\text{X}},\bm{Q}_{\text{Y}},\bm{Q}_{\text{Z}}} \left[\left(\eta_{\text{X}^{(1)}_\ell,\bm{Q}}\right)^2 - \left(\eta_{\text{X}^{(2)}_\ell,\bm{Q}}\right)^2\right],
\end{align}
with $\ell=3,4$, which corresponds to the sum of $G_u$ over the three X points.

Furthermore, the relation $\bm{Q}_{\text{X}}+\bm{Q}_{\text{Y}} + \bm{Q}_{\text{Z}}=\bm{G}$ allows for a third-order contribution $G^{(3)}_0(\bm{\eta})$. However, due to the difference in transformation properties under inversion and horizontal mirror operations, a finite $G^{(3)}_0(\bm{\eta})$ is allowed only for the X$_4$ mode, and can be written as
\begin{align}
G^{(3)}_0(\bm{\eta}) \propto&\,
\eta_{\text{X}^{(1)}_4,\bm{Q}_\text{X}}\eta_{\text{X}^{(1)}_4,\bm{Q}_\text{Y}}\eta_{\text{X}^{(1)}_4,\bm{Q}_\text{Z}} \notag\\
&-
\eta_{\text{X}^{(2)}_4,\bm{Q}_\text{X}}\eta_{\text{X}^{(2)}_4,\bm{Q}_\text{Y}}\eta_{\text{X}^{(2)}_4,\bm{Q}_\text{Z}}.
\label{eq:G0_diamond}
\end{align}
We note that replacing the minus sign with a plus sign in Eq.~\eqref{eq:G0_diamond} yields a quantity invariant in the achiral phase, and therefore contributes to the free energy $\mathcal{F}$. Since this term is third order in $\eta_{\text{X}^{(m)}_4,\bm{Q}}$, it reflects the inequivalence between the structures with $\bm{\eta}$ and $-\bm{\eta}$. This can be understood by comparing the contracted bond at the lower-left corner in Fig.~\ref{fig:diamond_image}(c) with the elongated bond obtained by reversing the displacements shown in Fig.~\ref{fig:diamond_image}(c). These observations are consistent with the coexistence of $G_0^{(2)}$ and $G_0^{(3)}$ for the X$_4$ mode.

 Although one can construct analogous expressions for the X$_3$ modes, such quantities do not break all mirror and inversion symmetries of the system and therefore do not correspond to $G_0$. Instead, they correspond to the electric (E) octupole $Q_{xyz}$ and the ET octupole $G_{xyz}$, given by
\begin{align}
Q^{(3)}_{xyz}(\bm{\eta}) \propto&\,
\eta_{\text{X}^{(1)}_3,\bm{Q}_\text{X}}\eta_{\text{X}^{(1)}_3,\bm{Q}_\text{Y}}\eta_{\text{X}^{(1)}_3,\bm{Q}_\text{Z}} \notag\\
&+
\eta_{\text{X}^{(2)}_3,\bm{Q}_\text{X}}\eta_{\text{X}^{(2)}_3,\bm{Q}_\text{Y}}\eta_{\text{X}^{(2)}_3,\bm{Q}_\text{Z}},\\
G^{(3)}_{xyz}(\bm{\eta}) \propto&\,
\eta_{\text{X}^{(1)}_3,\bm{Q}_\text{X}}\eta_{\text{X}^{(1)}_3,\bm{Q}_\text{Y}}\eta_{\text{X}^{(1)}_3,\bm{Q}_\text{Z}} \notag\\
&-
\eta_{\text{X}^{(2)}_3,\bm{Q}_\text{X}}\eta_{\text{X}^{(2)}_3,\bm{Q}_\text{Y}}\eta_{\text{X}^{(2)}_3,\bm{Q}_\text{Z}},
\end{align}
which are invariant in the chiral phase P$2_1$3. This implies that the triple-$\bm{q}$ order of the X$_3$ modes induces not only $G_0$ but also $Q_{xyz}$ and $G_{xyz}$. In the point group O$_{\rm h}$, $Q_{xyz}$ and $G_{xyz}$ belong to the $A_{2\rm{u}}$ and $A_{2\rm{g}}$ irreps, respectively, and they are both odd under the fourfold screw operations. Since such operations disappear in the P$2_13$, $Q_{xyz}$ and $G_{xyz}$ emerge.

It is worth noting that the triple-$\bm{q}$ structure realized by the X$_3$ modes can be viewed as a superposition of RH and LH fcc sublattices. Nevertheless, a distinction between RH and LH remains, determined by which sublattice in the diamond structure corresponds to the RH or LH fcc configuration. 
This correspondence can be seen by comparing with the RH and LH fcc structures shown in Figs.~\ref{fig:fcc_image}(b) and \ref{fig:fcc_image}(c).

\subsection{\label{apdx:G0_6q}Third-order $G_{0}$ by the N point six-$\boldsymbol{q}$ order in the bcc lattice}
\begin{figure}
\includegraphics[width=0.9\linewidth]{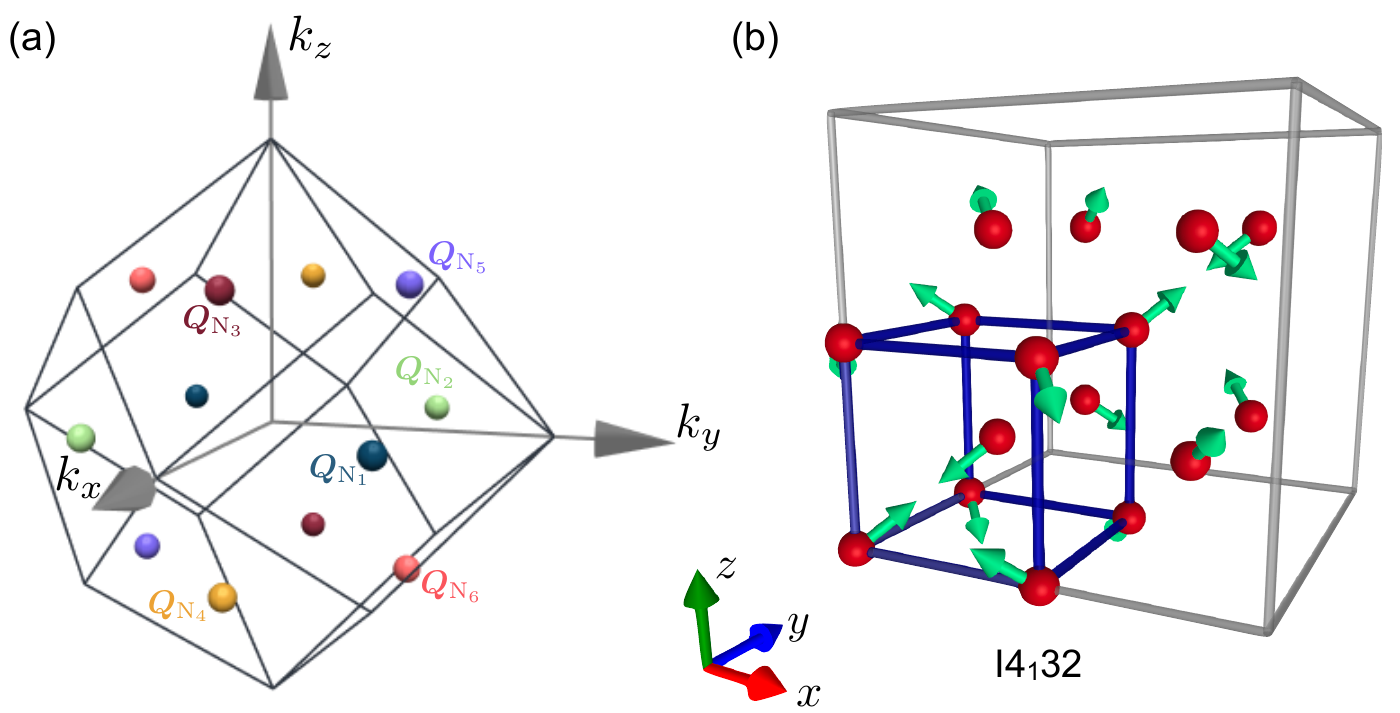}
\caption{\label{fig:6q}
(a) N points $\bm{Q}_{\text{N}_n}(n=1,2,\cdots,6)$ with $\bm{Q}_{\text{N}_n}\equiv -\bm{Q}_{\text{N}_n}$ in the first BZ of the bcc lattice. (b) A six-$\bm{q}$ configuration of $B_{1u}$ modes at the N points, which lowers the symmetry to I$4_132$. Spheres and arrows indicate the 2$a$ sites in the Im$\bar{3}$m and the directions of displacement, respectively. Bonds are drawn only for the shortest bonds.
}
\end{figure}

Let us show an example of $G^{(3)}_0$ for the transition from the bcc lattice (2$a$ site in Im$\bar{3}$m, No.~229) to I$4_132$ (No.~214), driven by a six-$\bm{q}$ order at the N point. The N points include six distinct wave vectors $\bm{p}=\bm{Q}_{\text{N}_n}(n=1,2,\cdots,6)$: 
$\bm{Q}_{\text{N}_1} = \left(\frac12,\frac12,0\right)$,
  $\bm{Q}_{\text{N}_2} = \left(-\frac12,\frac12,0\right)$, 
  $\bm{Q}_{\text{N}_3} = \left(\frac12,0,\frac12\right)$, 
  $\bm{Q}_{\text{N}_4} = \left(\frac12,0,-\frac12\right)$, 
  $\bm{Q}_{\text{N}_5} = \left(0,\frac12,\frac12\right)$, 
  $\bm{Q}_{\text{N}_6} = \left(0,-\frac12,\frac12\right)$
  in units of $2\pi/a$ shown in Fig.~\ref{fig:6q}(a). %The reciprocal lattice vectors $\bm{b}_{1,2,3}$ are given as $\bm{b}_1=2\bm{q}_5$, $\bm{b}_2=2\bm{q}_3$, and $\bm{b}_3=2\bm{q}_1$. 
  Since the $\bm{k}$ group at the N points is isomorphic to D$_{2\rm{h}}$, the eigenmodes can be classified by the D$_{2\rm{h}}$ group. Among them,  
  the longitudinal displacement at each N point belongs to the $B_{1u}$ irrep of D$_{2\rm{h}}$ with the eigenvector 
  $\boldsymbol{\epsilon}_{B_{1u},\bm{Q}_{\text{N}_n}}=
   \bm{Q}_{\text{N}_n}/|\bm{Q}_{\text{N}_n}|$ and the local displacement $\bm{d}(\bm{r})$ at the site $\bm{r}$ is given by
  \begin{align}
  	\bm{d}(\bm{r})=\sum_n \eta_{B_{1u},\bm{Q}_{\text{N}_n}}\boldsymbol{\epsilon}_{B_{1u},\bm{Q}_{\text{N}_n}}\cos( \bm{Q}_{\text{N}_n}\cdot \bm{r}).
  \end{align}
  The six-$\bm{q}$ order of $B_{1u}$ modes leads to an AtC transition. The primitive unit cell in the symmetry-broken I$4_132$ phase contains $2\times 2\times 2$ sites as shown in Fig.~\ref{fig:6q}(b). 
  
  The leading-order ET monopole $G_0$ in this case is $G_0^{(3)}$. Since the $B_{1u}$ mode is a real one-dimensional irrep and the $\bm{k}$ group at the N points is isomorphic to D$_{2\rm{h}}$, there is no $G^{(2)}$ according to the discussion (i) in Sec.~\ref{subsubsec:nonuniform_condition}. The form of the third-order ET monopole turns out to be nontrivial and distinct from those found in the fcc lattice discussed in Sec.~\ref{subsec:fcc} and the diamond structure in Appendix~\ref{app:diamond}. Since $B_{1u}$ is a one-dimensional irrep, there are no degrees of freedom at each $\bm{k}$ arm. This means that the chirality must be constructed by six $\bm{k}$ arms similarly to the $\bm{k}$-arm chirality discussed in Sec.~\ref{subsubsec:nonuniform_condition}. We note that only four sets of triple-$\bm{Q}_{\text{N}_n}$ can satisfy Umklapp conditions as $\bm{Q}_{\text{N}_1}+\bm{Q}_{\text{N}_3}+\bm{Q}_{\text{N}_6} = (1,0,1)\equiv {\bf 0}$, 
  $\bm{Q}_{\text{N}_1}+\bm{Q}_{\text{N}_4}+\bm{Q}_{\text{N}_5} = (1,1,0)\equiv {\bf 0}$, 
  $\bm{Q}_{\text{N}_2}+\bm{Q}_{\text{N}_3}+\bm{Q}_{\text{N}_5} = (0,1,1)\equiv {\bf 0}$, and  
  $\bm{Q}_{\text{N}_2}+\bm{Q}_{\text{N}_4}+\bm{Q}_{\text{N}_6} = {\bf 0}$.
To make $G^{(3)}_0$ be odd (even) under any improper (proper) operations in the space group Im$\bar{3}$m, we find the following linear combination as a unique $G^{(3)}_0$ up to an overall factor 
\begin{align}
	G^{(3)}_0= \frac{1}{4}\Big(\eta_1\eta_3\eta_6
  + \eta_1\eta_4\eta_5
  + \eta_2\eta_3\eta_5
  + \eta_2\eta_4\eta_6\Big),
\end{align}
where we have denoted $\eta_n\equiv \eta_{B_{1u},\bm{Q}_{\text{N}_n}}=\pm |\eta|$ for simplicity. There are $2^6=64$ configurations and they are classified into three types: $G^{(3)}_0=|\eta|^3$ (8 configurations), $G^{(3)}_0=-|\eta|^3$ (8 configurations), and $G^{(3)}_0=0$ (48 configurations). The last one is not our target since the resulting space group is I$\bar{4}\text{m}2$ (No.~119). $G^{(3)}_{0}=\pm |\eta|^3$ corresponds to the I$4_132$ phase with two chiralities. Parameterizing 
$	\kappa_1=\eta_1\eta_3\eta_6, \ 	\kappa_2=\eta_1\eta_4\eta_5, \ 
	\kappa_3=\eta_2\eta_3\eta_5, \ 	\kappa_4=\eta_2\eta_4\eta_6,	
$
we find $\kappa_1\kappa_2\kappa_3\kappa_4=(\eta_1\eta_2\cdots \eta_6)^2=|\eta|^{12}>0$. Thus, to obtain $G^{(3)}_0=\pm |\eta|^3$, $(\kappa_1,\kappa_2,\kappa_3,\kappa_4)=\kappa(1,1,1,1)$ is required, where $\kappa = \pm |\eta|^3$ represents the chirality. This can be achieved by choosing the signs of $\eta_1$, $\eta_2$, and $\eta_3$, i.e., $2^3=8$ choices. The remaining $\eta_4$, $\eta_5$, and $\eta_6$ are given as 
\begin{align}
	\frac{\eta_4}{|\eta|}=\text{sgn}(\eta_1\eta_2\eta_3),\ \frac{\eta_5}{|\eta|}=\text{sgn}(\kappa \eta_2\eta_3), \ \frac{\eta_6}{|\eta|}=\text{sgn}(\kappa\eta_1\eta_3),
\end{align}
which leads to $G^{(3)}_0=\kappa$. 

\subsection{\label{apdx:G0_sixth}Sixth-order $G_{0}$ at the W point of the fcc lattice}
\begin{figure}
\includegraphics[width=1.0\linewidth]{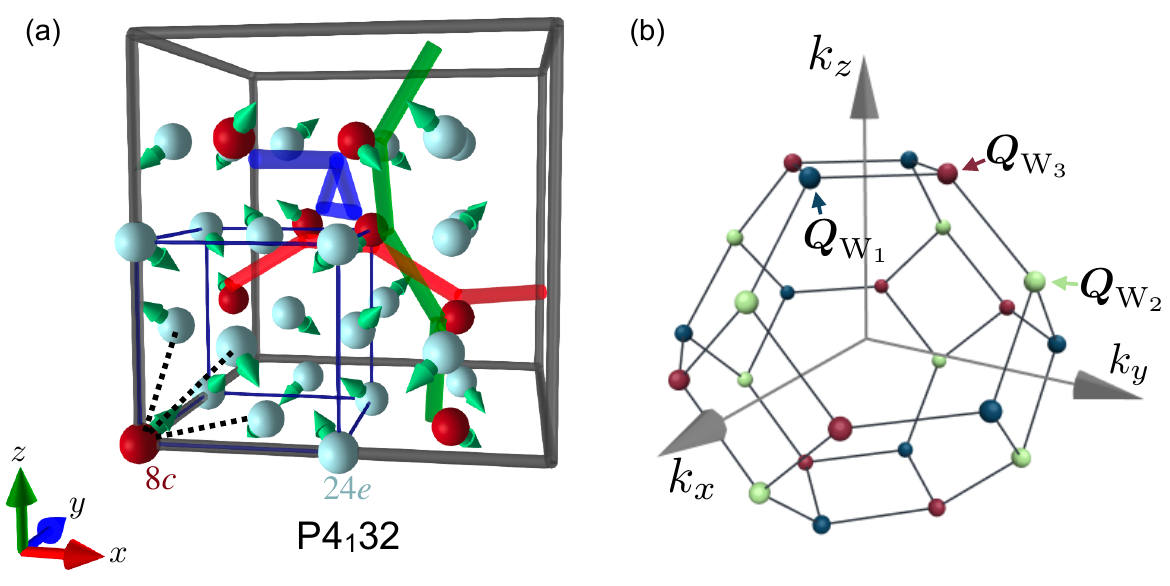}
\caption{\label{fig:fcc_W}
(a) Displacement pattern that transforms the fcc lattice into the chiral P$4_132$ structure, obtained from Eq.~\eqref{eq:Wpoint_disp}. The red and sky-blue spheres represent the Wyckoff $8c$ and $24e$ sites of the P$4_132$ phase, respectively. The red, blue, and green tubes represent the fourfold helices along the $x$, $y$, and $z$ axes consisting of the $8c$ sites, respectively. The nearest-neighbor bonds of the fcc lattice are indicated by the dotted lines.
(b) The three W points $\bm{Q}_{\text{W}_1}$ (blue), $\bm{Q}_{\text{W}_2}$ (green), and $\bm{Q}_{\text{W}_3}$ (red) in the first BZ of the fcc lattice. 
}
\end{figure}
Finally, we demonstrate that a sixth-order term of $G_{0,u}$ becomes the leading contribution when an AtC transition is driven by a one-dimensional irrep at the W point in the fcc lattice. We consider a transition from the simple fcc lattice to the chiral P$4_132$ (No.~213) or P$4_332$ (No.~212) structure. 
The fcc sites belong to the Wyckoff $4a$ site of the space group Fm$\bar{3}$m (No.~225), and they are split into $8c$ and $24e$ in the P$4_132$ (P$4_332$) phase, which are represented by the red and sky-blue spheres in Fig.~\ref{fig:fcc_W}(a). This transition occurs at the three W points of the first BZ of the fcc lattice [Fig.~\ref{fig:fcc_W}(b)], whose coordinates are given by $\bm{Q}_{\text{W}_1} = (1/2,1,0)$, $\bm{Q}_{\text{W}_2} = (1,0,1/2)$, and $\bm{Q}_{\text{W}_3} = (0,1/2,1)$ in units of $2\pi/a$. At the W points, the little group is isomorphic to the point group D$_{2\text{d}}$ and the one-dimensional irrep $B_2$ of D$_{2\text{d}}$ drives the transition. The eigenmodes $\bm{\epsilon}_{B_2,\bm{Q}}$ are given by
\begin{align}
&\bm{\epsilon}_{B_2,\bm{Q}_{\text{W}_1}} = (1,0,0),\\
&\bm{\epsilon}_{B_2,\bm{Q}_{\text{W}_2}} = (0,0,1),\\
&\bm{\epsilon}_{B_2,\bm{Q}_{\text{W}_3}} = (0,1,0).
\end{align}
With these eigenmodes, the displacement at the site $\bm{r}_i$ is given by
\begin{align}
\bm{d}(\bm{r}_i) =
\begin{pmatrix}
\eta_{B_2,\bm{Q}_{\text{W}_1}}e^{i\bm{Q}_{\text{W}_1}\cdot\bm{r}_i}+\eta^*_{B_2,\bm{Q}_{\text{W}_1}}e^{-i\bm{Q}_{\text{W}_1}\cdot\bm{r}_i}\\
\eta_{B_2,\bm{Q}_{\text{W}_3}}e^{i\bm{Q}_{\text{W}_3}\cdot\bm{r}_i}+\eta^*_{B_2,\bm{Q}_{\text{W}_3}}e^{-i\bm{Q}_{\text{W}_3}\cdot\bm{r}_i}\\
\eta_{B_2,\bm{Q}_{\text{W}_2}}e^{i\bm{Q}_{\text{W}_2}\cdot\bm{r}_i}+\eta^*_{B_2,\bm{Q}_{\text{W}_2}}e^{-i\bm{Q}_{\text{W}_2}\cdot\bm{r}_i}
\end{pmatrix},\label{eq:Wpoint_disp}
\end{align}
where $\eta_{B_2,\bm{Q}}$ is a complex number and satisfies the relation $\eta_{B_2,-\bm{Q}} = \eta_{B_2,\bm{Q}}^*$. The transition to the chiral P$4_132$ or P$4_332$ occurs when $\eta_{B_2,\bm{Q}}$ is chosen as $\eta_{B_2,\bm{Q}} = \eta e^{\pm i\frac{\pi}{4}}$, where $\eta$ is a real quantity representing the amplitude and the sign of the displacement. The phase part $e^{\pm i\frac{\pi}{4}}$ determines the space group of the resulting structure and $e^{-i\frac{\pi}{4}}$ ($e^{i\frac{\pi}{4}}$) leads to the chiral P$4_132$ (P$4_332$) structure. The displacement pattern for the P$4_132$ structure obtained from Eq.~\eqref{eq:Wpoint_disp}  is illustrated in Fig.~\ref{fig:fcc_W}(a). 

Now, let us construct the ET monopole $G_0$. Since $B_2$ is a one-dimensional irrep, the second-order term is prohibited as discussed in Sec.~\ref{subsubsec:nonuniform_condition}. Furthermore, the sum of the three wave vectors $\bm{Q}_{\text{W}_1}$, $\bm{Q}_{\text{W}_2}$, and $\bm{Q}_{\text{W}_3}$ is not equal to the reciprocal lattice vector $\bm{G}$. Instead, the relation $2\bm{Q}_{\text{W}_1} + 2\bm{Q}_{\text{W}_2} + 2\bm{Q}_{\text{W}_3} = \bm{G}$ holds, and thus $G_0$ must start at least at the sixth order to satisfy the Umklapp condition in Eq.~\eqref{eq:G0_n}. Keeping in mind that $G_0$ must be invariant under any rotation operation of the achiral phase, $G_0^{(6)}$ is given by
\begin{align}
G_0^{(6)} = \prod_{i=1,2,3}(\eta_{B_2,\bm{Q}_{\text{W}_i}}^2-\eta_{B_2,-\bm{Q}_{\text{W}_i}}^2).
\end{align}
Chirality-induced phenomena are governed by this sixth-order ET monopole, and thus, their temperature dependence within mean-field theory is $\propto |T_c-T|^{3}$.

\section{$G_{0,u}$ with multiple degrees of freedom}\label{app:BiTriG0u}
We discuss the form of $G_{0,u}$ with multiple degrees of freedom mentioned in Sec.~\ref{subsubsec:condition_uniform}. Table~\ref{tab:multi_copy_G0} lists nonzero bilinear and trilinear expressions for $G_{0,u}$ constructed from multiple structural order parameters $\boldsymbol{\eta}$, $\boldsymbol{\xi}$, and $\boldsymbol{\zeta}$. For arbitrary $\boldsymbol{\eta}$, $\boldsymbol{\xi}$, and $\boldsymbol{\zeta}$, there are seven bilinear and sixteen trilinear expressions. For example, the bilinear form $\eta_1\xi_2-\eta_2\xi_1$ appears for the $E_2$ irrep in the C$_{6{\rm v}}$ point group in Table~\ref{tab:multi_copy_G0}. This term is lower order than the single-copy contribution $3\eta_1^2\eta_2-\eta_2^3$ listed in Table~\ref{tab:G0uq_zero}. Which contribution becomes the leading term in $G_u$ depends on the directions of the order parameters $\boldsymbol{\eta}$ and $\boldsymbol{\xi}$. When $\boldsymbol{\eta}\parallel (0,1)$, the single-copy term $3\eta_1^2\eta_2-\eta_2^3$ is finite, whereas the bilinear term $\eta_1\xi_2-\eta_2\xi_1$ vanishes if $\boldsymbol{\eta}\parallel \boldsymbol{\xi}$. For general directions with both $\eta_1$ and $\eta_2$ finite and $\boldsymbol{\eta}\nparallel \boldsymbol{\xi}$, the bilinear term $\eta_1\xi_2-\eta_2\xi_1$ is nonzero and provides the leading contribution to $G_u$. More generally, when all components $\eta_1$, $\eta_2$, and $\eta_3$ are finite, several multi-copy contributions can appear at lower order than the corresponding single-copy entries in Table~\ref{tab:G0uq_zero}. When the symmetry of the ordered chiral phase $\mathcal H(\boldsymbol{\eta})$ is higher than that for a general $\boldsymbol{\eta}$, as listed in the seventh column of Table~\ref{tab:G0uq_zero}, the expressions listed in the sixth column of Table~\ref{tab:multi_copy_G0} appear at the same order as those in Table~\ref{tab:G0uq_zero}.
\begin{table}[t]
\centering
\scriptsize
\caption{
Bilinear $(n=2)$ and trilinear $(n=3)$ $G_{0,u}$ constructed from two or three independent copies of the same multidimensional irrep $\rho$ of an achiral crystallographic point group $\mathcal G$. For trilinear terms of two-dimensional irreps, we use
$\psi_\eta=\eta_1+i\eta_2$,
$\psi_\xi=\xi_1+i\xi_2$,
$\psi_\zeta=\zeta_1+i\zeta_2$,
$\mathcal R_3=\operatorname{Re}(\psi_\eta\psi_\xi\psi_\zeta)$,
and
$\mathcal I_3=\operatorname{Im}(\psi_\eta\psi_\xi\psi_\zeta)$. 
When $\eta=\xi=\zeta$, $\mathcal R_3\to\eta_1^3-3\eta_1\eta_2^2$ and $\mathcal I_3\to3\eta_1^2\eta_2-\eta_2^3$. 
For three-dimensional irreps, $\mathcal D=\det(\boldsymbol\eta,\boldsymbol\xi,\boldsymbol\zeta)$ and $\mathcal S$ denotes the fully symmetric mixed product
$\eta_1\xi_2\zeta_3 + \eta_3\xi_2\zeta_1 + \text{cyclic}$.
Only nonzero cases are listed. The sixth column indicates the $G_{0,u}$ for $\boldsymbol{\eta}\parallel \boldsymbol{\xi}~(\parallel 
\boldsymbol{\zeta})$, which is usually the case for second-order transitions. Note that the antisymmetric products $\eta_1\xi_2-\eta_2\xi_1$ and $\mathcal D$ vanish.
}
\label{tab:multi_copy_G0}
\resizebox{0.48\textwidth}{!}{%
\begin{tabular}{lllccc}
\hline
\hline
$n$ & $\mathcal G$ & $\rho$ & $G_{0,u}$ & $n$th order polynomials & 
$\boldsymbol{\eta}\parallel \boldsymbol{\xi}~(\parallel 
\boldsymbol{\zeta})$\\
\hline
2 &
C$_{3\rm{v}}$ &
$E$ &
$A_2$ &
$\eta_1\xi_2-\eta_2\xi_1$ & \\
2 &
S$_4$ &
$E$ &
$B$ &
$\eta_1\xi_1-\eta_2\xi_2,
\eta_1\xi_2+\eta_2\xi_1$ \hspace{2mm}\ \ & $\eta_1\xi_1-\eta_2\xi_2$\\
2 &
C$_{4\rm{v}}$ &
$E$ &
$A_2$ &
$\eta_1\xi_2-\eta_2\xi_1$ & \\
2 &
D$_{2\rm{d}}$ &
$E$ &
$B_1$ &
$\eta_1\xi_1-\eta_2\xi_2$ &  $\eta_1\xi_1-\eta_2\xi_2$\\
2 &
C$_{6\rm{v}}$ &
$E_1$ &
$A_2$ &
$\eta_1\xi_2-\eta_2\xi_1$ & \\
2 &
C$_{6\rm{v}}$ &
$E_2$ &
$A_2$ &
$\eta_1\xi_2-\eta_2\xi_1$ & \\
2 &
T$_{\rm{d}}$ &
$E$ &
$A_2$ &
$\eta_1\xi_2-\eta_2\xi_1$ & \\
\hline
3 &
C$_{3i}$ &
$E_{\rm u}$ &
$A_{\rm{u}}$ &
$\mathcal R_3, \mathcal I_3$ & $\mathcal R_3, \mathcal I_3$ \\
3 &
C$_{3\rm{h}}$ &
$E''$ &
$A''$ &
$\mathcal R_3, \mathcal I_3$ & $\mathcal R_3, \mathcal I_3$ \\
3 &
C$_{3\rm{v}}$ &
$E$ &
$A_2$ &
$\mathcal I_3$ & $\mathcal I_3$\\
3 &
D$_{3\rm{d}}$ &
$E_{\rm u}$ &
$A_{1\rm{u}}$ &
$\mathcal R_3$ & $\mathcal R_3$\\
3 &
D$_{3\rm{h}}$ &
$E''$ &
$A_1''$ &
$\mathcal I_3$ & $\mathcal I_3$\\
3 &
C$_{6\rm{h}}$ &
$E_{2\rm{u}}$ &
$A_{\rm{u}}$ &
$\mathcal R_3, \mathcal I_3$ & $\mathcal R_3, \mathcal I_3$\\
3 &
C$_{6\rm{v}}$ &
$E_2$ &
$A_2$ &
$\mathcal I_3$ & $\mathcal I_3$\\
3 &
D$_{6\rm{h}}$ &
$E_{2\rm{u}}$ &
$A_{1\rm{u}}$ &
$\mathcal I_3$ & $\mathcal I_3$\\
3 &
T$_{\rm{h}}$ &
$E_{\rm u}$ &
$A_{\rm{u}}$ &
$\mathcal R_3, \mathcal I_3$ & $\mathcal R_3, \mathcal I_3$\\
3 &
T$_{\rm{h}}$ &
$T_{\rm u}$ &
$A_{\rm{u}}$ &
$\mathcal D, \mathcal S$ & $\mathcal S$\\
3 &
T$_{\rm{d}}$ &
$E$ &
$A_2$ &
$\mathcal I_3$ & $\mathcal I_3$\\
3 &
T$_{\rm{d}}$ &
$T_1$ &
$A_2$ &
$\mathcal S$ & $\mathcal S$ \\
3 &
T$_{\rm{d}}$ &
$T_2$ &
$A_2$ &
$\mathcal D$ & \\
3 &
O$_{\rm{h}}$ &
$E_{\rm u}$ &
$A_{1\rm{u}}$ &
$\mathcal I_3$ & $\mathcal I_3$\\
3 &
O$_{\rm{h}}$ &
$T_{1\rm{u}}$ &
$A_{1\rm{u}}$ &
$\mathcal D$ & \\
3 &
O$_{\rm{h}}$ &
$T_{2\rm{u}}$ &
$A_{1\rm{u}}$ &
$\mathcal S$ & $\mathcal S$\\
\hline\hline
\end{tabular}%
}
\end{table}

\section{\label{apdx:karm}$\boldsymbol{k}$-arm chirality}
We list the candidate special $\bm{k}$ points and space groups that realize the $\bm{k}$-arm chirality discussed in Sec.~\ref{subsubsec:nonuniform_condition}, together with the resulting form of $G_u$ and the isotropy subgroup $\mathcal{H}(\boldsymbol{\eta})$. Tables~\ref{tab:karm_mono} and \ref{tab:karm_tetra} summarize the monoclinic/orthorhombic and tetragonal cases, respectively. We note that there are no candidates for systems with $C_3$ rotation at the special $\bm{k}$ points. At non-special $\bm{k}$ points, $\bm{k}$-arm chirality is generally allowed. For high-symmetry lines on a plane, one can also classify the possible form of $\bm{k}$-arm chirality, but the number of such candidates is too large, and thus, we do not discuss them in this paper.

\begin{table}[htbp]
\centering
\small
\caption{\label{tab:karm_mono}$\bm{k}$-arm chirality type quadratic pseudoscalar candidates for special $\bm{k}$ points (indicated by $\bm{p}$) in monoclinic and orthorhombic space group ${\mathscr G}$. The point group ${\mathcal G}$, $\bm{k}$ group ${\mathscr G}_{\bm p}$, the irrep(s), the form of $G_u$, and isotropy subgroup ${\mathcal H}(\boldsymbol{\eta})$ are listed. $\rho_\alpha=|\eta_\alpha|^2$ and $\rho_{\alpha\beta}=|\eta_\alpha|^2+|\eta_\beta|^2$ for a one-dimensional irrep, where $\alpha$ and $\beta$ are the indices of $\bm{k}$ arms. When there are four $\bm{k}$ arms, proper operations in ${\mathscr G}$ interchange $\rho_\alpha\leftrightarrow \rho_\beta$ and $\rho_\gamma\leftrightarrow \rho_\delta$ for Fmm2 at L and Fdd2 at L. For Cmc2$_1$ at $\mathrm{R}$, R$_{1,2}^*$  indicates that $\mathrm{R}_1$ and
$\mathrm{R}_2$ form a complex-conjugate pair, where
$\rho_\alpha$ denotes the total density of the physical real mode
on arm $\alpha$, i.e.,
$\rho_\alpha=\tfrac{1}{2}(|\eta_{\alpha,\mathrm R_1}|^2+
|\eta_{\alpha,\mathrm R_2}|^2)=|\eta_{\alpha,\mathrm R_1}|^2$.
$\sharp$: $\rho_\alpha=\rho_\beta\ne \rho_\gamma=\rho_\delta$, $\flat$: $\rho_\alpha\ne \rho_\beta$ or $\rho_\gamma\ne\rho_\delta$, $\dag$: $\rho_\alpha=0$ or $\rho_\beta=0$, $\ddag$: $\rho_\alpha $ and $\rho_\beta\ne \rho_\alpha$ are both finite.}
\begin{tabular}{lllllll}
\hline
\hline
${\mathscr G}$ & ${\mathcal G}$ & $\bm{p}$ & ${\mathscr G}_{\bm{p}}$ & irrep(s) & $G_u$ & $\mathcal{H}(\boldsymbol{\eta})$\\
\hline
(8) Cm & C$_{\rm s}$ & $\mathrm{L}$ & C$_1$ & $\mathrm{L}_1$ & $\rho_\alpha-\rho_\beta$ & (1) P1\\
 & C$_{\rm s}$ & $\mathrm{V}$ & C$_1$ & $\mathrm{V}_1$ & $\rho_\alpha-\rho_\beta$ & (1) P1\\
(9) Cc & C$_{\rm s}$ & $\mathrm{L}$ & C$_1$ & $\mathrm{L}_1$ & $\rho_\alpha-\rho_\beta$ & (1) P1\\
 & C$_{\rm s}$ & $\mathrm{V}$ & C$_1$ & $\mathrm{V}_1$ & $\rho_\alpha-\rho_\beta$ & (1) P1\\
(35) Cmm2 & C$_{2{\rm v}}$ & $\mathrm{R}$ & C$_2$ & $\mathrm{R}_{1,2}$ & $\rho_\alpha-\rho_\beta$ & (5) C2\\
    & C$_{2{\rm v}}$ & $\mathrm{S}$ & C$_2$ & $\mathrm{S}_{1,2}$ & $\rho_\alpha-\rho_\beta$ &(3) P2\\
(36) Cmc2$_1$ & C$_{2{\rm v}}$ & $\mathrm{R}$ & C$_2$ & $\mathrm{R}_{1,2}^*$& $\rho_\alpha-\rho_\beta$ &(1) P1\\
 & C$_{2{\rm v}}$ & $\mathrm{S}$ & C$_2$ & $\mathrm{S}_{1,2}$ & $\rho_\alpha-\rho_\beta$ &(4) P2$_1$\\
(37) Ccc2 & C$_{2{\rm v}}$ & $\mathrm{R}$ & C$_2$ & $\mathrm{R}_{1,2}$ & $\rho_\alpha-\rho_\beta$ &(5) C2\\
    & C$_{2{\rm v}}$ & $\mathrm{S}$ & C$_2$ & $\mathrm{S}_{1,2}$ & $\rho_\alpha-\rho_\beta$ &(3) P2\\
(42) Fmm2 & C$_{2{\rm v}}$ & $\mathrm{L}$ & C$_1$ & $\mathrm{L}_1$ & $\rho_{\alpha\beta}-\rho_{\gamma\delta}$ &(5) C2$^{\sharp}$, (1) P1$^{\flat}$\\
(43) Fdd2 & C$_{2{\rm v}}$ & $\mathrm{L}$ & C$_1$ & $\mathrm{L}_1$ & $\rho_{\alpha\beta}-\rho_{\gamma\delta}$ &(5) C2$^{\sharp}$, (1) P1$^{\flat}$\\
(44) Imm2 & C$_{2{\rm v}}$ & $\mathrm{T}$ & C$_2$ & $\mathrm{T}_{1,2}$ & $\rho_\alpha-\rho_\beta$ &(5) C2$^{\dag}$, (3) P2$^\ddag$\\
(45) Iba2 & C$_{2{\rm v}}$ & $\mathrm{T}$ & C$_2$ & $\mathrm{T}_{1,2}$ & $\rho_\alpha-\rho_\beta$ &(5) C2$^\dag$, (3) P2$^\ddag$\\
(46) Ima2 & C$_{2{\rm v}}$ & $\mathrm{T}$ & C$_2$ & $\mathrm{T}_{1,2}$ & $\rho_\alpha-\rho_\beta$ &(5) C2$^{\dag}$, (4) P2$_1$$^\ddag$\\
\hline
\hline
\end{tabular}
\end{table}

\begin{table*}[htbp]
\centering
\small
\caption{\label{tab:karm_tetra}$\bm{k}$-arm chirality type quadratic pseudoscalar candidates for special $\bm{k}$ points (indicated by $\bm{p}$) in tetragonal space group ${\mathscr G}$. The point group ${\mathcal G}$, $\bm{k}$ group ${\mathscr G}_{\bm p}$, the irrep(s), the form of $G_u$, and isotropy subgroup ${\mathcal H}(\boldsymbol{\eta})$ are listed.  $\rho_\alpha=|\eta_\alpha|^2$ and $\rho_{\alpha\beta}=|\eta_\alpha|^2+|\eta_\beta|^2$ for a one-dimensional irrep, where the $\alpha$ and $\beta$ are the indices of $\bm{k}$ arms. For P$\bar{4}2_1$m and P$\bar{4}2_1$c, the order parameters belong to two-dimensional irreps and are denoted by $w_\alpha=|\boldsymbol{\eta}_\alpha|^2$ with $\boldsymbol{\eta}_{\alpha(\beta)}=(\eta_{\alpha(\beta),1},\eta_{\alpha(\beta),2})$. When there are four $\bm{k}$ arms, proper operations in ${\mathscr G}$ interchange $\rho_\alpha\leftrightarrow \rho_\beta$ and $\rho_\gamma\leftrightarrow \rho_\delta$ for I$\bar{4}$ at N, I$\bar{4}2$m at N, and I$\bar{4}2$d at N. $\sharp$: $\rho_\alpha=\rho_\beta\ne \rho_\gamma=\rho_\delta$, $\flat$: $\rho_\alpha\ne \rho_\beta$ or $\rho_\gamma\ne\rho_\delta$, $\dag$: $\rho_\alpha=0$ or $\rho_\beta=0$, $\ddag$: $\rho_\alpha $ and $\rho_\beta\ne \rho_\alpha$ are both finite. $\S$: $|\eta_{\alpha,1}|^2=|\eta_{\alpha,2}|^2$ and $|\eta_{\beta,1}|^2=|\eta_{\beta,2}|^2$. $\P$: Only one of the four among $\eta_{\alpha,1}$, $\eta_{\alpha,2}$, $\eta_{\beta,1}$, and $\eta_{\beta,2}$ is finite. $\%$: $\eta_{\alpha,1}\ne \eta_{\alpha,2}$ or $\eta_{\beta,1}\ne \eta_{\beta,2}$.
$\spadesuit$: $|\eta_{\alpha}|^2=|\eta_{\beta}|^2$ and $\eta_{\gamma}=\eta_{\delta}=0$ or $(\alpha\beta\leftrightarrow\gamma\delta)$.
$\heartsuit$: there are three cases; $\heartsuit_1$: only one of the four among $\eta_{\alpha}$, $\eta_{\beta}$, $\eta_{\gamma}$, and $\eta_{\delta}$ is finite. $\heartsuit_2$: $|\eta_{\alpha}|^2=|\eta_{\beta}|^2\ne |\eta_{\gamma}|^2=|\eta_{\delta}|^2$. $\heartsuit_3$: $|\eta_{\alpha}|^2=|\eta_{\beta}|^2\ne |\eta_{\gamma}|^2$,$\eta_{\delta}=0$ or ($\alpha\beta\leftrightarrow \gamma\delta$ and $\gamma\leftrightarrow \delta$).
$\diamondsuit$: $|\eta_{\alpha}|^2\ne |\eta_{\beta}|^2$ and $\eta_{\gamma}=\eta_{\delta}=0$ or $(\alpha\beta\leftrightarrow\gamma\delta)$.
$\clubsuit$: general cases. For $\heartsuit_2$ in I$\bar{4}2m$, the order parameter sign difference leads to two possible $\mathcal H(\boldsymbol{\eta})$:   I2$_1$2$_1$2$_1$ and I222.
}
\begin{tabular}{lllllll}
\hline
\hline
${\mathscr G}$ & ${\mathcal G}$ & $\bm{p}$ & ${\mathscr G}_{\bm{p}}$ & irrep(s) & $G_u$ & $\mathcal{H}(\boldsymbol{\eta})$\\
\hline
(81) P$\bar{4}$ & S$_4$ & $\mathrm{R}$ & C$_2$ & $\mathrm{R}_{1,2}$ & $\rho_\alpha-\rho_\beta$ &(5) C2\\
   & S$_4$ & $\mathrm{X}$ & C$_2$ & $\mathrm{X}_{1,2}$ & $\rho_\alpha-\rho_\beta$ &(3) P2\\
(82) I$\bar{4}$ & S$_4$ & $\mathrm{N}$ & C$_1$ & $\mathrm{N}_1$ & $\rho_{\alpha\beta}-\rho_{\gamma\delta}$ &(5) C2$^{\sharp}$, (1) P1$^\flat$\\
 & S$_4$ & $\mathrm{X}$ & C$_2$ & $\mathrm{X}_{1,2}$ & $\rho_\alpha-\rho_\beta$ & (5) C2$^\dag$, (3) P2$^\ddag$\\
(111) P$\bar{4}$2m & D$_{2{\rm d}}$ & $\mathrm{R}$ & D$_2$ & $\mathrm{R}_{1,2,3,4}$ & $\rho_\alpha-\rho_\beta$ &(23) I222$^\dag$, (21) C222$^\ddag$\\
 & D$_{2{\rm d}}$ & $\mathrm{X}$ & D$_2$ & $\mathrm{X}_{1,4}$ & $\rho_\alpha-\rho_\beta$ &(16) P222\\
 & D$_{2{\rm d}}$ & $\mathrm{X}$ & D$_2$ & $\mathrm{X}_{2,3}$ & $\rho_\alpha-\rho_\beta$ &(17) P222$_1^\dag$, (18) P2$_1$2$_1$2$^\ddag$\\
(112) P$\bar{4}$2c & D$_{2{\rm d}}$ & $\mathrm{R}$ & D$_2$ & $\mathrm{R}_{1,2,3,4}$ & $\rho_\alpha-\rho_\beta$ &(21) C222$^\dag$, (24) I$2_12_12_1^\ddag$\\
  & D$_{2{\rm d}}$ & $\mathrm{X}$ & D$_2$ & $\mathrm{X}_{1,4}$ & $\rho_\alpha-\rho_\beta$ &(16) P222\\
  & D$_{2{\rm d}}$ & $\mathrm{X}$ & D$_2$ & $\mathrm{X}_{2,3}$ & $\rho_\alpha-\rho_\beta$ &(17) P22$2_1^\dag$, (18) P$2_12_12^\ddag$\\
(113) P$\bar{4}$2$_1$m  & D$_{2{\rm d}}$ & $\mathrm{R}$ & D$_2$ & $\mathrm{R}_1$ & $w_\alpha-w_\beta$ &(5) C2$^\S$, (4) P$2_1$$^\P$, (1) P1$^\%$\\
  & D$_{2{\rm d}}$ & $\mathrm{X}$ & D$_2$ & $\mathrm{X}_1$ & $w_\alpha-w_\beta$ &(3) P2$^\S$, (4) P$2_1$$^\P$, (1) P1$^\%$\\
(114) P$\bar{4}$2$_1$c & D$_{2{\rm d}}$ & $\mathrm{R}$ & D$_2$ & $\mathrm{R}_1$ & $w_\alpha-w_\beta$ &(5) C2$^\S$, (4) P$2_1^\P$, (1) P1$^\%$\\
 & D$_{2{\rm d}}$ & $\mathrm{X}$ & D$_2$ & $\mathrm{X}_1$ & $w_\alpha-w_\beta$ &(3) P2$^\S$, (4) P$2_1^\P$, (1) P1$^\%$\\
(119) I$\bar{4}$m2 & D$_{2{\rm d}}$ & $\mathrm{X}$ & D$_2$ & $\mathrm{X}_{1,3}$ & $\rho_\alpha-\rho_\beta$ &(21) C222$^\dag$, (16) P222$^\ddag$\\
  & D$_{2{\rm d}}$ & $\mathrm{X}$ & D$_2$ & $\mathrm{X}_{2,4}$ & $\rho_\alpha-\rho_\beta$ &(20) C222$_1^\dag$, (18) P2$_1$2$_1$2$^\ddag$\\
(120) I$\bar{4}$c2 & D$_{2{\rm d}}$ & $\mathrm{X}$ & D$_2$ & $\mathrm{X}_{1,3}$ & $\rho_\alpha-\rho_\beta$ &(21) C222$^\dag$, (16) P222$^\ddag$\\
  & D$_{2{\rm d}}$ & $\mathrm{X}$ & D$_2$ & $\mathrm{X}_{2,4}$ & $\rho_\alpha-\rho_\beta$ &(20) C222$_1^\dag$, (18) P2$_1$2$_1$2$^\ddag$\\
(121) I$\bar{4}$2m  & D$_{2{\rm d}}$ & $\mathrm{N}$ & C$_2$ & $\mathrm{N}_{1,2}$ & $\rho_{\alpha\beta}-\rho_{\gamma\delta}$ &(24) I2$_1$2$_1$2$_1$$^{\heartsuit_2}$, (23) I222$^{\heartsuit_2}$, (21) C222$^\spadesuit$, (5) C2$^{\heartsuit_1}$, (3) P2$^\diamondsuit$, (1) P1$^\clubsuit$\\
(122) I$\bar{4}$2d & D$_{2{\rm d}}$ & $\mathrm{N}$ & C$_2$ & $\mathrm{N}_{1,2}$ & $\rho_{\alpha\beta}-\rho_{\gamma\delta}$ &(20) C222$_1^\spadesuit$, (5) C2$^\heartsuit$, (4) P2$_1^\diamondsuit$, (1) P1$^\clubsuit$\\
\hline
\hline
\end{tabular}
\end{table*}

\section{\label{apdx:irrep}Irreducible representations of the rutile and the diamond structures}
\begin{table*}%[H] add [H] placement to break table across pages
\caption{\label{tab:rutile}Symmetry operations of the little group at the Z point of the rutile structure and the corresponding representation matrices of the Z$_4$ mode. $\lbrace \alpha|\bm{\tau}\rbrace$ represents the combination of the point group operation $\alpha$ at the sublattice $s=1$ and the translational operation within the unit cell $\bm{\tau} = (a/2,a/2,c/2)$ relating the two sublattices. $C_2'$ and $C^{''}_2$ are the twofold rotation about the $[110]$ and $[100]$ directions, respectively. }
\begin{ruledtabular}
\begin{tabular}{ccccccccccc}
& $\lbrace E|\bm{0}\rbrace$ & $\lbrace C_2|\bm{0}\rbrace$ & $\lbrace C'_{2}|\bm{0}\rbrace$ & $\lbrace I|\bm{0}\rbrace$ & $\lbrace \sigma_h|\bm{0}\rbrace$ & $\lbrace \sigma_d|\bm{0}\rbrace$ & $\lbrace C_4|\bm{\tau}\rbrace$ & $\lbrace S_4|\bm{\tau}\rbrace$ & $\lbrace C^{''}_2|\bm{\tau}\rbrace$ & $\lbrace \sigma_v|\bm{\tau}\rbrace$ \\
\noalign{\vskip 3pt}
\hline\noalign{\vskip 3pt}
  Z$_4$   
&  $\begin{pmatrix} 1&0\\ 0&1\end{pmatrix}$    
&   $\begin{pmatrix} -1&0\\ 0&-1\end{pmatrix}$   
&  $\begin{pmatrix} 1&0\\ 0&1\end{pmatrix}$    
&  $\begin{pmatrix} 0&-1\\ -1&0\end{pmatrix}$    
& $\begin{pmatrix} 0&1\\ 1&0\end{pmatrix}$     
& $\begin{pmatrix} 0&1\\ 1&0\end{pmatrix}$     
&$\begin{pmatrix} 1&0\\ 0&-1\end{pmatrix}$      
& $\begin{pmatrix} 0&-1\\ 1&0\end{pmatrix}$     
& $\begin{pmatrix} -1&0\\ 0&1\end{pmatrix}$      
& $\begin{pmatrix} 0&-1\\ 1&0\end{pmatrix}$      \\
\end{tabular}
\end{ruledtabular}
\end{table*}

\begin{table*}%[H] add [H] placement to break table across pages
\caption{\label{tab:diamond}Symmetry operations of the little group at the X point of the diamond structure and the corresponding representation matrices of the X$_3$ and X$_4$ modes. $\lbrace \alpha|\bm{\tau}\rbrace$ represents the combination of the point group operation $\alpha$ at the sublattice $s=1$ and the translational operation within the unit cell $\bm{\tau} = (a/4,a/4,a/4)$ relating the two sublattices. $C_2'$ and $C^{''}_2$ are the twofold rotations about the $[110]$ and $[100]$ directions, respectively. }
\begin{ruledtabular}
\begin{tabular}{ccccccccccc}
& $\lbrace E|\bm{0}\rbrace$ & $\lbrace C_2|\bm{0}\rbrace$ & $\lbrace C^{''}_{2}|\bm{0}\rbrace$ & $\lbrace S_4|\bm{0}\rbrace$ & $\lbrace \sigma_d|\bm{0}\rbrace$ & $\lbrace C_4|\bm{\tau}\rbrace$ & $\lbrace C_2'|\bm{\tau}\rbrace$ & $\lbrace I|\bm{\tau}\rbrace$ & $\lbrace \sigma_h|\bm{\tau}\rbrace$ & $\lbrace \sigma_v|\bm{\tau}\rbrace$ \\
\noalign{\vskip 3pt}
\hline\noalign{\vskip 3pt}
  X$_3$   
&  $\begin{pmatrix} 1&0\\ 0&1\end{pmatrix}$    
&   $\begin{pmatrix} -1&0\\ 0&-1\end{pmatrix}$   
&  $\begin{pmatrix} 1&0\\ 0&-1\end{pmatrix}$    
&  $\begin{pmatrix} 0&1\\ -1&0\end{pmatrix}$    
& $\begin{pmatrix} 0&1\\ 1&0\end{pmatrix}$     
& $\begin{pmatrix} 1&0\\ 0&-1\end{pmatrix}$     
&$\begin{pmatrix} 1&0\\ 0&1\end{pmatrix}$      
& $\begin{pmatrix} 0&-1\\ -1&0\end{pmatrix}$     
& $\begin{pmatrix} 0&1\\ 1&0\end{pmatrix}$      
& $\begin{pmatrix} 0&-1\\ 1&0\end{pmatrix}$      \\
\noalign{\vskip 3pt}
\hline\noalign{\vskip 3pt}
X$_4$   
&  $\begin{pmatrix} 1&0\\ 0&1\end{pmatrix}$    
&   $\begin{pmatrix} -1&0\\ 0&-1\end{pmatrix}$   
&  $\begin{pmatrix} 1&0\\ 0&-1\end{pmatrix}$    
&  $\begin{pmatrix} 0&1\\ -1&0\end{pmatrix}$    
& $\begin{pmatrix} 0&1\\ 1&0\end{pmatrix}$     
& $\begin{pmatrix} -1&0\\ 0&1\end{pmatrix}$     
&$\begin{pmatrix} -1&0\\ 0&-1\end{pmatrix}$      
& $\begin{pmatrix} 0&1\\ 1&0\end{pmatrix}$     
& $\begin{pmatrix} 0&-1\\ -1&0\end{pmatrix}$      
& $\begin{pmatrix} 0&1\\ -1&0\end{pmatrix}$ \\
\end{tabular}
\end{ruledtabular}
\end{table*}
We summarize the transformation properties of the Z$_4$ mode of the rutile structure and the X$_3$ and X$_4$ modes of the diamond structure discussed in Appendix~\ref{app:rutile} and \ref{app:diamond}. For further details not covered here, see Refs.~\cite{BradleyCracknell1972,gunron}.

Table~\ref{tab:rutile} presents the representation matrices of the Z$_4$ mode $(\eta_{1,\bm{Q}_\text{Z}}, \eta_{2,\bm{Q}_\text{Z}})$ for the symmetry operations of the little group at the Z point. Here, $\{\alpha|\bm{t}\}$ is the Seitz symbol and denotes a symmetry operation composed of a point-group operation $\alpha$ and a translation $\bm{t}$. In Table~\ref{tab:rutile}, there are four nonsymmorphic operations $\lbrace\alpha|\bm{\tau}\rbrace$ with $\bm{\tau} = (a/2,a/2,c/2)$ relating the two sublattices. The operations $C_2'$ and $C_2^{''}$ represent twofold rotations about the $[110]$ and $[100]$ directions, respectively. $\sigma_d$ and $\sigma_v$ correspond to the mirror operations with respect to the $(\bar{1}10)$ and $(010)$ planes, respectively. Table~\ref{tab:diamond} lists the symmetry operations for the X$_3$ and X$_4$ irreps at the X point of the diamond structure, where $\bm{\tau}=(a/4,a/4,a/4)$ and $C_2'$ ($C_2''$) denotes the twofold rotation about the [110] ([100]) axis.

From Tables~\ref{tab:rutile} and \ref{tab:diamond}, it is found that the representation matrices of the Z$_4$, X$_3$, and X$_4$ modes are diagonal for rotation and screw operations, whereas they are off-diagonal for mirror, inversion, glide, and rotoinversion operations for both systems. 

This indicates that the characters of all the improper operations for the Z$_4$, X$_3$, and X$_4$ modes are zero, and thus, these modes satisfy the condition for the finite second-order term in the ET monopole $G_0$ given in Sec.~\ref{subsec:conditions} even though the little group is isomorphic to D$_{4\text{h}}$, where the second-order term is prohibited for symmorphic systems. Explicit expressions for the second-order term are constructed in Appendices~\ref{app:rutile} and \ref{app:diamond}.

%The key difference between the X$_3$ and X$_4$ modes at the X point of the diamond structure lies in the transformation properties related to the nonsymmorphic operations including $\boldsymbol{\tau}$ as shown in Table~\ref{tab:diamond}. This is related to the difference in the expression of $G_{0}$ between the fcc and diamond systems. 

\section{\label{apdx:D2}$\boldsymbol{k}$ dependence of dynamical matrix in the D$_2$ structure}
\begin{figure*}[t]
\includegraphics[width=1.0\linewidth]{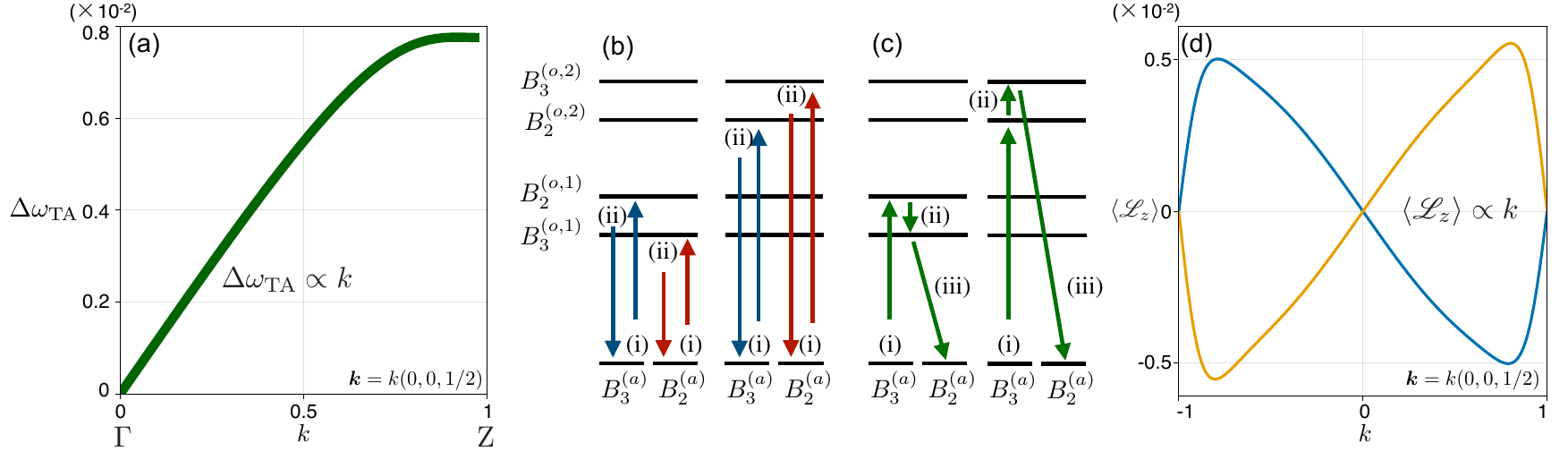}
\caption{\label{fig:ortho_split} 
(a) The $\bm{k}$ dependence of the splitting of the TA modes in the D$_2$ phase along the $\Gamma$-Z path. Schematic illustrations of (b) the second-order and (c) the third-order perturbation processes, where (i), (ii), and (iii) correspond to the sequence of perturbation processes. (d) The $k$ dependence of the angular momentum of the TA modes, where the blue and yellow lines represent the higher- and lower- energy branches, respectively.}
\end{figure*}

In order to clarify the origin of phonon band splittings for the AtC transition from the D$_{\rm 2d}$ to the D$_2$ structure, we examine the matrix elements of the dynamical matrix $\mathcal{D}(\bm{k})$ at infinitesimal $\bm{k} = (0,0,k)$ ($|k|\ll 1$) in the chiral phase with fixed $\eta = 0.05$. As briefly discussed in Sec.~\ref{sec:mechanism}, the sound velocities of the two chiral phonons coincide in systems with high symmetry. In such systems, the splitting between the two chiral-phonon dispersions is known to be proportional to $|\bm{k}|^3$~\cite{tsunetsugu_kusunose}. In the D$_2$ point group, however, the modes propagating along $\bm{k}=(0,0,k)$ with displacements in the $x$ and $y$ directions are inequivalent. This leads to a nonchiral splitting in their phonon velocities.

We expand $\mathcal{D}(\bm{k})$ with respect to $k$ up to the second
order and analyze the perturbation processes.
For clarity, we use the eigenstates at the $\Gamma$ point as the
basis for finite $k$.
The twelve modes at the $\Gamma$ point split into
$3A \oplus 3B_1 \oplus 3B_2 \oplus 3B_3$ in the D$_2$ symmetry.
Among them, one each of $B_1$, $B_2$, and $B_3$ corresponds to the
acoustic modes, where $B_1$ is the longitudinal and $B_2$ and $B_3$
are the transverse modes.
The remaining irreps constitute the optical modes.
In the following, we denote $B_\gamma$ ($\gamma = 1,2,3$) for the
acoustic modes as $B^{(a)}_{\gamma}$ and those for the optical modes
as $B^{(o,n)}_{\gamma}$, where $n = 1, 2$ is the index for
distinguishing the optical modes of the same irrep.

The phonon dispersion of the D$_2$ structure with $\eta=0.05$ along $\bm{k}=k(0,0,1/2)$ is already given in Fig.~\ref{fig:CPS_D2d}(b) and we have evaluated the $k$ dependence of the splitting of the TA branches $\Delta\omega_{\text{TA}}(\bm{k}) = |\omega_{\bm{k},2}-\omega_{\bm{k},1}|$ from the dispersion. The result is shown in Fig.~\ref{fig:ortho_split}(a). At the $\Gamma$ point, the TA branches are degenerate, whereas, at a finite $\bm{k}$, mixing of the two branches  gives rise to a finite splitting, where $\Delta\omega_{\text{TA}}\propto k$ for $k\sim0$. This indicates that the sound velocities of the TA modes are different reflecting the D$_2$ symmetry.

%The phonon dispersion along $\bm{k}=(0,0,k)$ is shown in Fig.~\ref{fig:ortho_split}(a), where the superscript $a$ and $o$ on the name of the irreps represent the acoustic and the optical modes, respectively. The number on the superscript denotes 
%the index for distinguishing the same irrep of the optical modes. At the $\Gamma$ point, the acoustic modes are degenerate, whereas, at a finite $\bm{k}$, mixing of the two branches  gives rise to a finite splitting as indicated in the inset of Fig.~\ref{fig:ortho_split}(a). Reflecting the D$_2$ symmetry, the two sound velocities are different.  

To analyze the origin of the sound-velocity difference and the properties of chiral phonons, we expand the dynamical matrix as
\begin{align}
\mathcal{D}(\bm{k}) \sim \mathcal{D}^{(0)} + \mathcal{D}^{(1)}k + \frac{1}{2}\mathcal{D}^{(2)}k^2,
\end{align}
where $\mathcal{D}^{(n)} \equiv \partial^n\mathcal{D}(\bm{k})/\partial k^n\big|_{\bm{k}=0}$. As noted above, $\mathcal{D}^{(0)}$ is diagonal in our basis. In general, $\mathcal{D}^{(1)}$ and $\mathcal{D}^{(2)}$ have off-diagonal elements that can induce splittings. However, due to symmetry constraints, only specific off-diagonal elements are allowed. For $\mathcal{D}^{(1)}$, the TA modes $B_{2(3)}^{(a)}$ can mix only with the optical modes $B_{2(3)}^{(o,n)}$, since $k$ ($=k_z$) belongs to the $B_1$ representation and the direct product $B_1\otimes B_2 \otimes B_3$ is invariant in the D$_2$ point group. Therefore, the matrix elements 
\begin{align}
&\mel{B_2^{(o,n)}}{\mathcal{D}^{(1)}k}{B_3^{(a)}},\label{eq:B2oDB3a}\\
&\mel{B_3^{(o,n)}}{\mathcal{D}^{(1)}k}{B_2^{(a)}},\label{eq:B3oDB2a}\\
&\mel{B_3^{(o,n)}}{\mathcal{D}^{(1)}k}{B_2^{(o,m)}},
\end{align}
with $m,n=1$ or $2$ are finite. Here, $|B_{2,3}^{(a),(o,n)}\rangle$ denotes the eigenvector at the $\Gamma$ point. 

The sound-velocity difference between the TA modes arises from second-order perturbation processes involving Eqs.~(\ref{eq:B2oDB3a}) and (\ref{eq:B3oDB2a}), as schematically shown in Fig.~\ref{fig:ortho_split}(b). Since the TA modes $B^{(a)}_2$ and $B^{(a)}_3$ transform as $u_{y,\bm{k}}$ and $u_{x,\bm{k}}$, respectively,  we consider Eq.~\eqref{eq:ortho_dyna} with $\lambda^{(1)}_{\bm{k}}$ and $\lambda^{\prime (1)}_{\bm{k}}$. Here, the superscript ``(1)'' corresponds to the couplings in Eqs.~(\ref{eq:gu1}) and (\ref{eq:gu_prime1}). In addition to the difference in the matrix elements (\ref{eq:B2oDB3a}) and (\ref{eq:B3oDB2a}), the optical-mode energies differ between the $B_2^{(o,n)}$ and $B_3^{(o,n)}$ ($n=1,2$) modes. These contributions lead to a finite $\lambda^{\prime(1)}_{\bm{k}}\propto k^2$ in Eq.~(\ref{eq:ortho_dyna}), resulting in inequivalence between the $x$ and $y$ directions, i.e., a diagonal coupling between the two TA modes described by $g_u^{\prime(1)}(\bm{k})$ in Eq.~(\ref{eq:gu_prime1}).

In contrast, the off-diagonal coupling $g_u^{(1)}(\bm{k})$ in Eq.~(\ref{eq:gu1}) appears at order $k^3$ via third- or higher-order processes involving transitions between optical modes:
\begin{align}
\sim \mel{B_2^{(a)}}{\mathcal{D}^{(1)}}{B_3^{(o,m)}}&\mel{B_3^{(o,m)}}{\mathcal{D}^{(1)}}{B_2^{(o,n)}}\notag\\
&\times\mel{B_2^{(o,n)}}{\mathcal{D}^{(1)}}{B_3^{(a)}}
\end{align}
as schematically illustrated in Fig.~\ref{fig:ortho_split}(c). Since $\mathcal{D}^{(1)}$ is purely imaginary, this contribution to $\lambda_{\bm{k}}^{(1)}$ in Eq.~(\ref{eq:ortho_dyna}) is proportional to $k^3$.
 
 %\red{\sout{since $B_2^{(o,n)}$ and $B_3^{(o,n)}$ are not degenerate and second-order processes such as $\mel{B_2^{(a)}}{\mathcal{D}^{(1)}}{B_3^{(o,n)}}\mel{B_2^{(o,n)}}{\mathcal{D}^{(1)}}{B_3^{(a)}}$ are prohibited.}} \sout{Group theoretically, the matrix elements of $\mathcal{D}^{(1)}$ can be complex as long as it remains Hermitian since all the irreps. are one-dimensional. Then, the imaginary part corresponds to $\lambda^{(1)}_{\bm{k}}$. For the present model, $\bm{k}$ dependence is incorporated through the phase factor $e^{i\bm{k}\cdot\bm{r}_i}$ arises from the Fourier transformation of Eq.~\eqref{eq:Hami_phonon}. Therefore, the matrix elements of $\mathcal{D}^{(1)}$ become pure imaginary. }
 
For contributions involving $\mathcal{D}^{(2)}$, a diagonal coupling becomes finite because $k^2$ belongs to the $A$ irrep, leading to finite matrix elements
\begin{align}
\mel{B_2^{(a)}}{\mathcal{D}^{(2)}k^2}{B_2^{(a)}}\ne 
\mel{B_3^{(a)}}{\mathcal{D}^{(2)}k^2}{B_3^{(a)}},	
\end{align}
where the inequality arises from the inequivalence between the $x$ and $y$ directions in the D$_2$ point group. This $k^2$ term contributes to $\lambda^{\prime (1)}_{\bm{k}}$. Similarly, acoustic-optical couplings within the same irreps become finite for the same reason. Second-order perturbation processes involving both $\mathcal{D}^{(1)}$ and $\mathcal{D}^{(2)}$, such as $\mel{B_2^{(a)}}{\mathcal{D}^{(1)}k}{B_3^{(o,n)}}\mel{B_3^{(o,n)}}{\mathcal{D}^{(2)}k^2}{B_3^{(a)}}$, lead to a finite off-diagonal coupling between the TA modes proportional to $k^3$, which contributes to $\lambda_{\bm{k}}^{(1)}$ in Eq.~(\ref{eq:ortho_dyna}).

Since the phonon dispersion is given by the square root of the eigenvalues of the dynamical matrix, the splitting of the TA modes is proportional to $k$ for $|k|\ll 1$ due to $\lambda^{\prime(1)}_{\bm{k}}\propto k^2$. It should be emphasized that this originates from the inequivalence of the $x$ and $y$ axes, whereas the splitting associated with the emergence of chiral phonons is encoded in $\lambda^{(1)}_{\bm{k}}$ and starts from order $k^3$, as discussed in previous studies~\cite{tsunetsugu_kusunose,kusunose_cubic}.

So far, we have obtained $\lambda^{(1)}_{\bm{k}}\propto k^3$ and 
$\lambda^{\prime (1)}_{\bm{k}}\propto k^2$ for 
$|k|\ll 1$, which leads to 
${\lambda}^{\prime (1)}_{\bm{k}}/{\lambda}^{(1)}_{\bm{k}}\simeq  A/k$ with $A$ being a constant. Substituting this into Eq.~(\ref{eq:tan}), one finds, for $0<k\ll 1$,
\begin{align}
\tan\frac{\theta_{\pm,\bm{k}}}{2}
&\simeq \begin{cases}
	\frac{1}{2}A^{-1}k\\[2mm]
	2Ak^{-1}
\end{cases}.
\end{align}
Thus, $\theta_{+,\bm{k}}\simeq k/A$ and $\theta_{-,\bm{k}}\simeq \pi-k/A$. From Eq.~\eqref{eq:ortho_Lz}, one finds $L_z^{\pm}=\pm\sin\theta_{\pm,\bm{k}}\simeq \pm k/A$. To confirm this behavior, we numerically evaluate the expectation values of the phonon angular momentum [Eq.~(\ref{eq:phonon_angular_def})], $\ev{\mathscr{L}_z}$, for the two TA modes as a function of $k$. The result is shown in Fig.~\ref{fig:ortho_split}(d), where the blue and yellow lines represent $\ev{\mathscr{L}_z}$ for the $B_2^{(a)}$ and $B_3^{(a)}$ modes, respectively. For $|k|\ll 1$, the linear relation $\ev{\mathscr{L}_z}\propto k$ is observed, consistent with the above analysis.

\begin{acknowledgments}
The authors would like to thank Hiroaki Kusunose, Takayuki Ishitobi, and Takuya Nomoto for fruitful discussions. This work is supported by a Grant-in-Aid for Transformative Research Areas (A) ``Asymmetric Quantum Matters'', JSPS KAKENHI  (Grant No.~JP23H04866 and No.~JP23H04869). 
\end{acknowledgments}

\bibliography{reference}

@book{Kelvin,
	author = {W. Thomson},
	publisher = {Cambridge University Press},
	title = {Baltimore Lectures on Molecular Dynamics and the Wave Theory of Light},
	year = {1904}}

@article{WATSON1953,
  author  = {Watson, J. D. and Crick, F. H. C.},
  title   = {{Molecular Structure of Nucleic Acids: A Structure for Deoxyribose Nucleic Acid}},
  journal = {Nature},
  year    = {1953},
  volume  = {171},
  number  = {4356},
  pages   = {737--738},
  doi     = {10.1038/171737a0},
  url     = {https://doi.org/10.1038/171737a0}
}

@article{Huggins_quartz,
  title = {{The Crystal Structure of Quartz}},
  author = {Huggins, Maurice L.},
  journal = {Phys. Rev.},
  volume = {19},
  issue = {4},
  pages = {363--368},
  numpages = {0},
  year = {1922},
  month = {Apr},
  publisher = {American Physical Society},
  doi = {10.1103/PhysRev.19.363},
  url = {https://link.aps.org/doi/10.1103/PhysRev.19.363}
}

@article{Bragg_quartz,
    author = {Bragg, William Lawrence and Gibbs, Reginald Edmund},
    title = {{The structure of $\alpha$ and $\beta$ quartz}},
    journal = {Proc. R. Soc. Lond. A},
    volume = {109},
    number = {751},
    pages = {405-427},
    year = {1925},
    month = {11},
    issn = {0950-1207},
    doi = {10.1098/rspa.1925.0135},
    url = {https://doi.org/10.1098/rspa.1925.0135},
    }

@article{Bradley,
author = {A. J. Bradley},
title = {{The crystal structures of the rhombohedral forms of selenium and tellurium}},
journal = {Philos. Mag.},
volume = {48},
number = {285},
pages = {477--496},
year = {1924},
publisher = {Taylor \& Francis},
doi = {10.1080/14786442408634511},
URL = {https://doi.org/10.1080/14786442408634511}}

@article{Pasteur1848,
  author  = {Pasteur, Louis},
  title   = {{M{\'e}moire sur la relation qui peut exister entre la forme cristalline et la composition chimique, et sur la cause de la polarisation rotatoire}},
  journal = {C. R. Acad. Sci. Paris},
  volume  = {26},
  pages   = {535--538},
  year    = {1848}
}

@article{Pasteur1848_full,
  author  = {Pasteur, Louis},
  title   = {{M{\'e}moire sur les relations qui peuvent exister entre la forme cristalline et la composition chimique, et sur la cause de la polarisation rotatoire}},
  journal = {Ann. Chim. Phys.},
  series  = {3},
  volume  = {24},
  number  = {6},
  pages   = {442--459},
  year    = {1848}
}

@article{Kusunose_2022,
doi = {10.1088/1361-648X/ac9209},
url = {https://doi.org/10.1088/1361-648X/ac9209},
year = {2022},
month = {sep},
publisher = {IOP Publishing},
volume = {34},
number = {46},
pages = {464002},
author = {Kusunose, Hiroaki and Hayami, Satoru},
title = {{Generalization of microscopic multipoles and cross-correlated phenomena by their orderings}},
journal = {J. Phys.: Condens. Matter}
}

@article{Hayami2018PRB,
  title = {Classification of atomic-scale multipoles under crystallographic point groups and application to linear response tensors},
  author = {Hayami, Satoru and Yatsushiro, Megumi and Yanagi, Yuki and Kusunose, Hiroaki},
  journal = {Phys. Rev. B},
  volume = {98},
  issue = {16},
  pages = {165110},
  numpages = {35},
  year = {2018},
  month = {Oct},
  publisher = {American Physical Society},
  doi = {10.1103/PhysRevB.98.165110},
  url = {https://link.aps.org/doi/10.1103/PhysRevB.98.165110}
}

@article{CIM_nature,
  author  = {Furukawa, Tetsuya and Shimokawa, Yuri and Kobayashi, Kaya and Itou, Tetsuaki},
  title   = {{Observation of current-induced bulk magnetization in elemental tellurium}},
  journal = {Nat. Commun.},
  year    = {2017},
  volume  = {8},
  number  = {1},
  pages   = {954},
  doi     = {10.1038/s41467-017-01093-3},
  url     = {https://doi.org/10.1038/s41467-017-01093-3}
}

@article{CIM,
	author = {Furukawa, Tetsuya and Watanabe, Yuta and Ogasawara, Naoki and Kobayashi, Kaya and Itou, Tetsuaki},
	doi = {10.1103/PhysRevResearch.3.023111},
	issue = {2},
	journal = {Phys. Rev. Res.},
	month = {May},
	numpages = {14},
	pages = {023111},
	publisher = {American Physical Society},
	title = {{Current-induced magnetization caused by crystal chirality in nonmagnetic elemental tellurium}},
	url = {https://link.aps.org/doi/10.1103/PhysRevResearch.3.023111},
	volume = {3},
	year = {2021}}

@article{oiwa_rot,
  title = {{Rotation, Electric-Field Responses, and Absolute Enantioselection in Chiral Crystals}},
  author = {Oiwa, Rikuto and Kusunose, Hiroaki},
  journal = {Phys. Rev. Lett.},
  volume = {129},
  issue = {11},
  pages = {116401},
  numpages = {6},
  year = {2022},
  month = {Sep},
  publisher = {American Physical Society},
  doi = {10.1103/PhysRevLett.129.116401},
  url = {https://link.aps.org/doi/10.1103/PhysRevLett.129.116401}
}

@article{Dieny2020,
  author = {Dieny, B. and Prejbeanu, I. L. and Garello, K. and Gambardella, P. and Freitas, P. and Lehndorff, R. and Raberg, W. and Ebels, U. and Demokritov, S. O. and Akerman, J. and Deac, A. and Pirro, P. and Adelmann, C. and Anane, A. and Chumak, A. V. and Hirohata, A. and Mangin, S. and Valenzuela, Sergio O. and Onba\c{s}l{\i}, M. Cengiz and d'Aquino, M. and Prenat, G. and Finocchio, G. and Lopez-Diaz, L. and Chantrell, R. and Chubykalo-Fesenko, O. and Bortolotti, P.},
  title = {{Opportunities and challenges for spintronics in the microelectronics industry}},
  journal = {Nat. Electron.},
  year = {2020},
  volume = {3},
  number = {8},
  pages = {446--459},
  doi = {10.1038/s41928-020-0461-5},
  url = {https://doi.org/10.1038/s41928-020-0461-5}
}

@article{Fiebig2016,
  author = {Fiebig, Manfred and Lottermoser, Thomas and Meier, Dennis and Trassin, Morgan},
  title = {{The evolution of multiferroics}},
  journal = {Nat. Rev. Mater.},
  year = {2016},
  volume = {1},
  number = {8},
  pages = {16046},
  doi = {10.1038/natrevmats.2016.46},
  url = {https://doi.org/10.1038/natrevmats.2016.46}
}

@article{Zhang_EdH,
  title = {Angular Momentum of Phonons and the Einstein--de Haas Effect},
  author = {Zhang, Lifa and Niu, Qian},
  journal = {Phys. Rev. Lett.},
  volume = {112},
  issue = {8},
  pages = {085503},
  numpages = {5},
  year = {2014},
  month = {Feb},
  publisher = {American Physical Society},
  doi = {10.1103/PhysRevLett.112.085503},
  url = {https://link.aps.org/doi/10.1103/PhysRevLett.112.085503}
}

@article{tsunetsugu_kusunose,
author = {Tsunetsugu ,Hirokazu and Kusunose ,Hiroaki},
title = {{Theory of Energy Dispersion of Chiral Phonons}},
journal = {J. Phys. Soc. Jpn},
volume = {92},
number = {2},
pages = {023601},
year = {2023},
doi = {10.7566/JPSJ.92.023601},
URL = { https://doi.org/10.7566/JPSJ.92.023601}
}

@article{kusunose_cubic,
author = {Tsunetsugu ,Hirokazu and Kusunose ,Hiroaki},
title = {{Chiral Phonons in a Cubic Lattice}},
journal = {J. Phys. Soc. Jpn},
volume = {95},
number = {1},
pages = {013601},
year = {2026},
doi = {10.7566/JPSJ.95.013601},
URL = { https://doi.org/10.7566/JPSJ.95.013601}
}

@article{Hamada,
	author = {Hamada, Masato and Minamitani, Emi and Hirayama, Motoaki and Murakami, Shuichi},
	doi = {10.1103/PhysRevLett.121.175301},
	issue = {17},
	journal = {Phys. Rev. Lett.},
	month = {Oct},
	numpages = {5},
	pages = {175301},
	publisher = {American Physical Society},
	title = {{Phonon Angular Momentum Induced by the Temperature Gradient}},
	url = {https://link.aps.org/doi/10.1103/PhysRevLett.121.175301},
	volume = {121},
	year = {2018}}

@article{Zhu_ChiralPhonon,
	author = {Hanyu Zhu and Jun Yi and Ming-Yang Li and Jun Xiao and Lifa Zhang and Chih-Wen Yang and Robert A. Kaindl and Lain-Jong Li and Yuan Wang and Xiang Zhang},
	doi = {10.1126/science.aar2711},
	journal = {Science},
	number = {6375},
	pages = {579-582},
	title = {{Observation of chiral phonons}},
	url = {https://www.science.org/doi/abs/10.1126/science.aar2711},
	volume = {359},
	year = {2018}}

@article{Ishito,
	author = {Ishito, Kyosuke and Mao, Huiling and Kousaka, Yusuke and Togawa, Yoshihiko and Iwasaki, Satoshi and Zhang, Tiantian and Murakami, Shuichi and Kishine, Jun-	ichiro and Satoh, Takuya},
	date = {2023/01/01},
	doi = {10.1038/s41567-022-01790-x},
	id = {Ishito2023},
	isbn = {1745-2481},
	journal = {Nat. Phys.},
	number = {1},
	pages = {35--39},
	title = {{Truly chiral phonons in $\alpha$-HgS}},
	url = {https://doi.org/10.1038/s41567-022-01790-x},
	volume = {19},
	year = {2023}}

@article{Ueda2023,
  author  = {Hiroki Ueda and Mirian Garc\'{\i}a-Fern\'andez and Stefano Agrestini and Carl P. Romao and Jeroen van den Brink and Nicola A. Spaldin and Ke-Jin Zhou and Urs Staub},
  title   = {Chiral phonons in quartz probed by X-rays},
  journal = {Nature},
  year    = {2023},
  volume  = {618},
  number  = {7967},
  pages   = {946--950},
  doi     = {10.1038/s41586-023-06016-5},
  url     = {https://doi.org/10.1038/s41586-023-06016-5}
}

@article{murakami_chiral_weyl,
author = {Zhang, Tiantian and Huang, Zhiheng and Pan, Zitian and Du, Luojun and Zhang, Guangyu and Murakami, Shuichi},
title = {{Weyl Phonons in Chiral Crystals}},
journal = {Nano Lett.},
volume = {23},
number = {16},
pages = {7561-7567},
year = {2023},
doi = {10.1021/acs.nanolett.3c02132},
URL = {https://doi.org/10.1021/acs.nanolett.3c02132
}}

@article{Bousquet_2025_review,
doi = {10.1088/1361-648X/adb674},
url = {https://doi.org/10.1088/1361-648X/adb674},
year = {2025},
month = {mar},
publisher = {IOP Publishing},
volume = {37},
number = {16},
pages = {163004},
author = {Bousquet, Eric and Fava, Mauro and Romestan, Zachary and G{\'o}mez-Ortiz, Fernando and McCabe, Emma E and Romero, Aldo H},
title = {{Structural chirality and related properties in periodic inorganic solids: review and perspectives}},
journal = {J. Phys.: Condens. Matter}
}

@article{CCM,
author = {David Avnir  and Ofer Biham  and Daniel Lidar  and Ofer Malcai },
title = {{Is the Geometry of Nature Fractal?}},
journal = {Science},
volume = {279},
number = {5347},
pages = {39-40},
year = {1998},
doi = {10.1126/science.279.5347.39},
URL = {https://www.science.org/doi/abs/10.1126/science.279.5347.39}}

@article{Hausdorff,
author = {Buda, Andrzej B. and Mislow, Kurt},
title = {{A Hausdorff chirality measure}},
journal = {J. Am. Chem. Soc.},
volume = {114},
number = {15},
pages = {6006-6012},
year = {1992},
doi = {10.1021/ja00041a016},
URL = {https://doi.org/10.1021/ja00041a016}
}

@article{Toroidal,
	author = {Hayami ,Satoru and Kusunose ,Hiroaki},
	doi = {10.7566/JPSJ.87.033709},
	journal = {J. Phys. Soc. Jpn.},
	number = {3},
	pages = {033709},
	title = {{Microscopic Description of Electric and Magnetic Toroidal Multipoles in Hybrid Orbitals}},
	url = {https://doi.org/10.7566/JPSJ.87.033709},
	volume = {87},
	year = {2018}}

@article{Multipole_review,
	author = {S. Hayami and H. Kusunose},
	doi = {10.7566/JPSJ.93.072001},
	journal = {J. Phys. Soc. Jpn.},
	number = {7},
	pages = {072001},
	title = {{Unified Description of Electronic Orderings and Cross Correlations by Complete Multipole Representation}},
	url = {https://doi.org/10.7566/JPSJ.93.072001},
	volume = {93},
	year = {2024}}

@article{Inda,
	author = {Inda, A. and Oiwa, R. and Hayami, S. and Yamamoto, H. M. and Kusunose, H.},
	doi = {10.1063/5.0204254},
	issn = {0021-9606},
	journal = {J. Chem. Phys.},
	month = {05},
	number = {18},
	pages = {184117},
	title = {{Quantification of chirality based on electric toroidal monopole}},
	url = {https://doi.org/10.1063/5.0204254},
	volume = {160},
	year = {2024}}

@article{Oiwa_Kusunose_ET,
	author = {Oiwa, Rikuto and Kusunose, Hiroaki},
	doi = {10.1103/1zq8-pqh8},
	issue = {3},
	journal = {Phys. Rev. Res.},
	month = {Sep},
	numpages = {15},
	pages = {033250},
	publisher = {American Physical Society},
	title = {{Predominant electronic order parameter for structural chirality: Role of spinless electronic toroidal multipoles in Te and Se}},
	url = {https://link.aps.org/doi/10.1103/1zq8-pqh8},
	volume = {7},
	year = {2025}}

@ARTICLE{Hoshino_chiral,
  title    = "{{Spin-Derived Electric Polarization and Chirality Density Inherent in Localized Electron Orbitals}}",
  author   = "Hoshino, Shintaro and Suzuki, Michi-To and Ikeda, Hiroaki",
  journal  = "Phys. Rev. Lett.",
  volume   =  130,
  number   =  25,
  pages    =  256801,
  month    =  jun,
  year     =  2023,
  doi      = {10.1103/PhysRevLett.130.256801}
}

@article{Moffat1992,
    author = {Moffatt, Henry Keith and Ricca, Renzo L.},
    title = {{Helicity and the C\u{a}lug\u{a}reanu invariant}},
    journal = {Proc. R. Soc. London A},
    volume = {439},
    number = {1906},
    pages = {411-429},
    year = {1992},
    month = {11},
    doi = {10.1098/rspa.1992.0159},
    url = {https://doi.org/10.1098/rspa.1992.0159}
}

@article{Moffat2014,
author = {H. Keith Moffatt },
title = {{Helicity and singular structures in fluid dynamics}},
journal = {Proc. Natl. Acad. Sci. U.S.A.},
volume = {111},
number = {10},
pages = {3663-3670},
year = {2014},
doi = {10.1073/pnas.1400277111},
URL = {https://www.pnas.org/doi/abs/10.1073/pnas.1400277111}
}

@article{Helicity,
  title = {{Structural chirality measurements and computation of handedness in periodic solids}},
  author = {G\'omez-Ortiz, Fernando and Fava, Mauro and McCabe, Emma E. and Romero, Aldo H. and Bousquet, Eric},
  journal = {Phys. Rev. B},
  volume = {110},
  issue = {17},
  pages = {174112},
  numpages = {9},
  year = {2024},
  month = {Nov},
  publisher = {American Physical Society},
  doi = {10.1103/PhysRevB.110.174112},
  url = {https://link.aps.org/doi/10.1103/PhysRevB.110.174112}
}

@article{KNO_experi,
	author = {Duri{\v s}, Katarina and M{\"u}ller, Ulrich and Jansen, Martin},
	doi = {https://doi.org/10.1002/zaac.201100511},
	journal = {Z. Anorg. Allg. Chem.},
	number = {5},
	pages = {737-743},
	title = {{K$_3$NiO$_2$ Revisited, Phase Transition and Crystal Structure Refinement}},
	url = {https://onlinelibrary.wiley.com/doi/abs/10.1002/zaac.201100511},
	volume = {638},
	year = {2012}}

@article{CsCuCl,
doi = {10.1088/0953-8984/8/38/010},
url = {https://doi.org/10.1088/0953-8984/8/38/010},
year = {1996},
month = {sep},
publisher = {},
volume = {8},
number = {38},
pages = {7059},
author = {Koiso, T and Yamamoto, K and Hata, Y and Takahashi, Y and Kita, E and Ohshima, K and Okamura, F P},
title = {{Determination of the chiral structure of  using anomalous x-ray scattering near the Cs K absorption edge}},
journal = {J. Phys.: Condens. Matter}
}

@article{MgTiO,
  title = {{Spin Singlet Formation in ${\mathrm{M}\mathrm{g}\mathrm{T}\mathrm{i}}_{2}{\mathrm{O}}_{4}$: Evidence of a Helical Dimerization Pattern}},
  author = {Schmidt, M. and Ratcliff, W. and Radaelli, P. G. and Refson, K. and Harrison, N. M. and Cheong, S. W.},
  journal = {Phys. Rev. Lett.},
  volume = {92},
  issue = {5},
  pages = {056402},
  numpages = {4},
  year = {2004},
  month = {Feb},
  publisher = {American Physical Society},
  doi = {10.1103/PhysRevLett.92.056402},
  url = {https://link.aps.org/doi/10.1103/PhysRevLett.92.056402}
}

@article{Romao2024,
author = {Romao, Carl P. and Juraschek, Dominik M.},
title = {{Phonon-Induced Geometric Chirality}},
journal = {ACS Nano},
volume = {18},
number = {43},
pages = {29550-29557},
year = {2024},
doi = {10.1021/acsnano.4c05978},
URL = {https://doi.org/10.1021/acsnano.4c05978}
}

@article{photo_induced_chirality,
author = {Z. Zeng  and M. F{\"o}rst and M. Fechner  and M. Buzzi  and E. B. Amuah  and C. Putzke  and P. J. W. Moll  and D. Prabhakaran  and P. G. Radaelli  and A. Cavalleri },
title = {{Photo-induced chirality in a nonchiral crystal}},
journal = {Science},
volume = {387},
number = {6732},
pages = {431-436},
year = {2025},
doi = {10.1126/science.adr4713},
URL = {https://www.science.org/doi/abs/10.1126/science.adr4713}}

@article{Juraschek2025,
  author = {Juraschek, Dominik M. and
            Geilhufe, R. Matthias and
            Zhu, Hanyu and
            Basini, Martina and
            Baum, Peter and
            Baydin, Andrey and
            Chaudhary, Swati and
            Fechner, Michael and
            Flebus, Benedetta and
            Grissonnanche, Gael and
            Kirilyuk, Andrei I. and
            Lemeshko, Mikhail and
            Maehrlein, Sebastian F. and
            Mignolet, Maxime and
            Murakami, Shuichi and
            Niu, Qian and
            Nowak, Ulrich and
            Romao, Carl P. and
            Rostami, Habib and
            Satoh, Takuya and
            Spaldin, Nicola A. and
            Ueda, Hiroki and
            Zhang, Lifa},
  title   = {{Chiral phonons}},
  journal = {Nat. Phys.},
  year    = {2025},
  volume  = {21},
  number  = {10},
  pages   = {1532--1540},
  doi     = {10.1038/s41567-025-03001-9},
  url     = {https://doi.org/10.1038/s41567-025-03001-9},
  issn    = {1745-2481}
}

@article{Te_Hirayama,
	author = {Hirayama, Motoaki and Okugawa, Ryo and Ishibashi, Shoji and Murakami, Shuichi and Miyake, Takashi},
	doi = {10.1103/PhysRevLett.114.206401},
	issue = {20},
	journal = {Phys. Rev. Lett.},
	month = {May},
	numpages = {5},
	pages = {206401},
	publisher = {American Physical Society},
	title = {{Weyl Node and Spin Texture in Trigonal Tellurium and Selenium}},
	url = {https://link.aps.org/doi/10.1103/PhysRevLett.114.206401},
	volume = {114},
	year = {2015}}

@article{Bernu,
  title = {{Signature of N\'eel order in exact spectra of quantum antiferromagnets on finite lattices}},
  author = {Bernu, B. and Lhuillier, C. and Pierre, L.},
  journal = {Phys. Rev. Lett.},
  volume = {69},
  issue = {17},
  pages = {2590--2593},
  numpages = {0},
  year = {1992},
  month = {Oct},
  publisher = {American Physical Society},
  doi = {10.1103/PhysRevLett.69.2590},
  url = {https://link.aps.org/doi/10.1103/PhysRevLett.69.2590}
}

@article{Capriotti,
  title = {{Long-Range N\'eel Order in the Triangular Heisenberg Model}},
  author = {Capriotti, Luca and Trumper, Adolfo E. and Sorella, Sandro},
  journal = {Phys. Rev. Lett.},
  volume = {82},
  issue = {19},
  pages = {3899--3902},
  numpages = {0},
  year = {1999},
  month = {May},
  publisher = {American Physical Society},
  doi = {10.1103/PhysRevLett.82.3899},
  url = {https://link.aps.org/doi/10.1103/PhysRevLett.82.3899}
}

@article{MnSi,
	author = {S. M{\"u}hlbauer and B. Binz and F. Jonietz and C. Pfleiderer and A. Rosch and A. Neubauer and R. Georgii and P. B{\"o}ni},
	doi = {10.1126/science.1166767},
	journal = {Science},
	number = {5916},
	pages = {915-919},
	title = {{Skyrmion Lattice in a Chiral Magnet}},
	url = {https://www.science.org/doi/abs/10.1126/science.1166767},
	volume = {323},
	year = {2009}}

@article{FeGe,
	author = {Yu, X. Z. and Kanazawa, N. and Onose, Y. and Kimoto, K. and Zhang, W. Z. and Ishiwata, S. and Matsui, Y. and Tokura, Y.},
	date = {2011/02/01},
	doi = {10.1038/nmat2916},
	id = {Yu2011},
	isbn = {1476-4660},
	journal = {Nat. Mater.},
	number = {2},
	pages = {106--109},
	title = {{Near room-temperature formation of a skyrmion crystal in thin-films of the helimagnet FeGe}},
	url = {https://doi.org/10.1038/nmat2916},
	volume = {10},
	year = {2011}}

@article{CoSi,
doi = {10.1088/1361-648X/aab0ba},
url = {https://doi.org/10.1088/1361-648X/aab0ba},
year = {2018},
month = {mar},
publisher = {IOP Publishing},
volume = {30},
number = {13},
pages = {135501},
author = {Pshenay-Severin, D A and Ivanov, Y V and Burkov, A A and Burkov, A T},
title = {{Band structure and unconventional electronic topology of CoSi}},
journal = {J. Phys.: Condens. Matter}
}

@article{Barry_science,
author = {Barry Bradlyn  and Jennifer Cano  and Zhijun Wang  and M. G. Vergniory  and C. Felser  and R. J. Cava  and B. Andrei Bernevig },
title = {{Beyond Dirac and Weyl fermions: Unconventional quasiparticles in conventional crystals}},
journal = {Science},
volume = {353},
number = {6299},
pages = {aaf5037},
year = {2016},
doi = {10.1126/science.aaf5037},
URL = {https://www.science.org/doi/abs/10.1126/science.aaf5037}}

@article{Kitou_chiral,
	author = {Kitou, Shunsuke and Gen, Masaki and Nakamura, Yuiga and Tokunaga, Yusuke and Arima, Taka-hisa},
	doi = {10.1021/acs.chemmater.4c00118},
	journal = {Chem. Mater.},
	number = {6},
	pages = {2993-2999},
	title = {{Cluster Rearrangement by Chiral Charge Order in Lacunar Spinel GaNb$_4$Se$_8$}},
	url = {https://doi.org/10.1021/acs.chemmater.4c00118},
	volume = {36},
	year = {2024},
	}

@article{KNO,
	author = {Fava, Mauro and McCabe, Emma and Romero, Aldo H. and Bousquet, Eric},
	doi = {10.1103/PhysRevB.111.174102},
	issue = {17},
	journal = {Phys. Rev. B},
	month = {May},
	numpages = {9},
	pages = {174102},
	publisher = {American Physical Society},
	title = {{Phonon-driven mechanism for the chiral phase transition of ${\mathrm{K}}_{3}{\mathrm{NiO}}_{2}$}},
	url = {https://link.aps.org/doi/10.1103/PhysRevB.111.174102},
	volume = {111},
	year = {2025}}

@article{SrSi2,
author = {Pringle, G. E.},
title = {{The structure of SrSi$_2$: a crystal of class O (432)}},
journal = {Acta Cryst. B},
year = "1972",
volume = "28",
number = "8",
pages = "2326--2328",
month = "Aug",
doi = {10.1107/S0567740872006053},
url = {https://doi.org/10.1107/S0567740872006053},
}

@article{beta-Mn,
author = {G.D. Preston},
title = {{The crystal structure of $\beta$-manganese }},
journal = {Philos. Mag.},
volume = {5},
number = {33},
pages = {1207--1225},
year = {1928},
publisher = {Taylor ¥& Francis},
doi = {10.1080/14786440608564570},
URL = {https://doi.org/10.1080/14786440608564570}
}

@article{Talanov2016,
  author  = {Talanov, V. M. and Talanov, M. V. and Shirokov, V. B.},
  title   = {{Theory of the formation of $P4_132$ ($P4_332$) phase spinels}},
  journal = {Crystallogr. Rep.},
  volume  = {61},
  number  = {2},
  pages   = {159--169},
  year    = {2016},
  doi     = {10.1134/S1063774516020280}
}

@article{BornMayer1932,
  author  = {Born, Max and Mayer, Joseph E.},
  title   = {Zur Gittertheorie der Ionenkristalle},
  journal = {Z. Phys.},
  year    = {1932},
  volume  = {75},
  number  = {1},
  pages   = {1--18},
  doi     = {10.1007/BF01340511},
  url     = {https://doi.org/10.1007/BF01340511},
  issn    = {0044-3328},
}

@article{slater_koster,
	author = {Slater, J. C. and Koster, G. F.},
	doi = {10.1103/PhysRev.94.1498},
	issue = {6},
	journal = {Phys. Rev.},
	month = {Jun},
	numpages = {0},
	pages = {1498--1524},
	publisher = {American Physical Society},
	title = {{Simplified LCAO Method for the Periodic Potential Problem}},
	url = {https://link.aps.org/doi/10.1103/PhysRev.94.1498},
	volume = {94},
	year = {1954}}

@article{Kubo_murakami,
  title={{Current-induced orbital and spin magnetizations in crystals with helical structure}},
  author={Yoda, Taiki and Yokoyama, Takehito and Murakami, Shuichi},
  journal={Sci. Rep.},
  volume={5},
  number={1},
  pages={12024},
  year={2015},
  doi = {https://doi.org/10.1038/srep12024},
  publisher={{Nature Publishing Group UK London}}
}

@book{BradleyCracknell1972,
  author    = {C. J. Bradley and A. P. Cracknell},
  title     = {{The Mathematical Theory of Symmetry in Solids: Representation Theory for Point Groups and Space Groups}},
  publisher = {Oxford University Press},
  year      = {1972},
  address    = {Oxford},
  isbn      = {978-0199582587},
  doi       = {10.1093/oso/9780199582587.001.0001}
}

@book{gunron,
  title={{Group Theory and Its Applications in Physics}},
  author={Inui, T. and Tanabe, Y. and Onodera, Y.},
  isbn={978-3-540-60445-7},
  series={Springer series in solid-state sciences},
  year={1990},
  publisher={Springer Berlin, Heidelberg},
  doi = {10.1007/978-3-642-80021-4}
}

@article{Tokura_review,
  title = {{Metal-insulator transitions}},
  author = {Imada, Masatoshi and Fujimori, Atsushi and Tokura, Yoshinori},
  journal = {Rev. Mod. Phys.},
  volume = {70},
  issue = {4},
  pages = {1039--1263},
  numpages = {0},
  year = {1998},
  month = {Oct},
  publisher = {American Physical Society},
  doi = {10.1103/RevModPhys.70.1039},
  url = {https://link.aps.org/doi/10.1103/RevModPhys.70.1039}
}

@article{tokura_nagaosa_orbital,
author = {Y. Tokura  and N. Nagaosa },
title = {{Orbital Physics in Transition-Metal Oxides}},
journal = {Science},
volume = {288},
number = {5465},
pages = {462-468},
year = {2000},
doi = {10.1126/science.288.5465.462},
URL = {https://www.science.org/doi/abs/10.1126/science.288.5465.462},
}

@article{Santini,
  title = {{Multipolar interactions in $f$-electron systems: The paradigm of actinide dioxides}},
  author = {Santini, Paolo and Carretta, Stefano and Amoretti, Giuseppe and Caciuffo, Roberto and Magnani, Nicola and Lander, Gerard H.},
  journal = {Rev. Mod. Phys.},
  volume = {81},
  issue = {2},
  pages = {807--863},
  numpages = {0},
  year = {2009},
  month = {Jun},
  publisher = {American Physical Society},
  doi = {10.1103/RevModPhys.81.807},
  url = {https://link.aps.org/doi/10.1103/RevModPhys.81.807}
}

@article{Cowley,
author = {R.A. Cowley},
title = {{Structural phase transitions I. Landau theory}},
journal = {Adv. Phys.},
volume = {29},
number = {1},
pages = {1--110},
year = {1980},
publisher = {Taylor ¥& Francis},
doi = {10.1080/00018738000101346},
URL = {https://doi.org/10.1080/00018738000101346}
}

@misc{gomezortiz2026,
  title={Chirality Cannot Be Ferroic in Enantiomorphic Space-Groups},
  author={F. G{\'o}mez-Ortiz and S. Mamoudou Taganga and E. E. McCabe and A. H. Romero and E. Bousquet},
  year={2026},
  eprint={2603.22501},
  archivePrefix={arXiv},
  primaryClass={cond-mat.mtrl-sci},
  url={https://arxiv.org/abs/2603.22501},
}

@article{orbital_magne,
  title = {{Berry phase, hyperorbits, and the Hofstadter spectrum: Semiclassical dynamics in magnetic Bloch bands}},
  author = {Chang, Ming-Che and Niu, Qian},
  journal = {Phys. Rev. B},
  volume = {53},
  issue = {11},
  pages = {7010--7023},
  numpages = {0},
  year = {1996},
  month = {Mar},
  publisher = {American Physical Society},
  doi = {10.1103/PhysRevB.53.7010},
  url = {https://link.aps.org/doi/10.1103/PhysRevB.53.7010}
}

@article{Rikken1997MChD,
  author = {Rikken, G. L. J. A. and Raupach, E.},
  title = {{Observation of magneto-chiral dichroism}},
  journal = {Nature},
  volume = {390},
  number = {6659},
  pages = {493--494},
  year = {1997},
  doi = {10.1038/37323}
}

@article{Rikken2001EMChA,
  author = {Rikken, G. L. J. A. and F{\"o}lling, J. and Wyder, P.},
  title = {{Electrical Magnetochiral Anisotropy}},
  journal = {Phys. Rev. Lett.},
  volume = {87},
  pages = {236602},
  year = {2001},
  doi = {10.1103/PhysRevLett.87.236602}
}

@article{Pop2014EMChA,
  author = {Pop, Flavia and Auban-Senzier, Pascale and Canadell, Enric and Rikken, Geert L. J. A. and Avarvari, Narcis},
  title = {{Electrical magnetochiral anisotropy in a bulk chiral molecular conductor}},
  journal = {Nat. Commun.},
  volume = {5},
  pages = {3757},
  year = {2014},
  doi = {10.1038/ncomms4757}
}

@article{Naaman2012CISS,
  author = {Naaman, Ron and Waldeck, David H.},
  title = {{Chiral-Induced Spin Selectivity Effect}},
  journal = {J. Phys. Chem. Lett.},
  volume = {3},
  number = {16},
  pages = {2178--2187},
  year = {2012},
  doi = {10.1021/jz300793y}
}

@article{Kishine2020ChiralPhonon,
  author = {Kishine, J. and Ovchinnikov, A. S. and Tereshchenko, A. A.},
  title = {{Chirality-Induced Phonon Dispersion in a Noncentrosymmetric Micropolar Crystal}},
  journal = {Phys. Rev. Lett.},
  volume = {125},
  pages = {245302},
  year = {2020},
  doi = {10.1103/PhysRevLett.125.245302}
}

@article{Tateishi2025JPSJ,
  author = {Tateishi, Tomomi and Kato, Akihito and Kishine, Jun-ichiro},
  title = {{Electron-Chiral Phonon Coupling, Crystal Angular Momentum, and Phonon Chirality}},
  journal = {J. Phys. Soc. Jpn.},
  volume = {94},
  pages = {053601},
  year = {2025},
  doi = {10.7566/JPSJ.94.053601}
}

@article{roy2022long,
  title={{Long-range current-induced spin accumulation in chiral crystals}},
  author={Roy, Arunesh and Cerasoli, Frank T and Jayaraj, Anooja and Tenzin, Karma and Nardelli, Marco Buongiorno and S{\l}awi{\'n}ska, Jagoda},
  journal={npj Comput. Mater.},
  volume={8},
  number={1},
  pages={243},
  year={2022},
  doi = {https://doi.org/10.1038/s41524-022-00931-3},
  publisher={Nature Publishing Group UK London}
}

@article{Inui2020ChiralSpinPolarization,
  author = {Inui, Akito and Aoki, Ryo and Nishiue, Yasuhiro and Shiota, Yoichi and Kousaka, Yusuke and Shishido, Hiroaki and Hirobe, Daichi and Suda, Masayuki and Ohe, Jun-ichiro and Kishine, Jun-ichiro and Yamamoto, Hiroshi M. and Togawa, Yoshihiko},
  title = {{Chirality-Induced Spin-Polarized State of a Chiral Crystal}},
  journal = {Phys. Rev. Lett.},
  volume = {124},
  pages = {166602},
  year = {2020},
  doi = {10.1103/PhysRevLett.124.166602}
}

@book{BarronBook,
  author    = {Barron, Laurence D.},
  title     = {{Molecular Light Scattering and Optical Activity}},
  edition   = {2},
  publisher = {Cambridge University Press},
  address   = {Cambridge},
  year      = {2004},
  doi       = {10.1017/CBO9780511535468},
  isbn      = {9780511535468}
}

@article{Barron1986,
  author  = {Barron, L. D.},
  title   = {{True and false chirality and parity violation}},
  journal = {Chem. Phys. Lett.},
  volume  = {123},
  number  = {5},
  pages   = {423--427},
  year    = {1986},
  doi     = {10.1016/0009-2614(86)80035-5}
}

@article{Buda1992QuantifyingChirality,
  author  = {Buda, Andrzej B. and Auf der Heyde, Thomas and Mislow, Kurt},
  title   = {{On Quantifying Chirality}},
  journal = {Angew. Chem. Int. Ed.},
  volume  = {31},
  number  = {8},
  pages   = {989--1007},
  year    = {1992},
  doi     = {10.1002/anie.199209891}
}

@article{Zabrodsky1995CCM,
  author  = {Zabrodsky, Hagit and Avnir, David},
  title   = {{Continuous Symmetry Measures. 4. Chirality}},
  journal = {J. Am. Chem. Soc.},
  volume  = {117},
  number  = {1},
  pages   = {462--473},
  year    = {1995},
  doi     = {10.1021/ja00106a053}
}

@article{Kishine2022DefinitionChirality,
  author  = {Kishine, Jun-ichiro and Kusunose, Hiroaki and Yamamoto, Hiroshi M.},
  title   = {{On the Definition of Chirality and Enantioselective Fields}},
  journal = {Isr. J. Chem.},
  volume  = {62},
  number  = {11--12},
  pages   = {e202200049},
  year    = {2022},
  doi     = {10.1002/ijch.202200049}
}

@article{Hayashida2021Ferrochiral,
  author = {Hayashida, Takeshi and Kimura, Kenta and Urushihara, Daisuke and Asaka, Toru and Kimura, Tsuyoshi},
  title = {{Observation of Ferrochiral Transition Induced by an Antiferroaxial Ordering of Antipolar Structural Units in Ba(TiO)Cu$_4$(PO$_4$)$_4$}},
  journal = {J. Am. Chem. Soc.},
  volume = {143},
  number = {9},
  pages = {3638--3646},
  year = {2021},
  doi = {10.1021/jacs.1c00391}
}

@article{Ishito_Te,
  author = {Ishito, Kyosuke and Mao, Huiling and Kobayashi, Kaya and Kousaka, Yusuke and Togawa, Yoshihiko and Kusunose, Hiroaki and Kishine, Jun-ichiro and Satoh, Takuya},
  title = {{Chiral phonons: Circularly polarized Raman spectroscopy and ab initio calculations in a chiral crystal tellurium}},
  journal = {Chirality},
  volume = {35},
  number = {6},
  pages = {338--345},
  year = {2023},
  doi = {10.1002/chir.23544}
}

@article{PathwaysChirality2025,
  title = {Pathways to crystal chirality: An algorithm to identify displacive chiral phase transitions},
  author = {G\'omez-Ortiz, Fernando and Romero, Aldo H. and Bousquet, Eric},
  journal = {Phys. Rev. B},
  volume = {112},
  issue = {1},
  pages = {014107},
  numpages = {11},
  year = {2025},
  month = {Jul},
  publisher = {American Physical Society},
  doi = {10.1103/n4sh-7x7c},
  url = {https://link.aps.org/doi/10.1103/n4sh-7x7c}
}

@article{IshitobiHattori2026PEC,
  author = {Ishitobi, Takayuki and Hattori, Kazumasa},
  title = {{Purely Electronic Chirality without Structural Chirality}},
  journal = {Phys. Rev. Lett.},
  volume = {136},
  number = {5},
  year = {2026},
  pages = {056402},
  doi = {10.1103/dc1q-xzbd},
  url = {https://link.aps.org/doi/10.1103/dc1q-xzbd}
}

@ARTICLE{Matsubara2025,
  title     = {{Spin-flip scattering at a chiral interface of helical chains}},
  author    = {Matsubara, Keita and Hattori, Kazumasa},
  journal   = {J. Phys. Soc. Jpn.},
  volume    =  {94},
  number    =  {5},
  year      =  {2025},
  pages     = {054706},
  doi       = {10.7566/JPSJ.94.054706}
}

@ARTICLE{HattoriTsunetsugu2016,
  title     = {{Classical Monte Carlo Study for Antiferro Quadrupole Orders in a Diamond Lattice}},
  author    = {Hattori, Kazumasa and Hirokazu Tsunetsugu},
  journal   = {J. Phys. Soc. Jpn.},
  volume    =  {85},
  year      =  {2016},
  pages     = {094001},
  doi       = {10.7566/JPSJ.85.094001}
}

@ARTICLE{PrimarySecondaryOP2024,
  title     = {{Primary and secondary order parameters in the fully frustrated transverse-field Ising model on the square lattice}},
  author    = {Gabe Schumm and  Hui Shao and Wenan Guo and Fr\'ed\'ric Mila and Anders W. Sandvik},
  journal   = {Phys. Rev. B},
  volume    =  {109},
  year      =  {2024},
  pages     =  {L140408},
  doi       = {10.1103/PhysRevB.109.L140408}
}

@ARTICLE{Miki2025ElectronicAsymmetry,
  title     ={{Quantification of Electronic Asymmetry: Chirality and Axiality in Solids}},
  author    = {Tatsuya Miki and Hiroaki Ikeda and Michi-To Suzuki and Shintaro Hoshino},
  journal   = {Phys. Rev. Lett.},
  volume    =  {134},
  year      =  {2025},
  pages     =  {226401},
  doi       = {10.1103/PhysRevLett.134.226401}
}

@article{Harris1999ChiralParameters,
  author = {Harris, A. B. and Kamien, R. D. and Lubensky, T. C.},
  title = {{Molecular chirality and chiral parameters}},
  journal = {Rev. Mod. Phys.},
  volume = {71},
  pages = {1745},
  year = {1999},
  doi = {10.1103/RevModPhys.71.1745}
}

@article{Lee1984,
  author  = {Lee, D. H. and Joannopoulos, J. D. and Negele, J. W. and Landau, D. P.},
  title   = {{Discrete-Symmetry Breaking and Novel Critical Phenomena in an Antiferromagnetic Planar ({XY}) Model in Two Dimensions}},
  journal = {Phys. Rev. Lett.},
  volume  = {52},
  pages   = {433--436},
  year    = {1984},
  doi     = {10.1103/PhysRevLett.52.433}
}

@article{MiyashitaShiba1984,
  author  = {Miyashita, Seiji and Shiba, Hiroyuki},
  title   = {{Nature of the Phase Transition of the Two-Dimensional Antiferromagnetic Plane Rotator Model on the Triangular Lattice}},
  journal = {J. Phys. Soc. Jpn.},
  volume  = {53},
  pages   = {1145},
  year    = {1984},
  doi     = {10.1143/JPSJ.53.1145}
}

@article{ObuchiKawamura2012,
  author  = {Obuchi, Tomoyuki and Kawamura, Hikaru},
  title   = {{Spin and Chiral Orderings of the Antiferromagnetic {XY} Model on the Triangular Lattice and Their Critical Properties}},
  journal = {J. Phys. Soc. Jpn.},
  volume  = {81},
  pages   = {054003},
  year    = {2012},
  doi     = {10.1143/JPSJ.81.054003}
}

@ARTICLE{Hattori2024,
  title     = "{{Orbital moir\'e and quadrupolar triple-$q$ physics in a triangular lattice}}",
  author    = "Hattori, Kazumasa and Ishitobi, Takayuki and Tsunetsugu, Hirokazu",
  journal   = "Phys. Rev. Res.",
  publisher = "American Physical Society (APS)",
  volume    =  6,
  number    =  4,
  pages     =  {L042068},
  month     =  dec,
  year      =  2024,
  doi       = {10.1103/physrevresearch.6.l042068}
}

@article{G-Furuya,
  title = {Order parameter fluctuation effects on current-induced magnetization},
  author = {Furuya, Genta and Hattori, Kazumasa},
  journal = {Phys. Rev. B},
  volume = {112},
  issue = {3},
  pages = {035171},
  numpages = {10},
  year = {2025},
  month = {Jul},
  publisher = {American Physical Society},
  doi = {10.1103/3r8w-76lf},
  url = {https://link.aps.org/doi/10.1103/3r8w-76lf}
}

@article{PAM_Zhang,
  title = {Chiral Phonons at High-Symmetry Points in Monolayer Hexagonal Lattices},
  author = {Zhang, Lifa and Niu, Qian},
  journal = {Phys. Rev. Lett.},
  volume = {115},
  issue = {11},
  pages = {115502},
  numpages = {5},
  year = {2015},
  month = {Sep},
  publisher = {American Physical Society},
  doi = {10.1103/PhysRevLett.115.115502},
  url = {https://link.aps.org/doi/10.1103/PhysRevLett.115.115502}
}

@article{Miki_chiral,
  title = {Electron chirality and hydrodynamic helicity: Analysis in the atomic limit},
  author = {Miki, Tatsuya and Kakinuma, Yuta and Senami, Masato and Fukuda, Masahiro and Suzuki, Michi-To and Ikeda, Hiroaki and Hoshino, Shintaro},
  journal = {Phys. Rev. B},
  volume = {113},
  issue = {23},
  pages = {235127},
  numpages = {17},
  year = {2026},
  month = {Jun},
  publisher = {American Physical Society},
  doi = {10.1103/ghyy-mynx},
  url = {https://link.aps.org/doi/10.1103/ghyy-mynx}
}

@misc{piezochiral,
      title={The Piezochiral Effect}, 
      author={Z. Zeng and M. F\"{o}rst and M. Fechner and X. Deng and A. Cavalleri and P. G. Radaelli},
      year={2025},
      eprint={2510.21674},
      archivePrefix={arXiv},
      primaryClass={cond-mat.mtrl-sci},
      url={https://arxiv.org/abs/2510.21674},
}

@misc{ISODISTORT,
  author = {H. T. Stokes and D. M. Hatch and B. J. Campbell},
  title = {{ISODISTORT}, {ISOTROPY} {Software Suite}},
  howpublished = {\url{https://iso.byu.edu}}
}

@article{ISODISPLACE,
  author = {Campbell, B. J. and Stokes, H. T. and Tanner, D. E. and Hatch, D. M.},
  title = {{ISODISPLACE}: {An Internet Tool for Exploring Structural Distortions}},
  journal = {J. Appl. Cryst.},
  volume = {39},
  pages = {607--614},
  year = {2006},
  doi = {10.1107/S0021889806014075}
}

@article{REPRES,
  author = {Aroyo, M. I. and Kirov, A. and Capillas, C. and Perez-Mato, J. M. and Wondratschek, H.},
  title = {Bilbao Crystallographic Server. {II}. Representations of crystallographic point groups and space groups},
  journal = {Acta Crystallogr. A},
  volume = {62},
  pages = {115--128},
  year = {2006},
  doi = {10.1107/S0108767305040286}
}

@article{Okamoto2020,
  author    = {Okamoto, Yoshihiko and Amano, Haruki and Katayama, Naoyuki and Sawa, Hiroshi and Niki, Kenta and Mitoka, Rikuto and Harima, Hisatomo and Hasegawa, Takumi and Ogita, Norio and Tanaka, Yu and Takigawa, Masashi and Yokoyama, Yasunori and Takehana, Kanji and Imanaka, Yasutaka and Nakamura, Yuto and Kishida, Hideo and Takenaka, Koshi},
  title     = {{Regular-triangle trimer and charge order preserving the {Anderson} condition in the pyrochlore structure of {CsW$_2$O$_6$}}},
  journal   = {Nature Communications},
  volume    = {11},
  number    = {1},
  pages     = {3144},
  year      = {2020},
  doi       = {10.1038/s41467-020-16873-7}
}

@ARTICLE{Harter2017,
  title    = {{A parity-breaking electronic nematic phase transition in the
              spin-orbit coupled metal {Cd}$_{2}${Re}$_{2}${O}$_{7}$}},
  author   = "Harter, J W and Zhao, Z Y and Yan, J-Q and Mandrus, D G and Hsieh,
              D",
  journal  = "Science",
  volume   =  356,
  number   =  6335,
  pages    = "295--299",
  month    =  apr,
  year     =  2017,
  doi      =  {10.1126/science.aad1188}
}

@ARTICLE{Sodemann2015,
  title     = {{Quantum nonlinear Hall effect induced by Berry curvature dipole
               in time-reversal invariant materials}},
  author    = "Sodemann, Inti and Fu, Liang",
  journal   = "Phys. Rev. Lett.",
  publisher = "American Physical Society (APS)",
  volume    =  115,
  number    =  21,
  pages     =  216806,
  month     =  nov,
  year      =  2015,
  doi       = {10.1103/PhysRevLett.115.216806}
}

@ARTICLE{DiMatteo2017,
  title     = {{Nature of the tensor order in {Cd}$_{2}${Re}$_{2}${O}$_{7}$}},
  author    = "Di Matteo, S and Norman, M R",
  journal   = "Phys. Rev. B",
  publisher = "American Physical Society",
  volume    =  96,
  number    =  11,
  pages     =  115156,
  month     =  sep,
  year      =  2017,
  doi       =  {10.1103/PhysRevB.96.115156}
}

@ARTICLE{Hayami2019,
  title    = {{Electric Toroidal Quadrupoles in the Spin-Orbit-Coupled Metal
              Cd$_2$Re$_2$O$_7$}},
  author   = "Hayami, Satoru and Yanagi, Yuki and Kusunose, Hiroaki and Motome,
              Yukitoshi",
  journal  = "Phys. Rev. Lett.",
  volume   =  122,
  number   =  14,
  pages    =  147602,
  month    =  apr,
  year     =  2019,
  doi      =  {10.1103/PhysRevLett.122.147602}
}

\end{document}